\documentclass[graybox,vecphys,natbib]{svmult}

\usepackage{mathptmx}       
\usepackage{helvet}         
\usepackage{courier}        
\usepackage{type1cm}        
\usepackage{makeidx}         
\usepackage{graphicx}        
\usepackage{multicol}        
\usepackage[bottom]{footmisc}

\makeindex             
\def \lesssim {\mathrel{\vcenter
     {\offinterlineskip \hbox{$<$}\hbox{$\sim$}}}}
\def \gtrsim {\mathrel{\vcenter
     {\offinterlineskip \hbox{$>$}\hbox{$\sim$}}}}
\newcommand{\qvec}[1]{\textbf{\textit{#1}}}

\begin{document}
\bibliographystyle{spbasic}

\title*{Neutrino-driven Explosions}
\label{Chapter:NeutrinoDrivenExplosions}
\author{Hans-Thomas Janka}
\institute{Hans-Thomas Janka \at Max Planck Institute for Astrophysics, 
Karl-Schwarzschild-Str.~1, 85748 Garching, Germany\\ \email{thj@mpa-garching.mpg.de}
}
%
%
\maketitle

\abstract*{The question why and how core-collapse supernovae (SNe)
explode is one of the central and most long-standing riddles 
of stellar astrophysics. Solving this problem is crucial for
deciphering the SN phenomenon, for predicting its observable
signals such as light curves and spectra, nucleosynthesis
yields, neutrinos, and gravitational waves, for defining the
role of SNe in the dynamical and chemo-dynamical evolution of
galaxies, and for explaining the birth conditions and properties 
of neutron stars (NSs) and stellar-mass black holes. Since the 
formation of such compact remnants releases over hundred times 
more energy in neutrinos than the kinetic energy of the SN explosion,
neutrinos can be the decisive agents for powering the SN outburst.
According to the standard paradigm of the neutrino-driven 
mechanism, the energy transfer by the intense neutrino
flux to the medium behind the stagnating core-bounce shock,
assisted by violent hydrodynamic mass motions (sometimes subsumed
by the term ``turbulence''), revives the outward shock motion and
thus initiates the SN explosion. Because of the weak coupling of
neutrinos in the region of this energy deposition, detailed,
multi-dimensional hydrodynamic models including neutrino transport 
and a wide variety of physics are needed to assess the viability of
the mechanism. Owing to advanced numerical codes and increasing
supercomputer power, considerable progress has been achieved in
our understanding of the physical processes that have to act in 
concert for the success of neutrino-driven explosions. First
studies begin to reveal observational implications and avenues
to test the theoretical picture by data from individual SNe 
and SN remnants but also from population-integrated observables.
While models will be further refined, a real breakthrough is 
expected through the next Galactic core-collapse SN, when neutrinos 
and gravitational waves can be used to probe the conditions deep
inside the dying star.
}

\abstract{
The question why and how core-collapse supernovae (SNe)
explode is one of the central and most long-standing riddles
of stellar astrophysics. Solving this problem is crucial for
deciphering the SN phenomenon, for predicting its observable
signals such as light curves and spectra, nucleosynthesis
yields, neutrinos, and gravitational waves, for defining the
role of SNe in the dynamical and chemo-dynamical evolution of
galaxies, and for explaining the birth conditions and properties
of neutron stars (NSs) and stellar-mass black holes. Since the
formation of such compact remnants releases over hundred times 
more energy in neutrinos than the kinetic energy of the SN explosion,
neutrinos can be the decisive agents for powering the SN outburst.
According to the standard paradigm of the neutrino-driven
mechanism, the energy transfer by the intense neutrino
flux to the medium behind the stagnating core-bounce shock,
assisted by violent hydrodynamic mass motions (sometimes subsumed
by the term ``turbulence''), revives the outward shock motion and
thus initiates the SN explosion. Because of the weak coupling of
neutrinos in the region of this energy deposition, detailed,
multi-dimensional hydrodynamic models including neutrino transport
and a wide variety of physics are needed to assess the viability of
the mechanism. Owing to advanced numerical codes and increasing
supercomputer power, considerable progress has been achieved in
our understanding of the physical processes that have to act in
concert for the success of neutrino-driven explosions. First
studies begin to reveal observational implications and avenues
to test the theoretical picture by data from individual SNe
and SN remnants but also from population-integrated observables.
While models will be further refined, a real breakthrough is
expected through the next Galactic core-collapse SN, when neutrinos
and gravitational waves can be used to probe the conditions deep
inside the dying star.
}

\section{Introduction}
\label{sec:intro}

The term supernovae (SNe) was used for the first time in 1934 in three 
seminal papers by
\citet{BaadeZwicky1934a,BaadeZwicky1934b,BaadeZwicky1934c}, where 
the authors envisioned that ``the super-nova process might occur
to every star once in its lifetime, marking perhaps the cessation
of its existence as an ordinary star.'' Moreover, they hypothesized
that these extremely luminous and energetic outbursts might signal
the ``transition of an ordinary star to a neutron star 
(NS)\index{neutron star (NS)},
consisting mainly of neutrons\index{neutron}.'' 
Baade and Zwicky came forward with
this truly visionary proposal at a time when the detailed nuclear
processes in stars were not yet known. They based their arguments
on the insight that the observed SN phenomenon releases
an energy equal to a considerable fraction of the star's rest mass,
and they suggested that this energy originates from the gravitational
binding energy of a very compact star that ``may possess a very
small radius and an extremely high density'', which ``may far 
exceed the ordinary nuclear packing fractions'' 
\citep{BaadeZwicky1934a}.

This basic picture developed by Baade and Zwicky is still the
foundation of our present notion of stellar death and the birth of
NSs. The existence of NSs as well as their association with
SN remnants is meanwhile firmly established by many observations,
and the detection of two dozen neutrinos\index{neutrino}
from SN~1987A\index{SN~1987A} in the
underground experiments of Kamiokande~II\index{Kamiokande~II} 
\citep{Hirata1987},
Irvine-Michigan-Brookhaven\index{Irvine-Michigan-Brookhaven} 
\citep[IMB;][]{Bionta1987}, and
Baksan\index{Baksan}
\citep{Alexeyev1988} was a direct confirmation of the 
scenario of stellar core collapse and hot NS formation.
In fact, the huge binding energy\index{binding energy}
released during this process, amounting approximately to
\begin{equation}
E_\mathrm{b} \sim E_\mathrm{g} \approx
\frac{3}{5}\,\frac{G M_\mathrm{ns}^2}{R_\mathrm{ns}}
\approx 3.6\times 10^{53}\,\left(\frac{M_\mathrm{ns}}{1.5\,M_\odot}\right)^{\! 2}
\left(\frac{R_\mathrm{ns}}{10\,\mathrm{km}}\right)^{\! -1}\,\,\mathrm{erg}
\label{eq:ebind}
\end{equation}
($M_\mathrm{ns}$ is the NS mass and $R_\mathrm{ns}$ the NS
radius in this Newtonian estimate for a homogeneous sphere),
is mostly radiated in an intense pulse of neutrinos and
antineutrinos, of which a few electron antineutrinos could be
captured in the laboratories on earth.
Less than one percent of the gravitational 
energy\index{gravitational energy} was
sufficient to power the observed stellar explosion.

\runinhead{Initial Idea: Energy Arguments}
Because neutrinos carry away 100 times more energy than the
kinetic energy of typical core-collapse SNe, these elementary
particles were proposed by \citet{Colgate1966} and 
\citet{Arnett1966,Arnett1967} as possible agents to drive the SN 
explosion. Although being trapped in the supra-nuclear interior 
of the newly formed NS for several seconds on average, 
neutrinos interact only by the weak force\index{weak force}
and therefore are able to leak out much faster
than photons, whose electromagnetic reaction cross sections with
matter are roughly 20 orders of magnitude larger. It is this
property that allows neutrinos to transport energy out from 
deeper regions inside the hot NS on relevant time scales and
to transfer some of this energy to the gravitationally more loosely
bound surrounding layers. If this energy deposition is strong 
enough, the infalling matter of the stellar core can be 
lifted out of the gravitational trough of the NS to be expelled
in the SN blast.

\runinhead{A Long and Winding Road of Learning}
It took more than 20 years of gradual improvements in the numerical 
treatments, new discoveries and insights on the microphysics
side, and major revisions in many aspects of the early
understanding of stellar core collapse, to advance the initially 
rather sketchy picture of the role of neutrinos in the SN
explosion to a more consistent framework that is close to our
modern view. On this way lay, for example, the introduction of 
multi-energy-group neutrino diffusion\index{neutrino diffusion}
in the numerics sector;
the discoveries of weak neutral currents and coherent 
scattering of neutrinos off nuclei; the recognition of the 
importance of an evolving, density and entropy dependent
composition of heavy nuclei including their
excited states, alpha particles, and free nucleons in nuclear
statistical equilibrium\index{nuclear statistical equilibrium}
during core collapse;
the finding that neutrinos get trapped\index{neutrino trapping}
in the stellar plasma
at a density of about $10^{12}$\,g\,cm$^{-3}$ during infall,
which renders the further evolution until core bounce 
essentially adiabatic; the insight that neutrino-electron
scattering\index{neutrino-electron scattering}  
as well as electron captures\index{electron captures}
on heavy nuclei besides
free protons are important for determining the electron fraction
$Y_e$\index{electron fraction}
(i.e., the electron-to-baryon ratio\index{electron-to-baryon ratio}) 
and the entropy\index{entropy} per
baryon that the infalling matter carries into the trapping regime;
and the understanding that the core collapse proceeds 
self-similarly\index{self-similarity}
with a homologous velocity profile\index{homologous velocity profile}
($v(r) \propto r$, $r$ being the
radius) in the subsonically contacting inner core, whose mass 
depends on the electron fraction ($M_\mathrm{ic}\propto Y_e^2$)
and determines the location at which the bounce-shock forms.

The space in this chapter is not sufficient to describe the
theoretical arguments and evidence that have led to progress
on these important aspects, whose evolution did not always
follow a straight path and simple chronological order. The
reader is referred to the excellent and exhaustive review by 
\citet{Bethe1990} for the developments from the early steps
of stellar core-collapse theory and explosion modeling 
in the mid 1960's until about 1990, and also to the later 
review by \citet{Janka2007}, which attempts to report the
new directions pursued afterwards and the corresponding 
updates of our knowledge.

\runinhead{The Delayed Neutrino-heating Mechanism}
With gradually improved numerical schemes and continuously
increasing sophistication of the microphysics treatment used in
core-collapse SN simulations, it became clear that the prompt 
bounce-shock mechanism, where the hydrodynamic shock front 
from the core bounce directly initiates the
SN explosion, cannot succeed for the progenitor models provided
by stellar evolution calculations. Instead of accelerating outward
through the star, the shock experiences severe energy losses 
by nuclear photodisintegration\index{nuclear photodisintegration}
of iron-group nuclei to free
nucleons. Additional energy losses occur via a very
short ($\sim$10\,milliseconds) but extremely luminous flash
of electron neutrinos at the moment when the shock
breaks out of the neutrinosphere\index{neutrinosphere}
into the lower-density, outer
regions of the iron core. As a consequence, the shock front stalls
at only 100--200\,km even before being able to reach
the surface of the stellar iron core. Some extra source of energy
must come to the aid of the stagnating shock\index{shock stagnation}. 

At this point, as a revival of the idea that was born in the 
mid 1960's, neutrino energy transfer came back into play through
simulations by \citet{wilson_85}, which provided new hints to
the possibility that neutrino heating could indeed lead to a
rejuvenation of the stalled shock, but at a much later stage
after core bounce than expected. The scenario of the ``delayed 
neutrino-heating mechanism''\index{delayed neutrino-heating mechanism} 
was born and was worked out conceptually
and quantitatively in some detail in a paper by \citet{BetheWilson1985}.
Some hundred milliseconds after shock 
stagnation\index{shock stagnation}, the
conditions between the newly formed NS and the stalled shock
front become favorable for efficient 
neutrino-energy deposition\index{neutrino-energy deposition},
because the temperatures in this region decrease, while the
NS heats up during its contraction and radiates neutrinos with 
increasingly harder spectra. If neutrinos are able to transfer
enough energy to the postshock medium, the rise in pressure is
able to accelerate the shock outwards against the ram pressure
of the surrounding, still collapsing stellar shells. 
In this mechanism the shock expansion is driven by the
neutrino-energy deposition (and not by momentum transfer!) 
to the stellar plasma. This is consistent with the fact that
the neutrino luminosities radiated by the nascent NS stay 
considerably below the Eddington limit\index{Eddington limit}. 

\runinhead{New Lessons from Supernova 1987A}
Wilson's delayed neutrino-driven mechanism has become the 
paradigm for explaining the explosions of the majority of 
all core-collapse SNe, although his results depended
on a number of uncertain assumptions and could not be 
reproduced by modern simulations with more refined numerical
tools and state-of-the-art input physics. Moreover, SN~1987A
has led to a major revision of the simple picture of 
spherically symmetric explosions, which Wilson and others
(even much later) tried to establish with 
one-dimensional simulations (assuming that all physical
variables depend only on the radius). SN~1987A, being the
closest SN for hundreds of years and thus being observed with
unprecedented detailedness, revealed
the spectacle of a stellar death from the earliest moments
of radiation emission on. Therefore, SN~1987A provided, for the
first time, unambiguous evidence that strong hydrodynamic
mixing processes\index{hydrodynamic mixing processes} 
played a role even already during the first
second of the explosion, when radioactive elements such as
$^{56}$Ni (which provides the energy for the long-lasting high 
luminosity of a SN) are nucleosynthesized. 

Indeed, the first multi-dimensional simulations, which could
be carried out shortly later in two spatial dimensions 
\citep[i.e.\
assuming, for simplicity, axisymmetry around an arbitrary
direction;][]{Herantetal1994,Burrowsetal1995,JankaMueller1994,JankaMueller1995,JankaMueller1996}, demonstrated that violent 
convective overturn develops in the neutrino-heated postshock
layer because of negative entropy gradients. These
hydrodynamic instabilities lead to non-radial flows that may
explain the observed asymmetries and radial mixing effects
in SN~1987A. Furthermore, they can also provide crucial support 
to the neutrino-driven mechanism, potentially solving the dilemma
that Wilson's first successes could not be confirmed by later
spherical models, in which the neutrino heating remained
too weak to revive the stalled SN shock.
 
\runinhead{A Challenging Problem Until Today}
The unexpected scientific insight, triggered by SN~1987A, that 
SN explosions are genericly non-spherical, still poses major
challenges for today's numerical modeling. It implies that 
three-dimensional (3D) simulations are needed to capture the
true nature of the phenomenon of stellar core collapse and 
explosion, and, in particular, of the physical mechanism that
initiates the SN blast. For 20 years since the mid 1990's, 
SN computations have struggled with the enormous demands of
multi-dimensional neutrino transport\index{neutrino transport}, 
and only very recently
the first simulations have become possible in full three
dimensions due to the growing supercomputer power and newly
developed, massively parallel simulation codes. 

The main algorithmic as well as computational challenges
are connected to 
the neutrino propagation in the six-dimensional phase space
(made up of three spatial and three momentum variables) and
to the complexity of the neutrino-matter interactions, which
require the coupling of neutrinos and antineutrinos over the 
whole phase space. Both aspects together turn neutrino
transport into an integro-differential 
problem\index{integro-differential problem} with high 
demands on numerical efficiency and computing capabilities.
Even at the present time, all SN simulations therefore still
use approximations in various aspects, and the solution of
the crucial problem of neutrino transport in six-dimensional 
phase space and full generality will require supercomputing
on the exascale level. Ultimately, this step will have to be
taken, shall numerical models convincingly demonstrate 
the viability of the neutrino-heating mechanism and make
quantitative predictions of its observational implications.

\runinhead{Lack of Convincing Alternatives}
Even today, after appreciable improvements on the modeling 
side and considerable progress in our understanding of 
the processes that play a role during stellar collapse
and explosion, the neutrino-driven mechanism is not yet
finally established as the solution of the SN problem.
Success has been reported for
stars near the low-mass end of the SN progenitors (i.e., stars
between roughly 8\,$M_\odot$ and about 10\,$M_\odot$
with O-Ne-Mg\index{O-Ne-Mg core} or Fe cores\index{Fe core}), 
which develop neutrino-driven
explosions quite readily. The predicted explosion properties
such as energy, radioactive nickel yield, 
and intermediate-mass nucleosynthesis, seem to agree with 
those of the Crab SN\index{Crab} and some extragalactic
subluminous SNe. For more massive progenitors, however,
the situation is still ambiguous. Although modern 3D 
simulations could obtain explosions and thus provide support
for the viability of the neutrino-heating mechanism in principle,
the models are not yet able to demonstrate that the mechanism
is robust and that it is able to explain SN explosions with 
the observed energies.

These results should be considered only as a preliminary step,
because more work is needed to further improve the modeling.
Interpretation of the remaining shortcomings and still open 
questions as fundamental weaknesses or even failure of the
neutrino-driven mechanism is not justified in view of the
extreme complexity of the problem and the enormous challenges
to perform realistic simulations in all details. In this
context it is also important to note that 
convincing alternatives based on 
well-justified assumptions concerning the relevant physics, 
do not exist. The most interesting alternative possibility
are magnetohydrodynamic 
explosions\index{magnetohydrodynamic explosion}, in which 
strong magnetic fields\index{magnetic field} 
play a crucial role in pushing the
SN shock. Magnetic fields that
thread a rapidly spinning stellar core
are amplified during collapse by compression, rotational
winding, and the magnetorotational 
instability\index{magnetorotational instability}
\citep[e.g.,][and references therein]{Akiyamaetal2003}, 
tapping the gravitational and rotational energy of the 
newly formed NS. In the shear layer between the
surface of the NS and the stalled SN shock, magnetic
pressure can build up to accelerate the shock front and 
to launch magnetic jets along the rotation axis. This 
magnetorotational mechanism\index{magnetorotational mechanism} 
can work on relevant time scales only in
stars with stellar cores that spin much more rapidly than
in the far majority of SN progenitors, where angular momentum
transport and loss through 
magnetic fields\index{magnetic field} and stellar
winds\index{stellar wind} 
lead to efficient core deceleration during the
evolution through the red-giant phase 
\citep{Hegeretal2005}. 

\runinhead{Connection to Observations}
Ultimately, observational evidence is needed to decipher the 
secrets of the explosion mechanism of core-collapse SNe for a
final solution of this long-standing and nagging problem of
stellar astrophysics. For this reason, SN modeling has to 
strive for explanations of observational phenomena and has
to predict discriminating diagnostic effects. 
Before the next Galactic SN will offer the unique chance 
to perform high-precision measurements of neutrino and 
gravitational-wave signals as direct probes from the very
center of a dying star, SN theory must confront its 
predictions with the huge wealth of data available from
well studied individual SNe and SN remnants. Also the messages
communicated by the entire population of such events and
their relic objects should be exploited for constraints 
of the explosion physics. This Chapter will also review current
efforts in this direction.


\begin{figure}[!]
\sidecaption[t]
\includegraphics[scale=0.35]{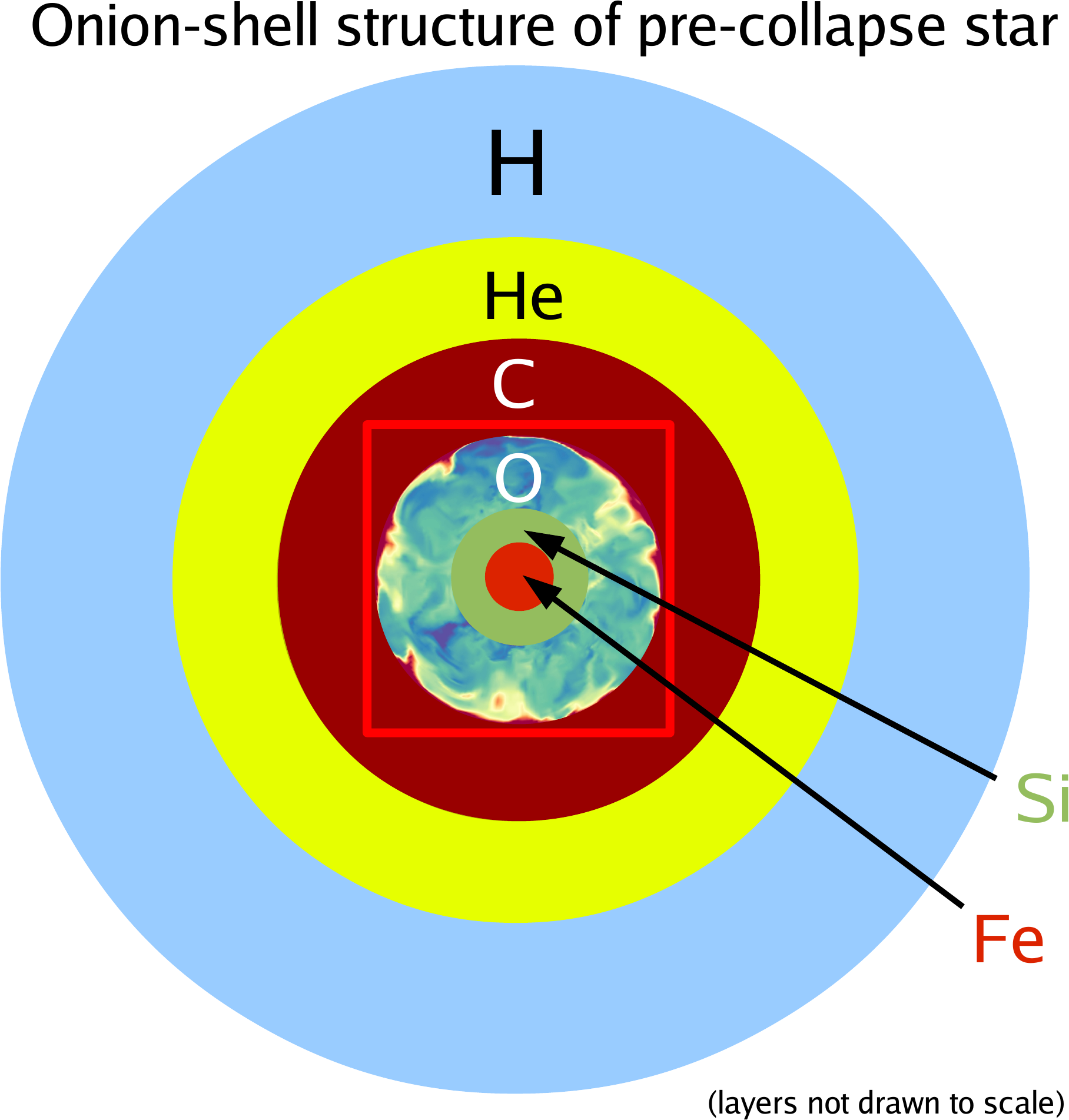}
\caption{Onion-shell structure of a SN progenitor star prior
to the onset of stellar core collapse. Shells of progressively
heavier elements contain the ashes of a sequence of nuclear
burning stages, which finally build up a degenerate core of
oxygen, neon and magnesium or iron-group elements at the center.
Convective burning can lead to large-scale velocity and density
perturbations in the oxygen and silicon layers (as indicated for
the O-shell). The red box marks the volume that is zoomed into
in Fig.~\ref{fig:janka-dynphases}
}
\label{fig:janka-onion}
\end{figure}

\begin{figure}[b]
\sidecaption
\caption{Density profiles ({\em top}) and compactness values
$\xi_{2.5}$ (Eq.~\ref{eq:compactness} for $M = 2.5\,M_\odot$;
{\em bottom}) for selected progenitor stars with ZAMS masses 
between 8.8\,$M_\odot$ and 30\,$M_\odot$. N8.8 denotes an
8.8\,$M_\odot$ O-Ne-Mg-core progenitor from \citet{Nomoto1984,Nomoto1987},
N20 a 20\,$M_\odot$ SN~1987A progenitor from \citet{NomotoHashimoto1988},
z9.6 a 9.6\,$M_\odot$ zero-metallicity progenitor provided by A.~Heger 
(2012, private communication), and all s-models are solar metallicity
progenitors from \citet{SukhboldWoosley2014}, \citet{WoosleyHeger2015},
and \citet{Sukhboldetal2016}. The progenitors with the smallest 
compactness values are classified as ``Crab-like'', the more compact
iron-core progenitors as ``SN1987A-like''. (Figures courtesy of 
Thomas Ertl)
}
\hspace{4.5cm}
\vbox{\vspace{-9.50cm}
\includegraphics[scale=0.95]{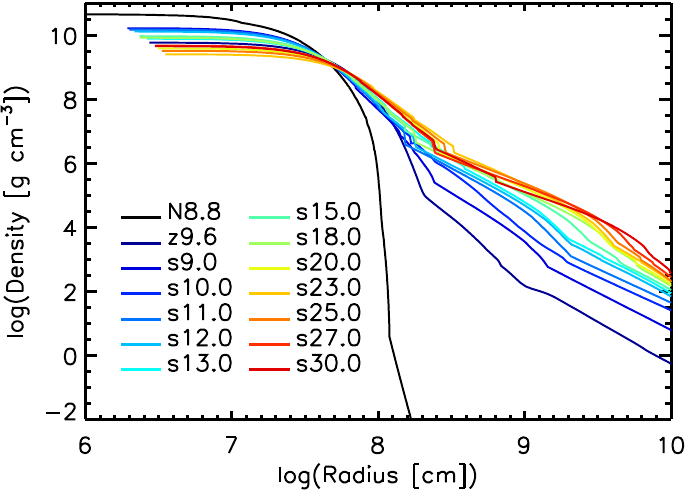}\vspace{5pt}\\
\includegraphics[scale=0.95]{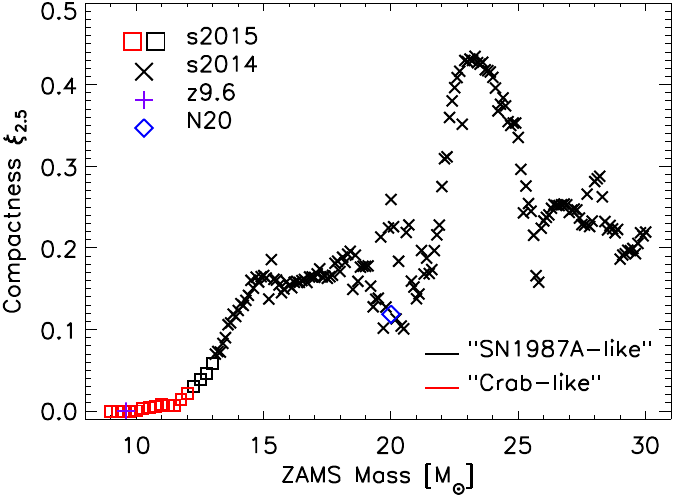}
}
\label{fig:janka-compactness}
\end{figure}

\section{From Stellar Core Collapse to Explosion}
\label{sec:phases}

Massive stars, when approaching the end of their hydrostatic
evolution, develop the so-called 
onion-shell structure\index{star!onion-shell structure}
(Fig.~\ref{fig:janka-onion}), 
where shells of lighter chemical elements surround layers of 
successively heavier elements, which contain the ashes of a long 
sequence of nuclear burning stages starting with hydrogen
burning on the main sequence of the color-magnitude diagram.
At the end of the stellar life,
the central core is composed of oxygen, neon and magnesium
or of iron-group elements and is stabilized against gravity
mainly by the pressure of degenerate electrons. 
Before its collapse
sets in, the baryonic mass of such a degenerate core ranges
between slightly less than 1.3\,$M_\odot$ and roughly 2\,$M_\odot$
with central densities between several $10^9$\,g\,cm$^{-3}$ and
more than $\sim$$10^{10}$\,g\,cm$^{-3}$, central temperatures 
around $10^{10}\,$K (roughly 1\,MeV in energy units), entropies
per nucleon of typically about 1\,$k_\mathrm{B}$, and 
electron-to-baryon ratios between $\sim$0.45 and $\sim$0.50.

\subsection{Core Structure of Stars at Collapse}
\label{sec:stellarcores}

The density structure at the time of stellar core collapse
exhibits considerable variability between different stars
(Fig.~\ref{fig:janka-compactness}, upper panel).
When evolved to the same central density, however, the density
profiles of the degenerate cores become nearly identical, simply
because the conditions are still close to hydrostatic equilibrium
and the equation-of-state is dominated by relativistic electrons,
therefore is well described by a polytrope with adiabatic index 
of 4/3.
In contrast, the density profiles of the surrounding shells
exhibit a wide spread of gradients, from abrupt
``cliffs'' bordering O-Ne-Mg cores in 
super-AGB stars\index{star!super-AGB}, to steep
slopes in progenitors with masses below about 
10\,$M_\odot$, to much shallower declines in more
massive stars (Fig.~\ref{fig:janka-compactness}, upper panel).

In order to characterize these differences of the stellar 
density profiles, \citet{OConnorOtt2011} introduced a
``compactness'' parameter\index{compactness parameter},
\begin{equation}
\xi_M = \frac{M/M_\odot}{R(M)/(1000\,\mathrm{km})}\,,
\label{eq:compactness}
\end{equation}
which is defined as the ratio of a chosen mass to the 
radius that encloses this mass. Figure~\ref{fig:janka-compactness}
(lower panel) displays $\xi_M$ for $M = 2.5\,M_\odot$ as a 
function of the zero-age-main-sequence (ZAMS)
mass\index{zero-age-main-sequence (ZAMS) mass} for a 
state-of-the-art set of SN progenitors. Any other choice of 
$M/M_\odot$ between $\sim$1.5 and $\sim$2.5 yields a similar 
pattern. The functional behavior is also independent of whether
$\xi_M$ is measured at the onset of the core collapse, at a 
moment when the collapsing stellar cores reach a certain
central density, or at core bounce, despite some quantitative
evolution taking place between these stages for the smaller
values of $M/M_\odot$ in the mentioned mass interval.
The non-monotonic variations above a ZAMS mass of about 
15\,$M_\odot$ are a consequence of a complex interplay
of carbon and oxygen shell burning, which can cause large
differences in the compactness for stars of very nearly the 
same mass \citep{SukhboldWoosley2014}.

Low-mass progenitors of less than roughly 12\,$M_\odot$
possess the lowerst core-compactness values, whereas above 
$\sim$12\,$M_\odot$ the compactness, e.g.\ $\xi_{2.5}$,
increases steeply (Fig.~\ref{fig:janka-compactness}, lower
panel). Observations and theoretical models suggest
that the Crab Nebula\index{Crab} is probably the remnant of 
a low-mass progenitor, most likely of a star with less
than about 10\,$M_\odot$, which exploded with a very low
energy ($\lesssim$10$^{50}$\,erg) and little $^{56}$Ni 
production ($\lesssim$10$^{-2}\,M_\odot$)
\citep[e.g.,][]{Nomotoetal1982,Kitauraetal2006,YangChevalier2015,OwenBarlow2015}.
Progenitors with a similarly small core compactness and
with SN explosion energies and $^{56}$Ni production 
below average are therefore denoted
as ``Crab-like'' in Fig.~\ref{fig:janka-compactness}, 
whereas stars more massive than 12\,$M_\odot$ are classified
as ``SN1987A-like'', because their explosion parameters
are closer to those of SN~1987A as a more canonical iron-core
SN. This distinction, however, is
not strict. Considering a more rigorous constraint
on the decline of the mass infall rate during stellar core
collapse, for example, \citet{Mueller2016} delimits SN
progenitors above $\sim$10\,$M_\odot$ from stars below this
mass limit, whose extremely tenuous envelope around the 
degenerate iron core defines a particularly close structural
proximity to O-Ne-Mg-core progenitors.
 
The compactness of the layers surrounding the
degenerate central core has important consequences for the
way how neutrino-driven explosions develop. This will be 
discussed in a later section.

\begin{figure}[!]
\sidecaption[t]
\includegraphics[scale=.25,angle=0]{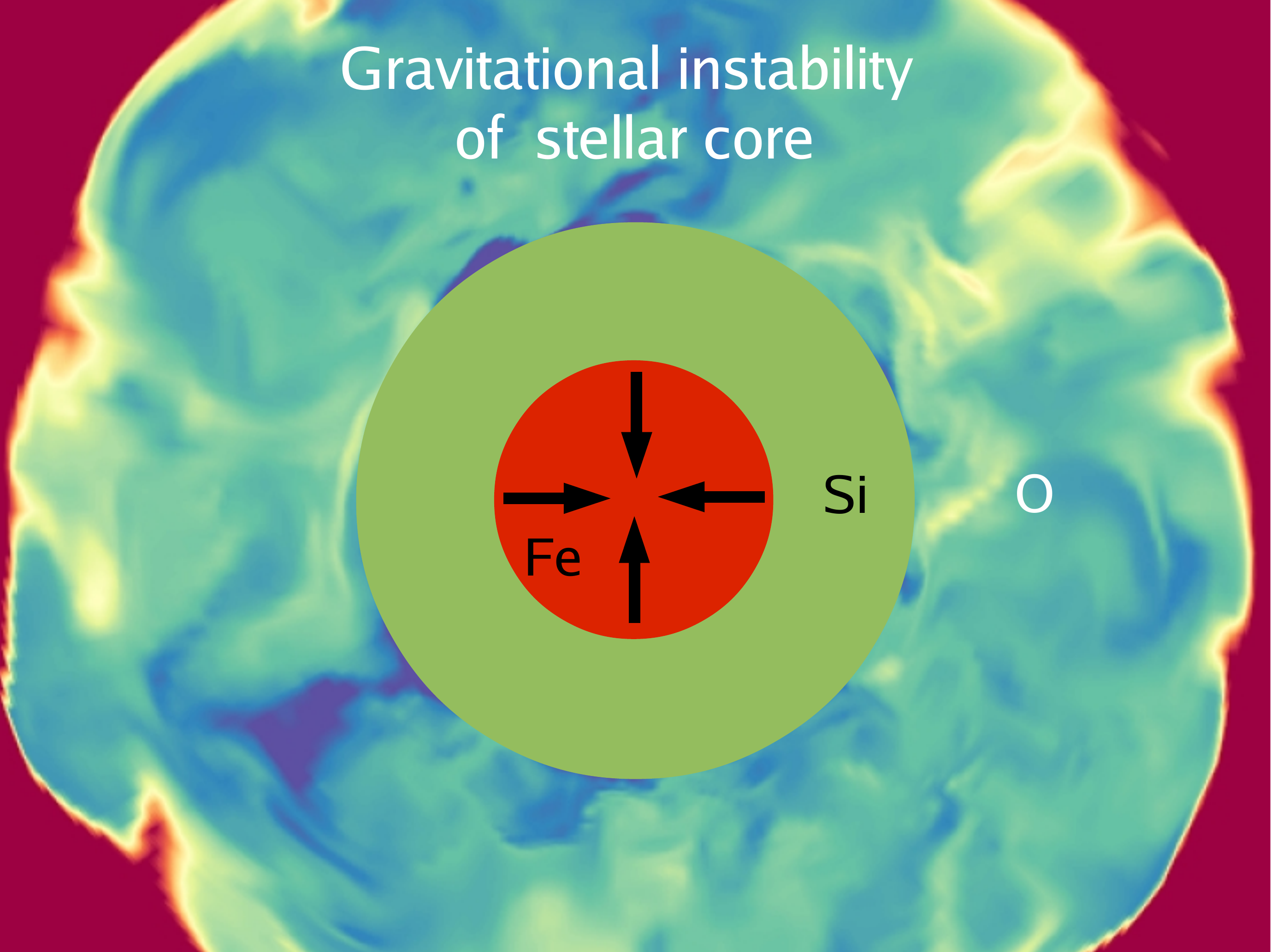}\hspace{2pt}
\includegraphics[scale=.25,angle=0]{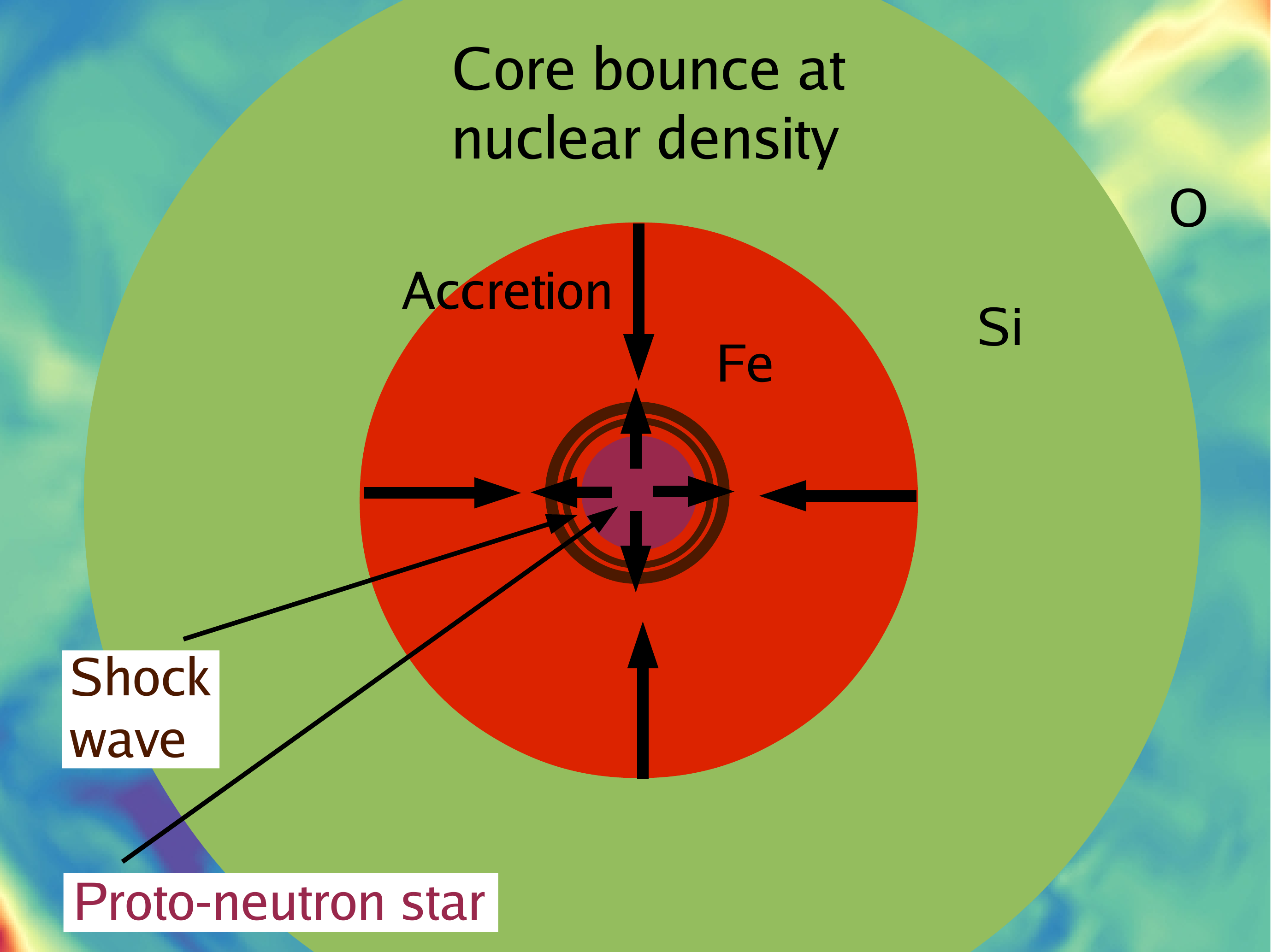}\vspace{2pt}\\
\includegraphics[scale=.25,angle=0]{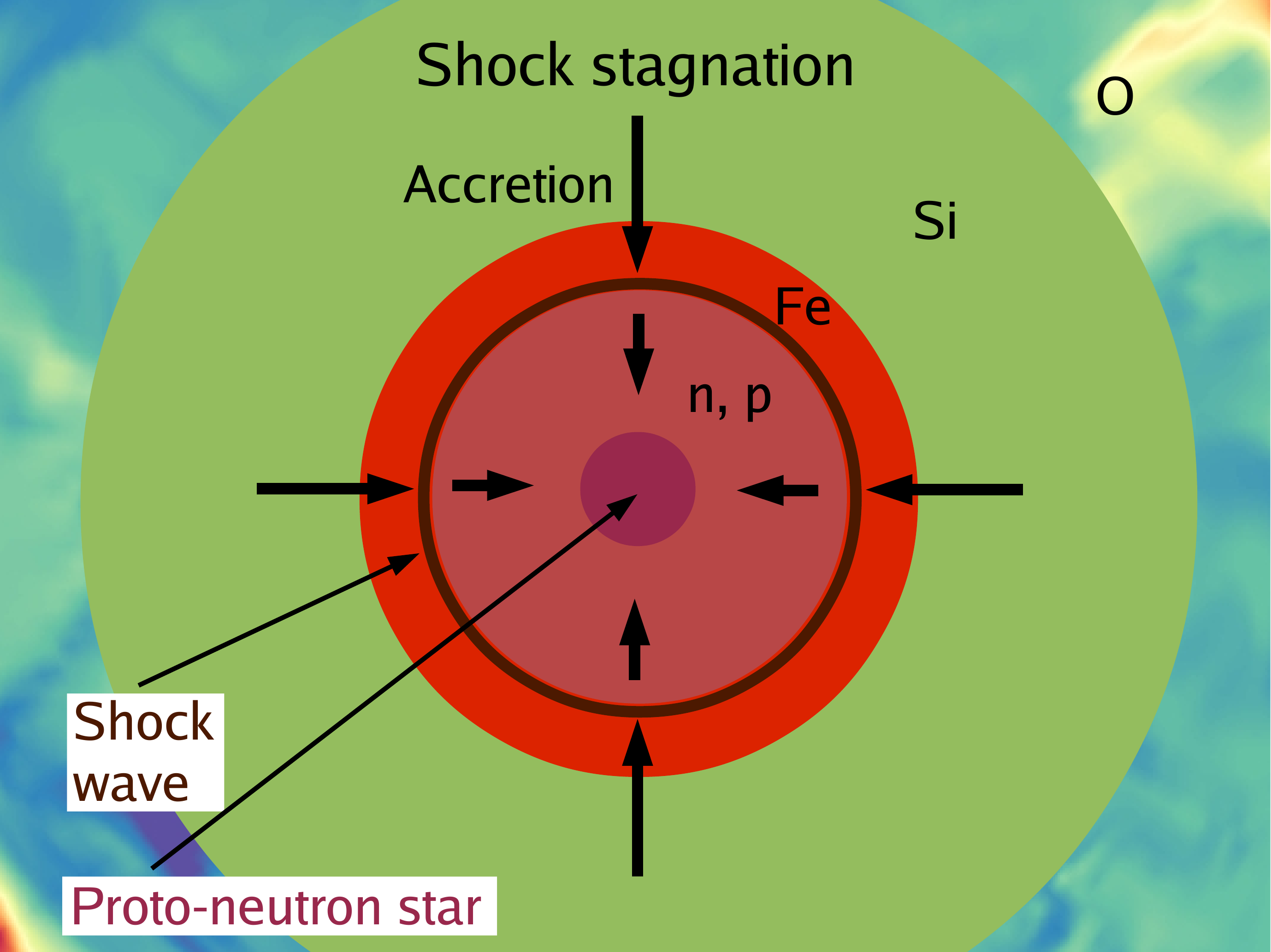}\hspace{2pt}
\includegraphics[scale=.25,angle=0]{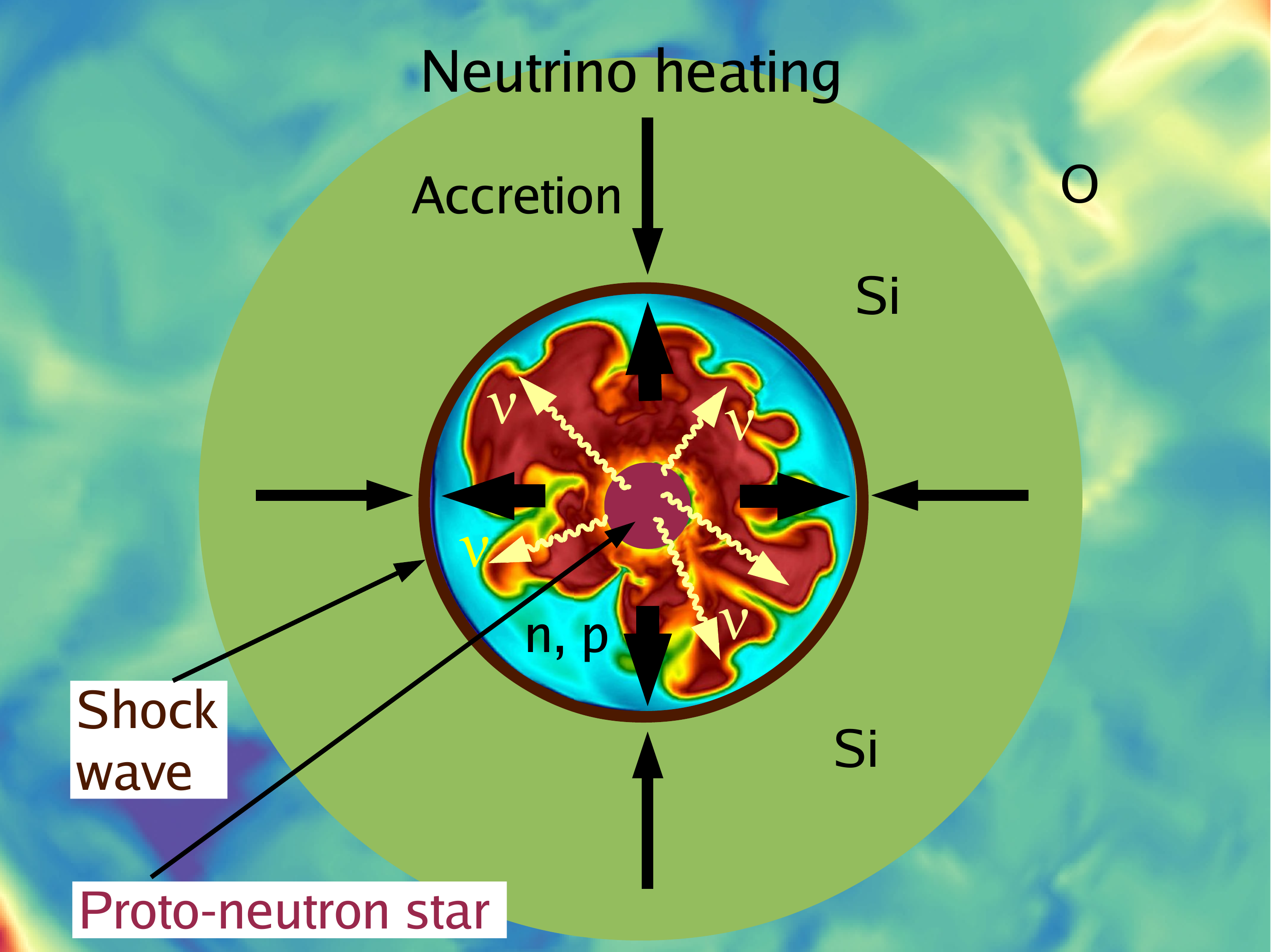}\vspace{2pt}\\
\includegraphics[scale=.25,angle=0]{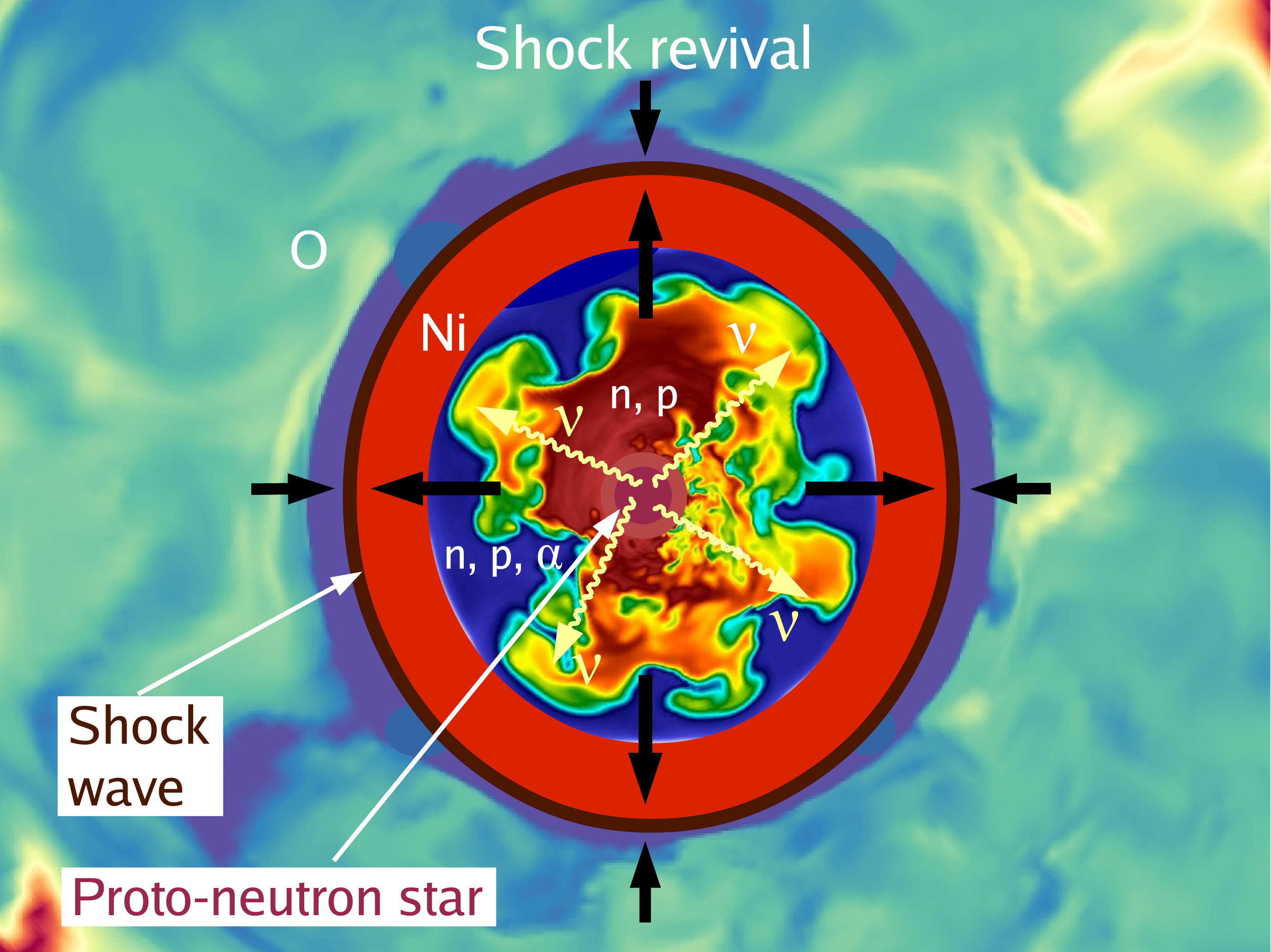}\hspace{2pt}
\includegraphics[scale=.25,angle=0]{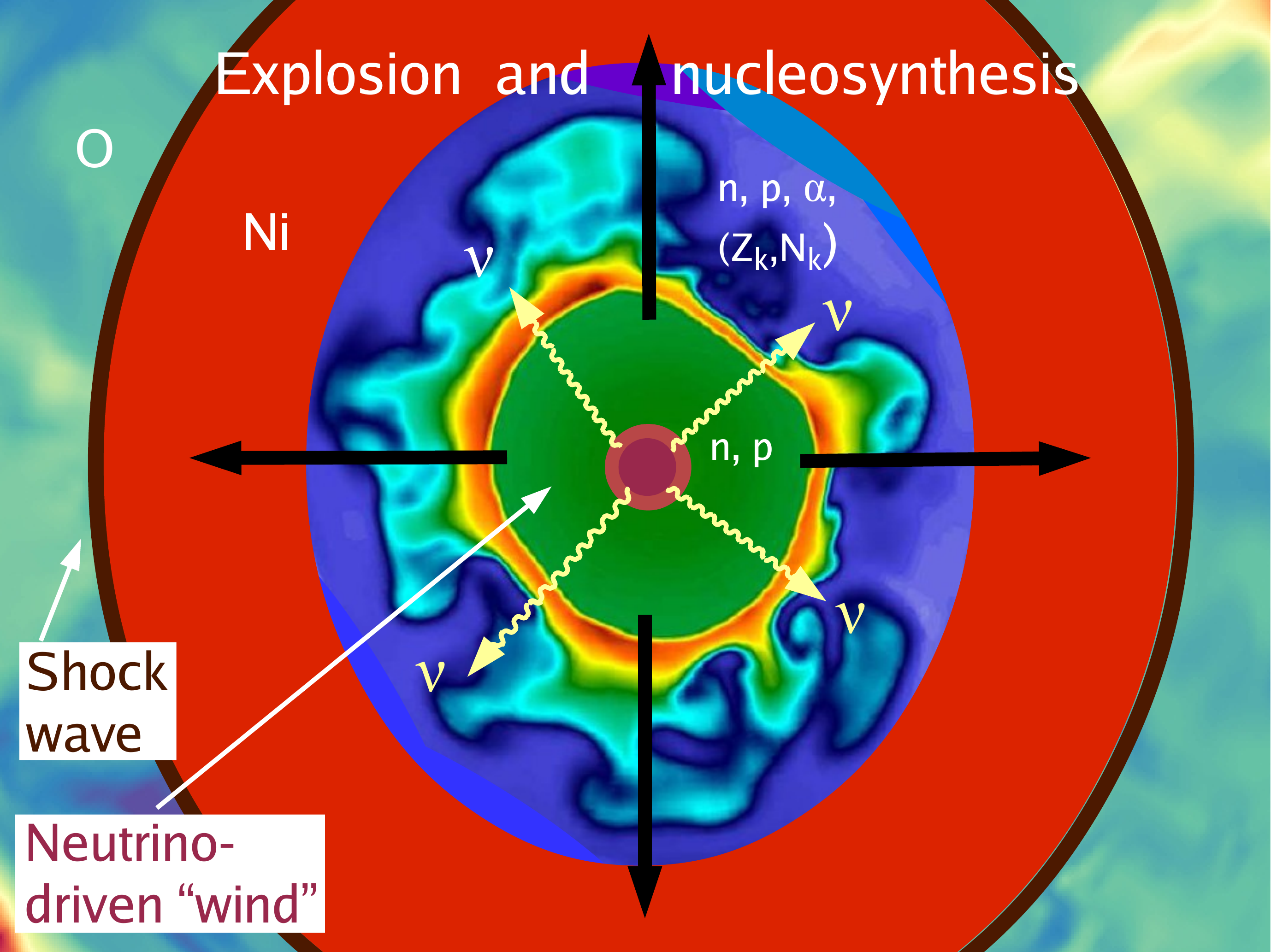}
\caption{Dynamical phases of stellar core collapse and explosion by
the neutrino-driven mechanism: infall, core bounce, stagnation of
the bounce shock, accretion and neutrino heating, shock revival,
and outward acceleration of the neutrino-powered SN shock front
({\em from top left to bottom right}). The horizontal scales of
the plots are roughly 15,000\,km, 4000\,km, 4000\,km, 2000\,km,
2000\,km, and 10,000\,km, respectively. The different shells are
not displayed to scale, but the central regions and the new-born,
hot NS (the ``proto-NS'') are
(approximately logarithmically) enlarged compared to the
outer layers. Superimposed on the graphical elements are
results from 3D simulations of pre-collapse convective O-shell
burning by \citet{Muelleretal2016} and of postshock asymmetries
(buoyant plumes of high-entropy matter and accretion cooler 
downflows)
during the first second of the SN explosion as computed by
\citet{Wongwathanaratetal2013}. In the bottom-right image the
spherical neutrino-driven wind, composed of free neutrons
and protons ($n$, $p$), is visible in green. It recombines to
$\alpha$-particles and heavy nuclei $(Z_k,N_k)$ when the
temperatures decrease in the expanding outflow. The sharp,
nearly spherical discontinuity bounding the green area is the
reverse shock that terminates the supersonic expansion of the
neutrino-driven wind
}
\label{fig:janka-dynphases}
\end{figure}

\subsection{Evolution Phases of Neutrino-driven Explosions}
\label{sec:explosionphases}

The dynamical evolution from the onset of stellar core collapse
to the successful initiation of the SN outburst can be divided
into six stages, which are displayed in a graphical way in 
Fig.~\ref{fig:janka-dynphases}.

\runinhead{Gravitational Instability and Collapse of the Stellar Core}
The gravitational instability of the degenerate O-Ne-Mg or 
iron core (Fig.~\ref{fig:janka-dynphases}, top left)
is initiated by electron captures\index{electron capture} 
on nuclei and free protons,
\begin{eqnarray}
e^- + p     &\longrightarrow& \nu_e + n      \,,\label{eq:ecapture1}\\
e^- + (A,Z) &\longrightarrow& \nu_e + (A,Z-1)\,,\label{eq:ecapture2}
\end{eqnarray}
and by the partial photodissociation\index{photodissociation} of 
heavy nuclei to $\alpha$ particles and free nucleons
\citep[for a discussion in detail, see][]{Langankeetal2003,Hixetal2003,Jankaetal2007}.
Both lead to a reduction of the 
effective adiabatic index\index{effective adiabatic index} of
the equation of state below the critical value for gravitational
instability of about 4/3. Initially, the electron neutrinos 
($\nu_e$) produced by electron captures can escape freely, but
at a density of some $10^{11}$\,g\,cm$^{-3}$, the neutrino mean
free path for coherent neutrino 
scattering\index{coherent neutrino scattering} off heavy nuclei
becomes so short that the neutrinos begin to diffuse. Finally,
at a density of about $10^{12}$\,g\,cm$^{-3}$, the outward 
neutrino diffusion is slower than the accelerating infall of
the stellar plasma, and neutrino 
trapping\index{neutrino trapping} sets in.
With neutrinos being unable to escape, the entropy remains
conserved, and the subsequent infall proceeds adiabatically.
The subsonically collapsing inner core develops a homologous
velocity\index{homologous velocity}
profile, which reaches a maximum velocity near the
interface to a nearly free-falling, supersonic outer core.
\citep[An excellent text-book discussion of the physics
during this phase is provided by][Chapter~18.]{Shapiro1983} 

\runinhead{Core Bounce and Shock Formation}
The implosion of the inner core is stopped abruptly when nuclear
saturation density\index{nuclear saturation density} 
($\rho_0\approx 2.7\times 10^{14}$\,g\,cm$^{-3}$
or $n_0\approx 0.16$\,fm$^{-3}$) is reached at the center and 
the phase transition to homogeneous nuclear matter leads to a
sudden stiffening of the equation of state. The increase of
the adiabatic index above 4/3 allows for a new stable state
where the stellar matter can be supported against its own
gravitational attraction by the internal pressure of the 
nucleon gas, which is highly incompressible due to the repulsive
part of the nucleon-nucleon interaction 
potential\index{nucleon interaction potential}. Since the
collapsing inner core overshoots the new equilibrium state,
it bounces back and its expansion creates pressure waves
that steepen into a shock front\index{shock front}
at the transition to the
supersonically infalling outer core 
(Fig.~\ref{fig:janka-dynphases}, top right).
Because of the initial loss of electron neutrinos, the
electron fraction\index{electron (number) fraction} 
at the center has dropped to values between
0.25 and 0.27, and the inner core has shrunk to a mass
below 0.5\,$M_\odot$, which roughly defines the 
location where the bounce-shock forms, nearly independent of
the progenitor star and only moderately different (within
$\sim$10\%) for different models of the nuclear equation of
state \citep{Jankaetal2012}.

\runinhead{Shock Stagnation and Shock-breakout Neutrino Burst}
The newly formed shock\index{shock}
begins to propagate outwards in radius
as well as in mass. Dissipation of kinetic energy in the 
infalling matter swept up by the shock raises the entropy
and temperature within the shock, creating high-energy photons 
that lead to the photodissociation of iron nuclei to free 
nucleons. The iron disintegration is essentially complete
as long as the shock
radius is smaller than the 
``dissociation radius''\index{dissociation radius} of iron,
\begin{equation}
R_\mathrm{diss} = \frac{G\,M\,m_\mathrm{u}}{8.8\,\mathrm{MeV}}
\gtrsim 160\,\left(\frac{M}{M_\odot}\right)\,\,\mathrm{km}\,,
\label{eq:rdiss}
\end{equation}
which is roughly given by the radius where the nuclear binding
energy of nucleons (with average mass $m_\mathrm{u}$) in 
iron-group material equals the free-fall kinetic energy (with
velocity $v_\mathrm{ff}$) in the gravitational potential of 
the enclosed mass $M$:
$\frac{1}{2}m_\mathrm{u}v_\mathrm{ff}^2 = 
GMm_\mathrm{u}R_\mathrm{diss}^{-1} = 8.8$\,MeV.
The conversion of kinetic energy to rest-mass energy thus
drains about 8.8\,MeV per nucleon or 
$1.7\times 10^{51}$\,erg per 0.1\,$M_\odot$ of energy from
the thermal reservoir, reducing the postshock pressure. 
Within only about a millisecond and after having overrun
only about 0.5\,$M_\odot$ of iron-core matter, the bounce
shock comes to a stop still well inside of the collapsing
iron core (Fig.~\ref{fig:janka-dynphases}, middle left).
At about this time the density behind the shock
has decreased to a value where the electron neutrinos,
which are abundantly produced by electron 
captures onto free protons in the postshock medium
(Eq.~\ref{eq:ecapture1}), start to
escape freely. A luminous flash of $\nu_e$, the so-called 
shock-breakout neutrino burst\index{shock-breakout neutrino burst}, 
is radiated and takes 
away additional energy from the postshock layer.
Since the velocities everywhere behind the shock become
negative, the shock expansion stalls and the shock
converts into an accretion shock\index{accretion shock}.

\runinhead{Neutrino Heating and Accretion}
Shortly after core bounce neutrino emission carries
away energy from the postshock layer. The conditions, however,
change fundamentally at later post-bounce times, because the
postshock temperature decreases as the density drops and the
plasma becomes more radiation dominated. Parallel to that, 
the neutrino spectra radiated from the contracting and 
increasingly hotter NS harden. While the first
effect diminishes the neutrino cooling at the shock,
the second effect allows for an increasing fraction
of the electron neutrinos ($\nu_e$) and antineutrinos
($\overline{\nu}_e$) streaming away from the 
neutrinosphere\index{neutrinosphere}
to be reabsorbed by free neutrons and protons,
\begin{eqnarray}
\nu_e + n           &\longrightarrow& p + e^- \,,\label{eq:nueabs}\\
\overline{\nu}_e + p &\longrightarrow& n + e^+ \,,\label{eq:barnueabs}
\end{eqnarray}
closer to the shock front 
(Fig.~\ref{fig:janka-dynphases}, middle right). This situation
defines the phase of neutrino heating\index{neutrino heating}, 
when the stalled 
shock receives fresh energy from the neutrinos streaming up 
from the neutrinosphere\index{neutrinosphere}.
Since neutrino-energy deposition creates a negative entropy 
gradient, the heated layer can become 
convectively unstable\index{convectively unstable}
\citep{Herantetal1994,Burrowsetal1995,JankaMueller1996,Foglizzoetal2006}.
Also the standing-accretion-shock instability
\index{standing-accretion-shock instability (SASI)}
\citep[SASI;][]{Blondinetal2003,BlondinMezzacappa2007,Schecketal2008,Foglizzoetal2015} 
can grow in the mass-accretion
flow between shock and nascent NS, leading to large-scale,
non-radial deformation and violent sloshing and 
spiral motions of the shock front, thus stirring the 
whole layer enclosed by the shock and the NS.
Mushroom-like high-entropy structures indicative of 
buoyancy-driven 
Rayleigh-Taylor instability\index{instability!Rayleigh-Taylor}
can be seen in the postshock region in the
middle-right panel of Fig.~\ref{fig:janka-dynphases}.

\runinhead{Shock Revival}
Neutrino-energy transfer to the shock raises the postshock
pressure. If the heating by neutrinos is strong enough, the 
shock can be pushed outwards and the SN explosion can be
launched. The non-radial fluid instabilities (buoyancy,
convective overturn, and the SASI) assist the neutrino-heating
mechanism in several ways. Besides causing more expansion of
the shock and thus enlarging the layer of neutrino-energy
deposition, the non-radial flows carry hot, neutrino-heated 
matter outwards to the shock and cooler gas inward, closer
to the NS, where this material can absorb energy from the 
neutrino flux more effectively. In combination, the 
multi-dimensional effects increase the efficiency of the
neutrino-energy transfer compared to the case of spherical
conditions. If the thermal pressure behind the shock, 
supported by turbulent pressure, overcomes the
ram pressure of the infalling preshock layer, runaway
shock expansion can set in 
\citep[e.g.,][]{JankaMueller1996,MurphyBurrows2008B,Nordhausetal2010,Hankeetal2012,CouchOConnor2014,Fernandezetal2014,Fernandez2015}.
The outward acceleration of the shock begins to trigger 
explosive nucleosynthesis\index{explosive nucleosynthesis} 
in the postshock medium
(Fig.~\ref{fig:janka-dynphases}, bottom left), producing 
also radioactive iron-group
and intermediate-mass nuclei\index{radioactive nuclei}
(e.g., $^{56,57}$Ni, $^{55,60}$Co, $^{44}$Ti), which 
power the luminous, long-time electromagnetic radiation 
of the expanding SN debris for many years.

\runinhead{Explosion and Nucleosynthesis}
For a transient period of time, matter swept up by the
accelerating SN shock is still accreted towards the
nascent NS, absorbs energy from neutrinos, and is 
partly ejected outwards again. When this phase of 
simultaneous mass accretion and outflow 
\citep[which might last for several seconds;][]{Mueller2015}
ends, neutrino-energy deposition in the near-surface layers
of the new-born NS launches the so-called neutrino-driven 
wind\index{neutrino-driven wind}, 
which is an essentially spherical, tenuous outflow of 
baryonic matter from the NS surface 
(Fig.~\ref{fig:janka-dynphases}, bottom right).
This high-entropy (several 10 to over 100\,$k_\mathrm{B}$
per nucleon) medium is initially composed of free neutrons
and protons, which recombine to $\alpha$-particles and 
finally partly to heavy nuclei when the expanding outflow
cools. Depending on whether there is 
neutron\index{neutron excess} or proton excess\index{proton excess}
in the outflow, neutron- or proton-rich nuclei
may be assembled in the neutrino-driven wind and can add
interesting nucleosynthetic yields to the innermost SN
ejecta \citep[for a review of possible processes, 
see][]{Jankaetal2007,Janka2012}.
While the shock propagates outwards through the progenitor
star (and needs hours to over a day to reach the 
stellar surface), the compact remnant left behind at the
center cools and deleptonizes by radiating neutrinos and
antineutrinos of all flavors. With the decaying neutrino 
emission also the neutrino-driven wind gradually loses
power and finally dies off.

\subsection{State-of-the-art Multi-dimensional Models}
\label{sec:multidmodels}

Owing to the high performance of modern parallel computers 
and efficient application codes,
self-consistent simulations of stellar core
collapse with detailed neutrino transport have become 
possible in three spatial dimensions.
These models begin to lend support to the viability of
the neutrino-driven mechanism and thus confirm the promising
perspectives that have been drawn by a growing number of 
successful explosions in two-dimensional (2D) simulations.
The latter models, however, suffer from the enforced constraint
of symmetry around a chosen axis. While this assumption reduces 
the computational complexity and costs considerably,
the artificially imposed symmetry channels the flow along
the preferred direction and also implies that turbulent 
energy cascades inversely to the 3D case 
\citep{Hankeetal2012}. Models in 2D could therefore only
be a preliminary step before 3D simulations became feasible.

\begin{figure}[!]
\sidecaption[t]
\includegraphics[scale=0.35]{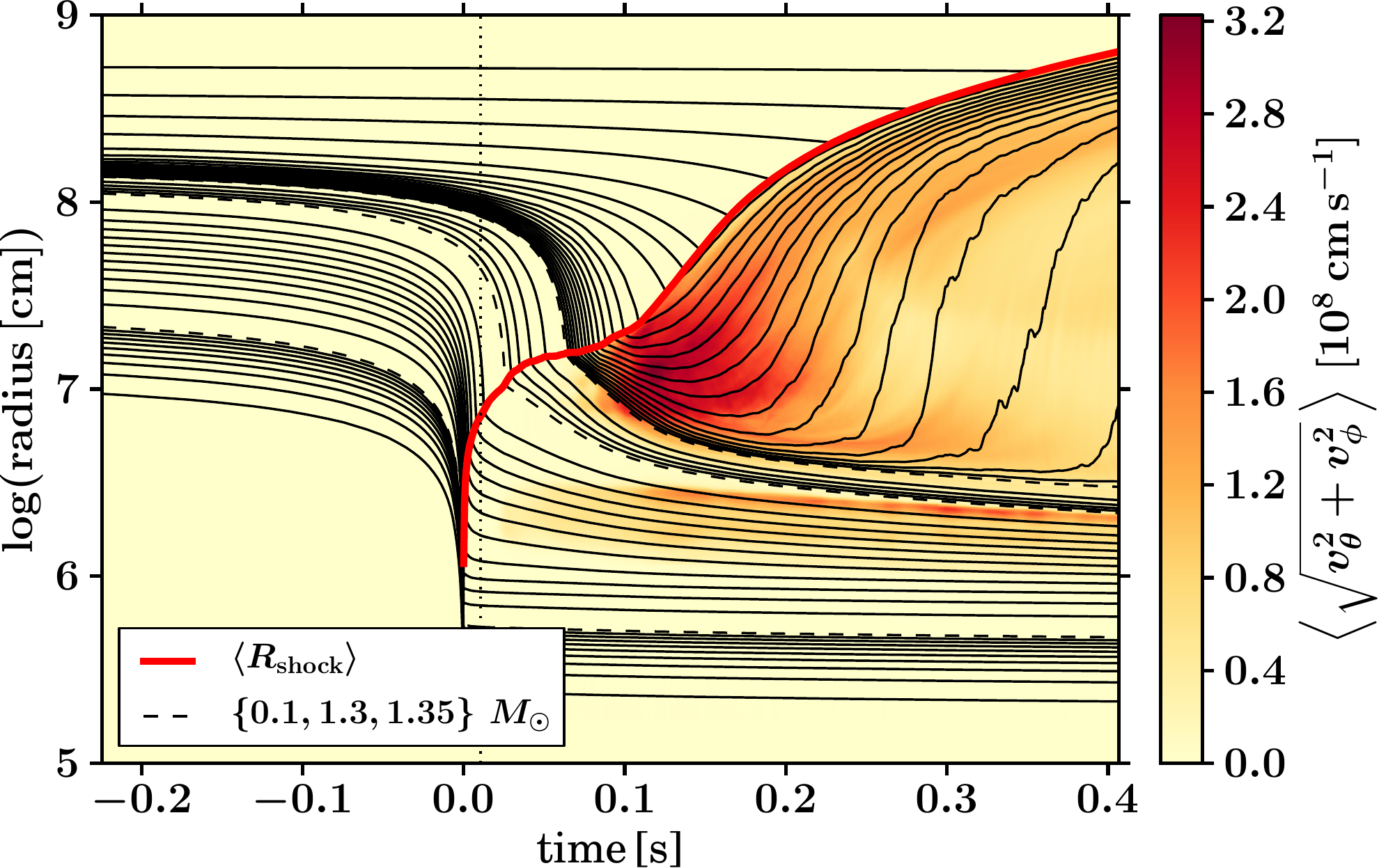}\hspace{5pt}
\includegraphics[scale=0.05]{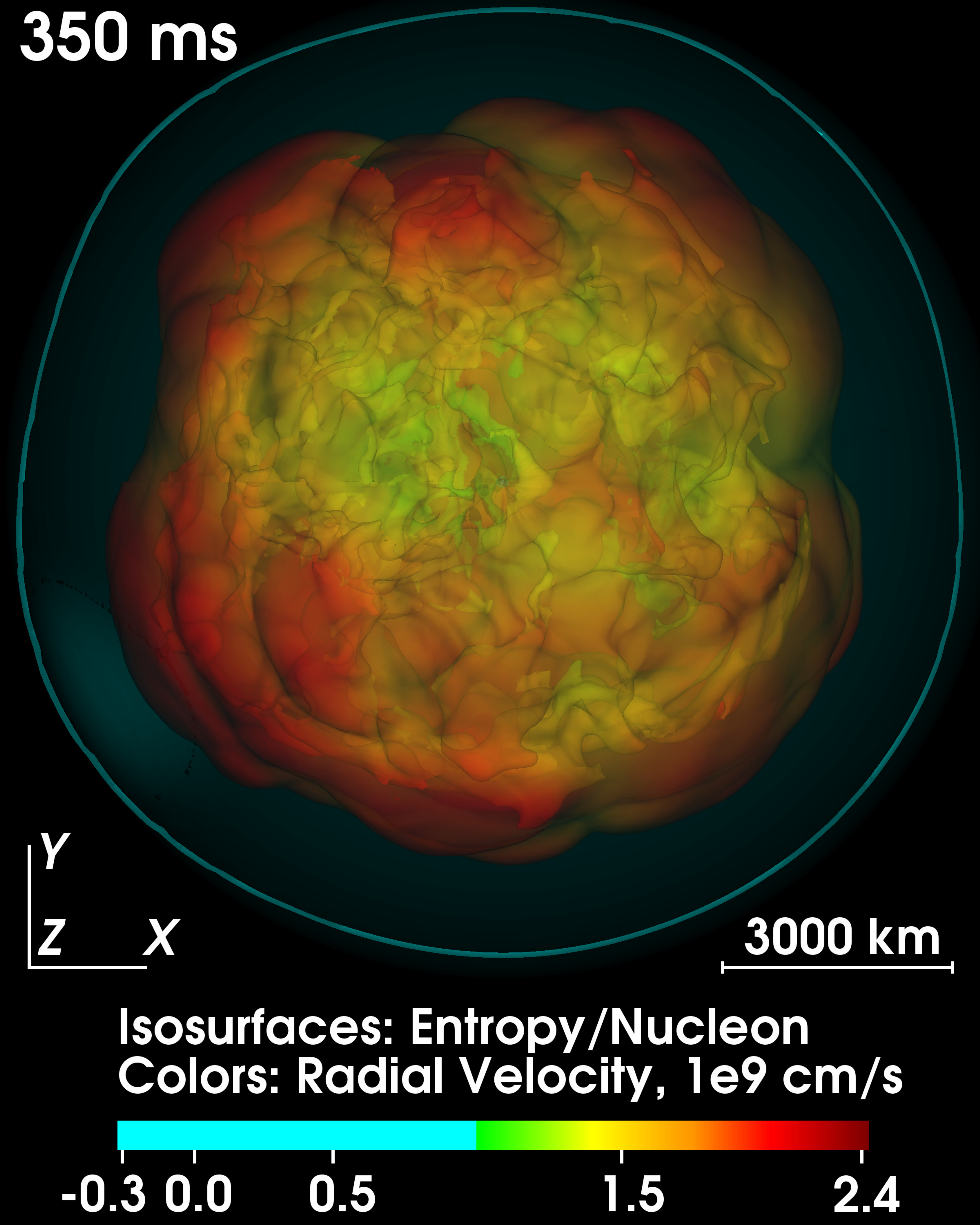}
\caption{Explosion of a 9.6\,$M_\odot$ progenitor by the 
neutrino-driven mechanism in a self-consistent 3D simulation
\citep{Melson2015a}.
{\em Left:} Time evolution displayed by a mass-shell plot
showing radii versus time (normalized to core bounce)
for selected values of the
enclosed mass, with the dashed curves corresponding to three
exemplary masses listed in the inset in the lower left corner.
The thick red line starting at core bounce marks the trajectory
of the SN shock. While the collapse of the stellar core is
visible by contracting mass shells, converging shells in the
lower part of the figure indicate the formation of the NS 
and outgoing shells signal the ejection of overlying matter
in the SN explosion and the onset of a neutrino-driven wind
from the surface of the new-born NS. 
The color coding depicts angular averages of the non-radial
velocities with red hues indicating strong convective
mass motions in the neutrino-heated region behind the SN shock.
{\em Right:} Ray-tracing image of the beginning explosion
(350\,ms after core bounce). The
SN shock is visible as a bluish, nearly circular line, enclosing
convectively perturbed high-entropy bubbles of neutrino-heated
matter (visualized by semi-transparent isosurfaces for chosen
values of the plasma entropy per nucleon). 
The color coding indicates radial velocities with red
corresponding to the highest values (see color bar). The shock
has a diameter of about 12,000\,km.
\citep[Figure from ][$\copyright$ The American Astronomical
Society]{Melson2015a}
}
\label{fig:janka-z96}
\end{figure}

\subsubsection{Electron-capture Supernovae and Low-mass Iron-core 
Explosions}
\label{sec:lowmassexp}

Low-mass SN progenitors with O-Ne-Mg cores 
(super-AGB stars\index{star!super-AGB})
or iron cores possess
particularly steep density profiles and extremely small values
of the core compactness\index{core compactness} 
(see Sect.~\ref{sec:stellarcores}). This
characteristic pre-collapse property facilitates neutrino-driven
explosions to develop fairly easily and early after core bounce.
The reason is a very rapid decline of the mass-accretion 
rate\index{mass-accretion rate},
\begin{equation}
\dot M \approx \frac{2M}{t_\mathrm{infall}}\,\frac{\rho}{\bar{\rho}}
= \frac{8\rho}{\sqrt{3}}\,\sqrt{G M r^3} \,,
\label{eq:mdot}
\end{equation}
as a consequence of a density decline $\rho(r)$ that is
much steeper than $r^{-3/2}$ 
\citep[$t_\mathrm{infall}$ is the collapse
time scale of a stellar shell with density $\rho(r)$ at initial
radius $r$, $\bar{\rho}$ the average density inside of $r$, and
$M$ the mass enclosed by $r$, which is essentially equal to the
NS mass in the case of low-mass progenitors;][]{Mueller2016}.
This leads to a fast drop of the ram pressure\index{ram pressure}
of matter falling into the shock, for which reason
the shock expands quickly and establishes favorable conditions for
neutrino-energy deposition around the newly formed 
NS \citep{Kitauraetal2006,Jankaetal2008}. The SN explosion
is thus powered by a neutrino-driven wind\index{neutrino-driven wind} 
absorbing energy from the
neutrinos leaving the NS. Due to their extremely steep density 
profiles, some low-mass SN progenitors (such as those with O-Ne-Mg
cores) are the only cases where explosions can be obtained even in
spherically symmetric (1D) simulations. Nevertheless, hydrodynamic 
instabilities develop in the neutrino-heated postshock layer and have 
important consequences for explosion asymmetries and the 
nucleosynthesis conditions in the innermost SN ejecta
\citep{Wanajoetal2011}.

In the case of O-Ne-Mg-core progenitors\index{O-Ne-Mg-core progenitor}
the shock propagates 
outwards continuously and accelerates rapidly at about 50--70\,ms
after core bounce \citep{Kitauraetal2006,Jankaetal2008,Fischeretal2010}.
Neutrino-heated plasma rises buoyantly in 
Rayleigh-Taylor fingers\index{instability!Rayleigh-Taylor},
but the structures freeze out on relatively small angular scales
because of their fast outward expansion. Also the 9.6\,$M_\odot$
iron-core progenitor of Fig.~\ref{fig:janka-compactness} explodes
in 1D, though only later than $\sim$300\,ms after bounce. In this
case convective overturn pushes the shock and speeds up the 
explosion considerably (Fig.~\ref{fig:janka-z96}), making it 
also more energetic \citep{Melson2015a}.
In both cases, however, multi-dimensional simulations predict
subenergetic SNe with explosion energies of
$\sim$(0.5--1.5)$\times 10^{50}$\,erg and fairly little production
of radioactive nickel (several $10^{-3}$\,$M_\odot$), which is
compatible with a growing sample of low-luminosity Type IIP
SNe and which has also been inferred for SN~1054\index{SN~1054}
as the birth event of the Crab\index{Crab}
Nebula \citep{YangChevalier2015}.

Because of their similar explosion dynamics in terms of a
rapid outward acceleration of the SN shock and a fast 
expansion of the postshock material, the nucleosynthesis in
low-mass SNe is very similar, irrespective of the different 
nature of the progenitor stars with O-Ne-Mg or iron cores.
This rapid expansion leads to $\nu_e$ and $\overline{\nu}_e$
absorption reactions (Eqs.~\ref{eq:nueabs}, \ref{eq:barnueabs}) 
and their inverse
processes to quickly freeze out in the ejecta, enabling matter,
in particular in the heads of the 
Rayleigh-Taylor mushrooms\index{instability!Rayleigh-Taylor}, to
be expelled with considerable neutron excess. This facilitates
the formation of neutron-rich nuclei\index{neutron-rich nuclei}. 
Low-mass SNe could thus
not only produce interesting amounts of $^{48}$Ca and $^{60}$Fe
but could also be dominant contributors to the Galactic repository
of light trans-iron elements\index{trans-iron elements} 
from zinc to zirconium, potentially
even of light r-process nuclei\index{r-process nuclei} 
up to palladium and silver 
\citep{Wanajoetal2011,Wanajoetal2013a,Wanajoetal2013b}.

Explosions of progenitor stars with O-Ne-Mg cores are often called
``electron-capture SNe''\index{electron-capture supernova}, 
because the collapse of the highly
degenerate stellar core is mainly initiated by electron captures
on magnesium and neon, in contrast to more massive progenitors
with iron cores, where nuclear photodisintegration plays the
dominant role. However, it is important to note that neither
the explosion mechanism of these SNe nor their nucleosynthesis
are distinct from those of low-mass progenitors with iron cores.

\begin{figure}[!]
\sidecaption[t]
\includegraphics[scale=.12]{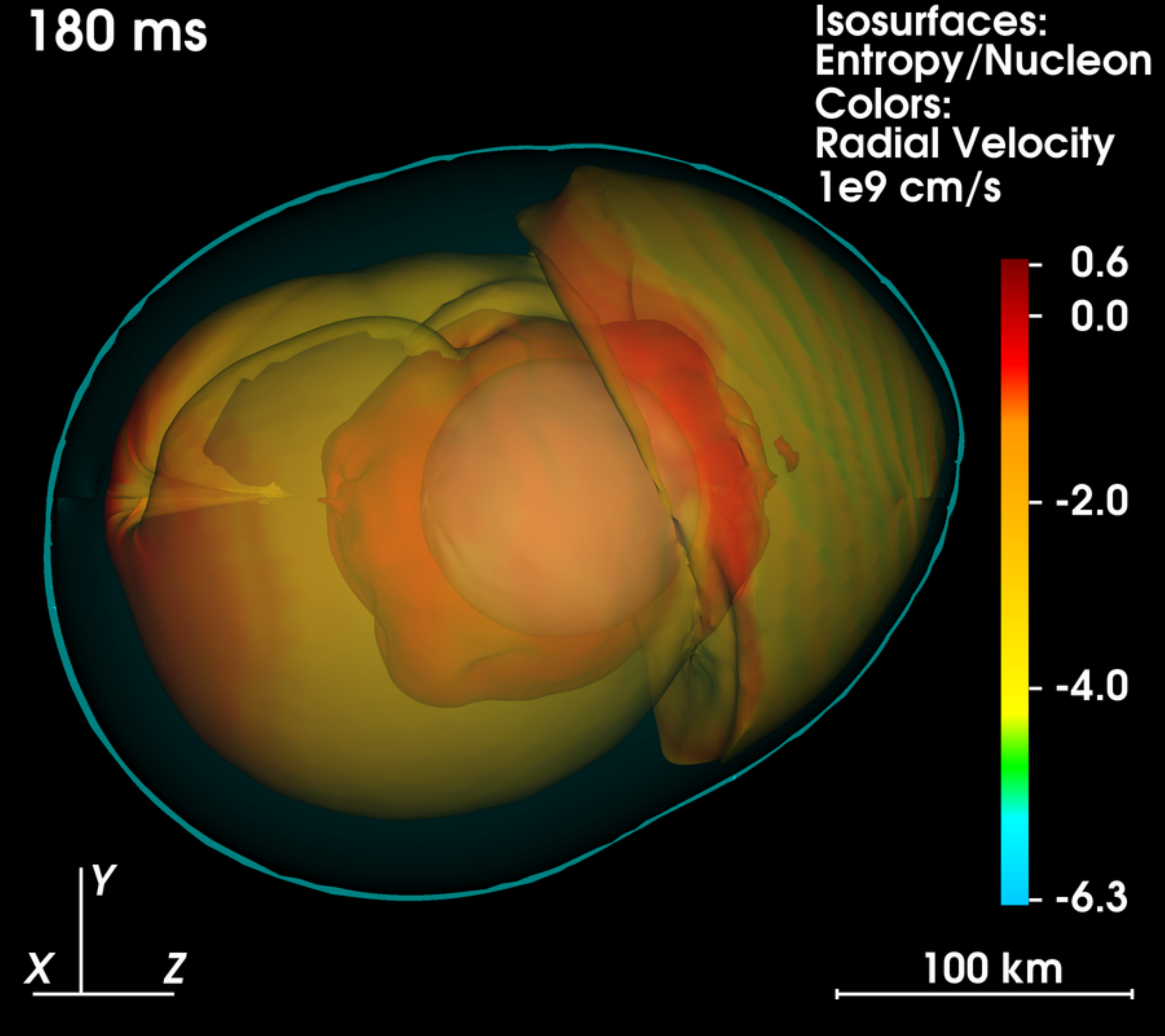}\hspace{0.05pt}
\includegraphics[scale=.12]{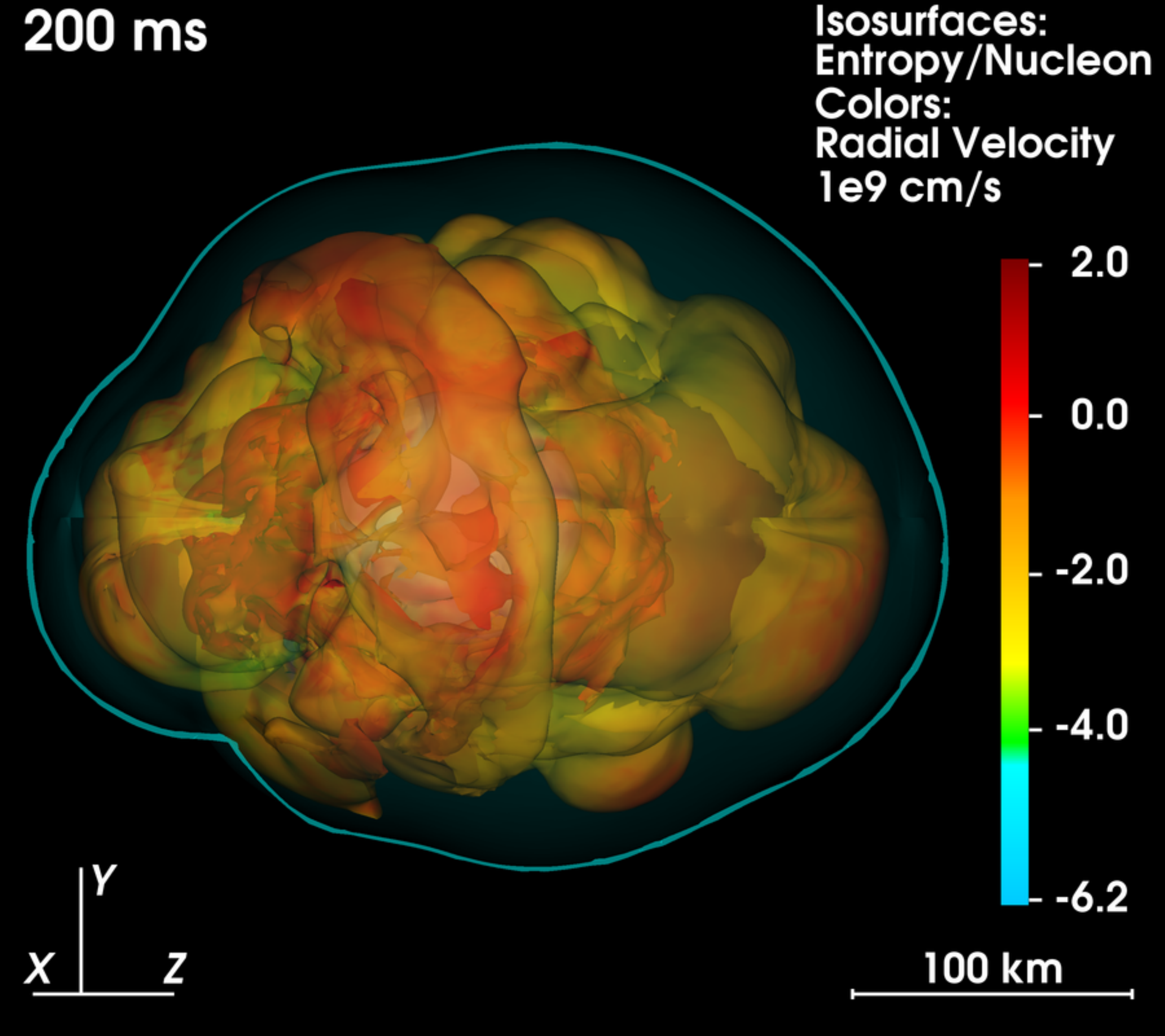}\hspace{0.05pt}
\includegraphics[scale=.12]{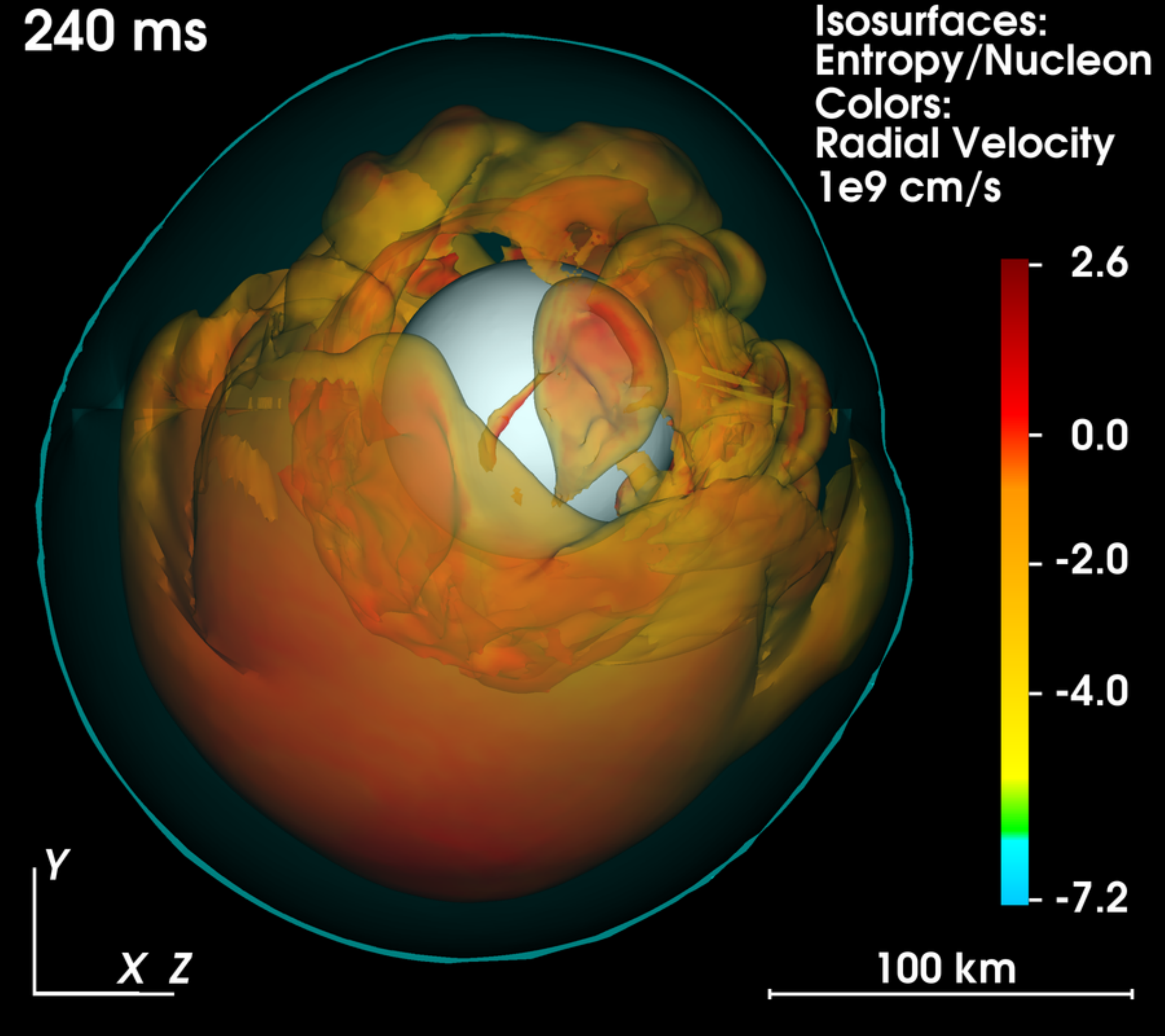}\vspace{1.20pt}\\
\includegraphics[scale=.12]{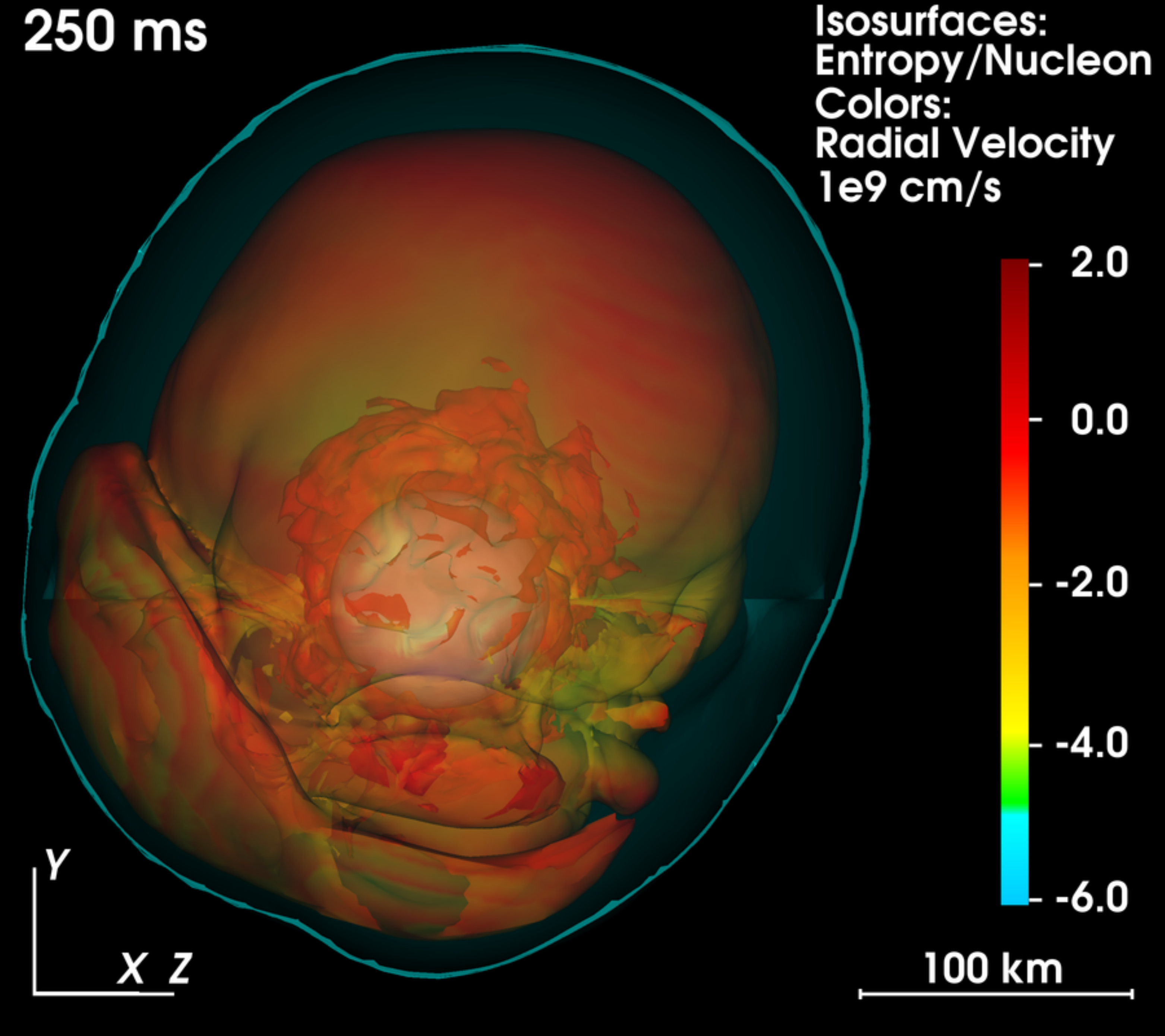}\hspace{0.05pt}
\includegraphics[scale=.12]{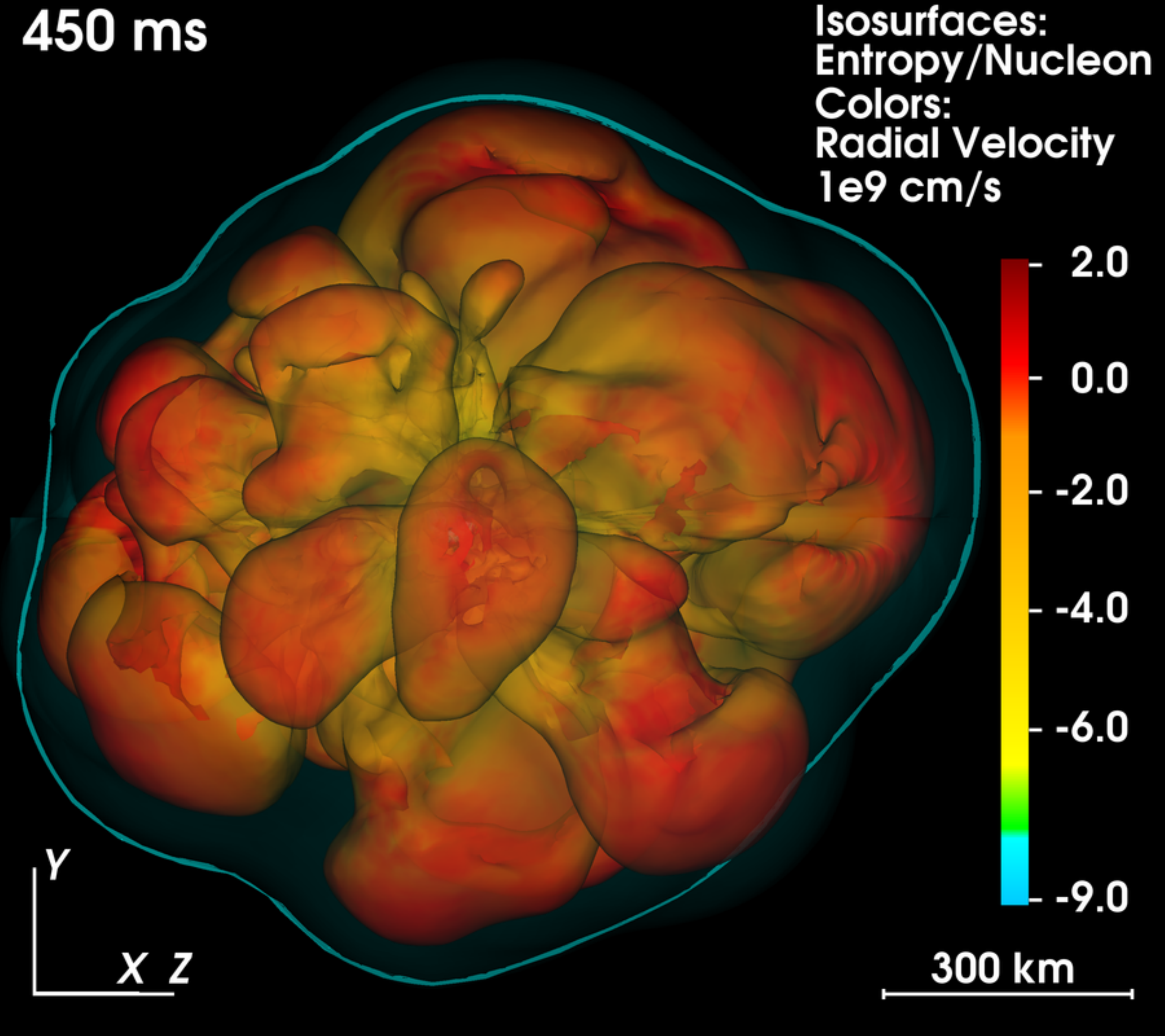}\hspace{0.05pt}
\includegraphics[scale=.12]{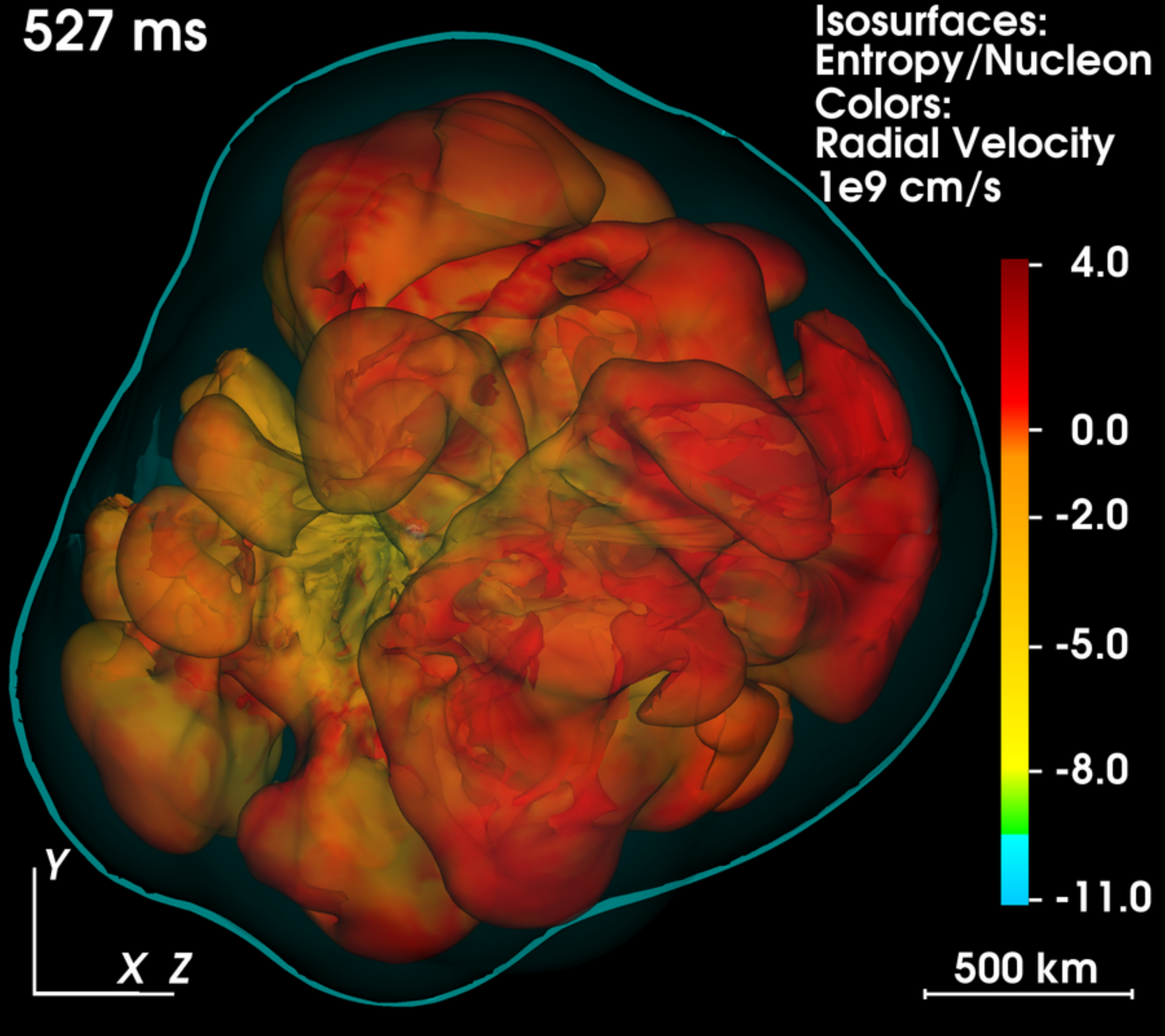}
\caption{Explosion of a 20\,$M_\odot$ progenitor by the neutrino-driven
mechanism in a self-consistent 3D simulation \citep{Melson2015b}.
The sequence of ray-tracing images (with post-bounce times given
in the upper left corners) shows that the postshock layer is 
massively stirred by phases of violent sloshing and spiral motions due to
the standing accretion shock instability (SASI), whose presence can be 
recognized by dominant dipolar and quadrupolar shock-deformation modes.
Runaway shock expansion sets in at about 400\,ms after core bounce.
The shock front is visible as a bluish, enveloping line, and the 
structures inside are semi-transparent isosurfaces for chosen values
of the plasma entropy per nucleon. The color coding of these
surfaces corresponds to the radial velocity as given by the color
bar. The white sphere visible in the upper right panel and faintly
also in the lower left panel is the surface of the new-born NS
corresponding to a density of $10^{11}$\,g\,cm$^{-3}$. At the end of 
the simulation the shock has reached a diameter of about 2000\,km
and expands with a velocity of $\sim$3500\,km\,s$^{-1}$
\citep[Figure from ][$\copyright$ The American Astronomical
Society]{Melson2015b}
}
\label{fig:janka-s20}
\end{figure}

\subsubsection{Supernovae of Massive Iron-core Progenitors}
\label{sec:massivesne}

Explosions develop considerably less readily in progenitor
stars more massive than, roughly, 10\,$M_\odot$. In particular,
neutrino-driven explosions for such stars cannot be obtained
in spherical symmetry, which means that the explosion mechanism
is genericly multi-dimensional. Non-radial hydrodynamic instabilities,
i.e., buoyant Rayleigh-Taylor 
plumes\index{instability!Rayleigh-Taylor}, convective overturn, 
SASI\index{standing-accretion-shock instability (SASI)}
mass motions, and turbulent flow
fragmentation in the postshock layer play a crucial role for
the success of the mechanism, instead of being just an accompanying 
phenomenon. They provide decisive support for the shock expansion
and thus can foster explosions even when 1D simulations fail
(see, e.g., \citealt{Herantetal1994,Burrowsetal1995,JankaMueller1996,MurphyBurrows2008B,Nordhausetal2010,Hankeetal2012}, and for reviews 
\citealt{Janka2012,Burrows2013,Jankaetal2016,Mueller2016}).

The higher mass-accretion rate\index{mass-accretion rate}
($\dot M$) of the collapsing core
of massive SN progenitors, which increases with higher values
of the core compactness\index{core compactness}, 
has two competing effects. On the
one hand a higher mass accretion rate hampers the outward
expansion of the stalled SN shock, because it has to be overcome
by the postshock pressure $P_\mathrm{s}$:
\begin{equation}
P_\mathrm{s} \ge \left( 1 - \beta_\rho^{-1}\right)\,\rho_0 v_0^2
= \frac{\dot M\sqrt{2G\,M}}{4\pi\,R_\mathrm{s}^{5/2}}\,,
\label{eq:postpress}
\end{equation}
where $R_\mathrm{s}$ is the 
shock-stagnation radius\index{shock-stagnation radius}, and
\begin{eqnarray}
v_0 &\approx& -\,\sqrt{\frac{2G\,M}{R_\mathrm{s}}}\,,
\label{eq:vfree}\\
\rho_0 &=& \frac{\dot M}{4\pi\,R_\mathrm{s}^2\,|v_0|} 
\label{eq:rhopre}
\end{eqnarray}
are the velocity (nearly free-fall) and density, respectively,
of the collapsing matter ahead of the shock, $M$ is the mass
of the accreting, nascent NS, and
$\beta_\rho = \rho_\mathrm{s}\rho_0^{-1}$
the ratio of postshock to preshock density
\citep{Janka2001A}. A higher mass
accretion rate therefore requires a bigger postshock
pressure to stabilize the shock at a given radius (equality
relation in Eq.~\ref{eq:postpress}) or to initiate outward
shock expansion according to the inequality condition in 
Eq.~(\ref{eq:postpress}).

On the other hand, a higher mass accretion rate enhances the 
total luminosity of electron neutrinos and antineutrinos,
\begin{equation}
L_{\nu_e} + L_{\bar\nu_e} = 2\beta_1 L_{\nu_x} +
\beta_2\,\frac{G M\,\dot M}{R_\mathrm{ns}}\,.
\label{eq:acclum}
\end{equation}
Here the first term on the rhs represents the contribution
from neutrinos diffusing out of the
high-density core of the NS. It is expressed in terms
of the luminosity of one species of 
heavy-lepton neutrinos\index{heavy-lepton neutrinos}
($\nu_x = \nu_\mu$, $\overline{\nu}_\mu$, $\nu_\tau$, or
$\overline{\nu}_\tau$). The second term stands for the
accretion component\index{accretion luminosity}, 
i.e., for energy that is radiated away by 
$\nu_e$ and $\overline{\nu}_e$ produced in the fresh inflow of
matter settling onto the NS surface at radius $R_\mathrm{ns}$.
Equation~(\ref{eq:acclum}) assumes steady-state conditions
(i.e., the mass-flow rate arriving at the NS surface is the
same as the accretion rate $\dot M$ ahead of the shock), 
and the parameters have the values of $\beta_1 = 1$ 
and $\beta_2 \approx 0.5$--1 \citep{MuellerJanka2014}. 
According to Eq.~(\ref{eq:acclum}), a higher mass-accretion
rate enhances the neutrino luminosity and therefore also
the heating behind the shock by the reabsorption of some of
the radiated neutrinos in the processes of Eqs.~(\ref{eq:nueabs})
and (\ref{eq:barnueabs}).

Two different effects associated with the mass infall of the
collapsing stellar core therefore rival each other concerning their
influence on the shock evolution: a higher mass accretion rate
means a larger ram pressure\index{ram pressure}, 
damping the shock expansion, and
it also leads to stronger neutrino heating, supporting shock
expansion. Which of these two effect wins, and whether the
neutrino-driven mechanism leads to a successful SN explosion,
is a quantitative question and sensitive to the detailed 
structure of each individual progenitor star.  
The neutrino-driven mechanism is therefore not ``robust'' in
the sense that success can be taken for granted for all 
massive SN progenitors once it has been 
demonstrated to work for selected cases. 

Recent self-consistent 3D simulations, which for the first time
include energy-dependent neutrino transport treatments, were
able to obtain successful explosions for some progenitors above
10\,$M_\odot$ \cite[namely for stars of 11.2, 15, 18, 20, 
27\,$M_\odot$, and 15\,$M_\odot$ with rotation 
by][respectively]{Takiwakietal2012,Takiwakietal2014,Lentzetal2015,Mueller2016,Melson2015b,Robertsetal2016,Jankaetal2016}, 
see Fig.~\ref{fig:janka-s20} for an example.
The results, though slightly less optimistic than 2D simulations,
still back up the neutrino-driven mechanism and suggest its
viability in principle. Even if it does not produce explosions
in other investigated cases, the neutrino mechanism seems to 
fail only marginally. 

Still, however, full-scale 3D models are 
not finally conclusive and require improvements in several
aspects. For example, they still lack sufficient (some more,
some less) numerical resolution for convergence on the description
of the turbulent flow in the postshock layer, employ a variety of
(different) approximations for the 3D neutrino transport, 
omit important microphysics, obtain explosions only with
special and not generally accepted assumptions such as rapid
rotation \citep{Jankaetal2016} or nonstandard modifications of
the neutrino opacities \citep{Melson2015b}, and start from
spherically symmetric initial conditions instead of the 
multi-dimensional flow that perturbs the convective burning 
shells\index{convective burning shells}
around the iron core already prior to collapse
\citep{Couchetal2015,Muelleretal2016,Mueller2016}.
It is well possible that the neutrino mechanism will 
work well and can explain the observational properties 
(e.g., explosion energies and radioactive yields) for a 
wide range of massive progenitors once all these
remaining deficiencies of current models have been removed.
It is also possible, however, that important
physics is still missing, for example new phenomena in the
nuclear sector (e.g., stronger nucleon-correlation effects
in the neutrino opacities than currently considered)
or associated with particle and neutrino physics
(e.g., neutrino-flavor oscillations\index{neutrino-flavor oscillations}
or a sterile neutrino\index{sterile neutrino}) 
inside of the new-born NS.

\section{Physics of the Neutrino-driven Mechanism}
\label{sec:mechanismphysics}

Meanwhile, based to a large extent on the education obtained
from continually improved numerical studies, a rather detailed
picture of the crucial ingredients and of their interplay in the 
neutrino-driven mechanism has been worked out. In this Section
the current understanding of the physics of neutrino-powered
explosions is briefly reviewed.

\subsection{Neutrino Heating}
\label{sec:heating}

Neutrinos deposit energy behind the stalled bounce shock
mainly through the absorption of $\nu_e$ and $\overline{\nu}_e$
by neutrons and protons, respectively (Eqs.~\ref{eq:nueabs}
and \ref{eq:barnueabs}). Energy transfer by scattering reactions
of neutrinos and antineutrinos of all flavors with electrons, 
positrons, and nucleons, by neutrino-antineutrino 
annihilation\index{neutrino-antineutrino annihilation},
and by neutrino interactions
with light nuclei\index{light nuclei}
(deuterons, tritons, $^3$He and 
$\alpha$-particles) altogether contribute to the neutrino
heating on a much lower level (much less than 10\%).

The formation of a postshock layer where neutrinos effectively
deposit energy can be easily understood by considering the
heating and cooling rates associated with the dominant 
processes. Neutrino absorptions (Eqs.~\ref{eq:nueabs} and 
\ref{eq:barnueabs}) dump energy in the stellar plasma outside
of the neutrinospheres\index{neutrinosphere} at a rate per
nucleon of
\begin{equation}
q_{\nu}^+ \approx  110\cdot \left(
{L_{\nu_e,52}\langle E_{\nu_e,15}^2
\rangle\over r_7^2\,\,s_{r,\nu_e} }
\,Y_n\,+\,
{L_{\bar\nu_e,52}\langle E_{\bar\nu_e,15}^2\rangle\over r_7^2\,\,
s_{r,\bar\nu_e} }\,Y_p \right)
\quad
\left\lbrack {{\rm MeV}\over {\rm s}\cdot {\rm nucleon}}\right\rbrack \,,
\label{eq:nuheating}
\end{equation}
while the inverse processes of $\nu_e$ and $\overline{\nu}_e$
emission by electron and position captures on free nucleons
extract energy from the stellar medium at a rate of per nucleon 
of
\begin{equation}
q_{\nu}^- \approx 145\,
\left( {k_{\mathrm{B}}T\over 2\,{\mathrm{MeV}}}\right)^{\! 6}
\ \ \left\lbrack{{\mathrm{MeV}}\over{\mathrm{s}}\cdot {\rm nucleon}}
\right\rbrack \,.
\label{eq:nucooling}
\end{equation}
In the latter expression the assumption is used
that the sum of neutron and proton abundances is unity,
$Y_n+Y_p = 1$. Moreover, it is assumed that the electron 
and positron degeneracy parameters\index{degeneracy parameter},
$\eta_{e^\pm} = \mu_{e^\pm}/(k_{\mathrm{B}}T)$ (where 
$\mu_{e^\pm}$ are the chemical potentials\index{chemical potential}), 
are small: $\eta_{e^-} = -\eta_{e^+}\approx 0$.
This approximation is good in the shock-heated layers because the
electron (number) fraction\index{electron (number) fraction} 
$Y_e = n_e/n_{\mathrm{b}}$ ($n_e$ and
$n_{\mathrm{b}}$ being the electron and baryon number density,
respectively) and thus the electron 
degeneracy\index{electron degeneracy} is rather low
and $e^\pm$ pairs are abundant. In Eq.~(\ref{eq:nucooling}), 
$k_{\mathrm{B}}$ is Boltzmann's constant and $T$ is measured
in Kelvin. In Eq.~(\ref{eq:nuheating}), $r_7$
is the radius in $10^7\,$cm, the neutrino luminosities $L_{\nu_i}$
are normalized to $10^{52}\,$erg$\,$s$^{-1}$, and
the mean squared neutrino energies\index{neutrino!mean squared energy}, 
$\langle E_{\nu_i}^2\rangle$,
are given in units of $(15\,{\rm MeV})^2$. The quantities 
$s_{r,\nu_i}$ in the denominators are the so-called
flux factors\index{flux factor}, 
which account for the fact that not all neutrinos
move radially outwards when they stream through the medium still
close outside of the neutrinospheres. The neutrino quantities
$L_{\nu_i}$, $\langle E_{\nu_i}^2\rangle$, and $s_{r,\nu_i}$
are calculated from the neutrino phase space occupation 
function\index{neutrino!phase space occupation function}
${\cal F}_{\nu_i}(\qvec{r},E,\qvec{n},t)$ at spatial point
$\qvec{r}$ at time $t$ ($\qvec{n}$ is the unit vector of the 
neutrino momentum corresponding to energy $E$) according to
\begin{eqnarray}
L_{\nu_i}(\qvec{r},t) &=& {c \over (hc)^3} \int_{A_r}\!\mathrm{d}A_r\,
\int\limits_0^\infty{\mathrm{d}}E\,E^3\!
\int_{4\pi}\!\mathrm{d}\Omega\,\,\qvec{n}\cdot\qvec{n}_r\,\,
{\cal F}_{\nu_i}(\qvec{r},E,\qvec{n},t)\,,\label{eq:nulum}\\
\langle E_{\nu_i}^2\rangle(\qvec{r},t) &=& {\cal E}^{-1}
\int_0^\infty\!\mathrm{d}E\,E^5\!
\int_{4\pi}\!\mathrm{d}\Omega\,{\cal F}_{\nu_i}(\qvec{r},E,\qvec{n},t)\,,
\label{eq:avenue}\\
s_{r,\nu_i}(\qvec{r},t) &=& {\cal E}^{-1} 
\int_0^\infty\!\mathrm{d}E\,E^3\!
\int_{4\pi}\!\mathrm{d}\Omega\,\,\qvec{n}\cdot\qvec{n}_r\,\,
{\cal F}_{\nu_i}(\qvec{r},E,\qvec{n},t)
\,, \label{eq:radfluxfactor}
\end{eqnarray}
where $\qvec{n}_r$ is the outward normal vector on a sphere of 
radius $r = |\qvec{r}|$, $A_r = 4\pi r^2$ is the area of the
sphere, $\mathrm{d}A_r$ a surface element on this sphere,
and ${\cal E}$ a normalization factor defined by
${\cal E} = \int_0^\infty\!\mathrm{d}E\,E^3\! \int_{4\pi}\!
\mathrm{d}\Omega\,{\cal F}_{\nu_i}(\qvec{r},E,\qvec{n},t)$.

Equations~(\ref{eq:nulum})--(\ref{eq:radfluxfactor}) are depicted for
the general 3D case. In a nonspherical situation $q_{\nu}^+$ and
$q_{\nu}^-$ of Eqs.~(\ref{eq:nuheating}) and (\ref{eq:nucooling}),
respectively, should be considered as averages over spheres at
radii $r$. The scalar $s_{r,\nu_i}$ is the radial component of the 
flux factor\index{flux factor} $\qvec{s}_{e,\nu_i}$, 
which relates the neutrino
energy density with the local 
neutrino energy-flux density\index{neutrino!energy-flux density}:
$\qvec{F}_{e,\nu_i} = c\,\qvec{s}_{e,\nu_i}\,\epsilon_{\nu_i}$,
where
\begin{eqnarray}
\epsilon_{\nu_i} &=& {1 \over (hc)^3}
\int_0^\infty\!\mathrm{d}E\,E^3\! \int_{4\pi}\!
\mathrm{d}\Omega\,{\cal F}_{\nu_i}(\qvec{r},E,\qvec{n},t)\,,
\label{eq:enerden} \\
\qvec{s}_{e,\nu_i}(\qvec{r},t) &=& {\cal E}^{-1}
\int_0^\infty\!\mathrm{d}E\,E^3\!
\int_{4\pi}\!\mathrm{d}\Omega\,\,\qvec{n}\,{\cal F}_{\nu_i}(\qvec{r},E,\qvec{n},t)
\,. \label{eq:fluxfactor}
\end{eqnarray}
The neutrino-heating rate depends on the local neutrino
energy density, $\epsilon_{\nu_i}$, and the mean squared energy 
$\langle E_{\nu_i}^2\rangle$. 
In order to coin the heating rate in
a form familiar from the 1D case, the neutrino luminosity 
(Eq.~\ref{eq:nulum}) through the sphere of radius $r$
was introduced in Eq.~(\ref{eq:nuheating})
by integrating the projected energy flux, 
$\qvec{F}_{e,\nu_i}\cdot \qvec{n}_r$, over the surface of the sphere.
The flux factor $s_{r,\nu_i}$ is typically close to 0.25 near
the neutrinosphere and gradually approaches unity when the neutrino 
distributions get more and more forward
peaked in the limit of free streaming with increasing distance from the
neutrinosphere\index{neutrinosphere}. 
Angle-dependent transport, i.e., solving the Boltzmann
transport equation\index{Boltzmann transport equation}, 
is necessary to accurately determine the spectral
and angular distributions of the neutrinos.

Since outside of the neutrinosphere\index{neutrinosphere}
the neutrino luminosities\index{neutrino!luminosity} and 
mean squared energies\index{neutrino!mean squared energy} 
are only weakly dependent on the radius and at
some distance away from the neutrinosphere also the flux factor is
already close to its asymptotic value of unity, Eq.~(\ref{eq:nuheating})
implies that the neutrino-heating rate\index{neutrino-heating rate}
drops roughly with $r^{-2}$. In contrast, the neutrino-cooling
rate\index{neutrino-cooling rate} (Eq.~\ref{eq:nucooling}) 
decreases much more steeply with approximately $r^{-6}$ because of  
a $T\propto r^{-1}$ decline in 
radiation-pressure dominated\index{radiation pressure}
low-density plasma behind the shock \citep{Janka2001A}. 
Since at a time later than a few
10\,ms after core bounce, the postshock medium becomes sufficiently
tenuous for radiation-dominated\index{radiation dominated}
conditions to prevail, the steep
profile of the cooling rate allows for the appearance of a 
crossing point
of $q_{\nu}^+$ and $q_{\nu}^-$, outside of which neutrino heating 
dominates neutrino cooling. The corresponding radius is called 
``gain radius''\index{gain radius}
\citep{BetheWilson1985}, $R_\mathrm{g}$, and 
satisfies the relation
\begin{equation}
R_\mathrm{g}\,T_\mathrm{g}^3\propto\sqrt{L_\nu\langle E_\nu^2\rangle}
\,,
\label{eq:gainradius}
\end{equation}
when $T_\mathrm{g} = T(R_\mathrm{g})$, and the total luminosity and 
mean squared energy of electron-flavor neutrinos are introduced
according to 
\begin{eqnarray}
L_\nu &=& L_{\nu_e} + L_{\bar\nu_e} \,, \label{eq:totlum}\\
\langle E_\nu^2\rangle &=& \frac{L_{\nu_e}\langle E_{\nu_e}^2\rangle
+ L_{\bar\nu_e}\langle E_{\bar\nu_e}^2\rangle}{L_{\nu_e}+L_{\bar\nu_e}}
\,,\label{eq:meane2}
\end{eqnarray}
using $Y_n \approx Y_p \approx 0.5$ and 
$s_{r,\nu_e}\approx s_{r,\bar\nu_e}\approx 1$ 
as reasonable approximations in Eq.~(\ref{eq:nuheating}). The
gain radius $R_\mathrm{g}$ thus separates a neutrino-cooling
layer between the neutrinosphere and this radius from a 
neutrino-heating layer (``gain layer''\index{gain layer}) 
between $R_\mathrm{g}$ and the shock front (at $R_\mathrm{s}$).

Appreciable amounts of energy can be transferred to the stellar
gas in the gain layer. Considering a mass 
$M_\mathrm{g} = \int_{R_\mathrm{g}}^{R_\mathrm{s}}
\mathrm{d}V\,\rho(\qvec{r})$ in this region and
employing Eqs.~(\ref{eq:nuheating}), (\ref{eq:totlum}), and
(\ref{eq:meane2}), one estimates a total energy deposition rate by
$\nu_e$ plus $\bar\nu_e$ absorption of
\begin{eqnarray}
Q_\nu^+ \!&=&\! q_\nu^+\,\frac{M_\mathrm{g}}{m_\mathrm{u}} \cr
\! &\sim& \! 6.3\times 10^{51}\,\frac{\mathrm{erg}}{\mathrm{s}}\,
\left ( \frac{\langle E_\nu^2\rangle}{(15\,\mathrm{MeV})^2} \right )\!
\left ( \frac{L_\nu}{6\times 10^{52}\,\mathrm{erg/s}} \right ) \cr
\! &\phantom{\sim}&\!
\phantom{6.3\times 10^{51}\,\frac{\mathrm{erg}}{\mathrm{s}}\,
\left ( \frac{\langle E_\nu^2\rangle}{(15\,\mathrm{MeV})^2} \right )\!}
\left ( \frac{M_\mathrm{g}}{0.01\,M_\odot} \right ) \!
\left ( \frac{R_\mathrm{g}}{100\,\mathrm{km}} \right )^{\! -2} \!, 
\label{eq:heatingrate}
\end{eqnarray}
which corresponds to a heating efficiency\index{heating efficiency}
$(Q_\nu^+ - Q_\nu^-)/L_\nu$
of typically 5--10\% when energy losses by the re-emission of neutrinos
from the gain layer (with an integrated rate of $Q_\nu^-$) are taken 
into account.

\subsection{Competition of Time Scales}
\label{sec:competition}

The energy deposited by neutrinos in the gain layer 
does not only depend on the product of luminosity and 
mean squared energy of the radiated neutrinos (the
``heating functional''\index{heating functional}) 
but also on the mass that is
neutrino-heated (Eq.~\ref{eq:heatingrate}) and on the 
duration matter is exposed to the neutrino heating. In the
spherically symmetric (1D) case, the matter accreted by
the stalled shock flows radially inward through the gain
layer. This implies that the residence time of the
matter in the gain layer is given by the 
advection time\index{advection time}
\begin{equation}
t_\mathrm{adv} = \int_{R_\mathrm{g}}^{R_\mathrm{s}} 
\frac{\mathrm{d}r}{|v_r(r)|} 
\sim \frac{R_\mathrm{s}-R_\mathrm{g}}{|v_1|} \,,
\label{eq:tadvec}
\end{equation}
which measures how long it takes the
accretion flow (with radial velocity $v_r(r)$) 
to move from $R_\mathrm{s}$ to
$R_\mathrm{g}$ \citep{Janka1998,Janka2001B,Thompsonetal2005}.
In Eq.~(\ref{eq:tadvec}) the last expression on the rhs side
is a rough approximation invoking the postshock velocity
$v_1 = v_0\,\beta_\rho^{-1}$ (with $\beta_\rho$ being the density 
jump in the shock and $v_0$ the preshock velocity of
Eq.~\ref{eq:vfree}). Instead of using Eq.~(\ref{eq:tadvec})
with (mass-weighted, angle-averaged)
mean values of the relevant quantities in the multi-dimensional 
case, where the postshock
layer is stirred by violent non-radial motions due to SASI
activity and buoyant flows, the dwell time\index{dwell time} 
of matter 
in the gain region can be represented by the more general
expression \citep{Burasetal2006B,MarekJanka2009}
\begin{equation}
t_\mathrm{dwell}\,\approx \,\frac{M_\mathrm{g}}{\dot M} \,,
\label{eq:tdwell}
\end{equation}
which relates the mass in the gain layer, $M_\mathrm{g}$, with
the mass accretion rate $\dot M$ through the shock and (for
conditions near steady state) through the gain radius. 
Equation~(\ref{eq:tdwell}) thus measures the time 
the (stationary) accretion flow needs to replace all the mass
in the gain layer by fresh matter, independent of the detailed
flow structure in this layer.

For discussing neutrino heating as a driver of the explosion,
the time matter is exposed to the heating must be compared to
the time that is required to deposit the energy that brings
matter to the threshold of becoming unbound in the gravitational
field of the NS. For this to be reached, neutrinos have to 
transfer the equivalent of the total 
binding energy\index{total binding energy} of the 
matter in the gain layer, which takes a time of
\begin{equation}
t_\mathrm{heat} = \frac{\int_{R_\mathrm{g}}^{R_\mathrm{s}}
\mathrm{d}V\,\rho(\qvec{r})\,|e_\mathrm{tot}(\qvec{r})|}{Q_\nu^+ - Q_\nu^-} 
= \frac{E_\mathrm{tot,g}}{Q_\nu^+ - Q_\nu^-}
\sim 
\frac{m_\mathrm{u}|e_\mathrm{tot}(R_\mathrm{g})|}{q_\nu^+-q_\nu^-}\,,
\label{eq:theat}
\end{equation}
where $e_\mathrm{tot}(\qvec{r})$ is the total (i.e., internal 
plus kinetic plus gravitational plus magnetic) specific 
(i.e., per unit of mass) energy of the stellar plasma at 
spatial point $\qvec{r}$ and $E_\mathrm{tot,g}$ the total
energy in the gain region. 

Runaway shock expansion\index{runaway shock expansion}
requires that $t_\mathrm{heat}
\lesssim t_\mathrm{dwell} \approx t_\mathrm{adv}$. This condition
implies that the matter behind the stalled shock can absorb
from neutrinos an energy equivalent to its net gravitational
binding energy during the time interval the gas stays
in the gain layer, i.e., before it is advected into the 
cooling layer and loses its energy again by efficient radiation
of neutrinos. Since the shock begins to expand due to the 
pressure increase associated with the growing thermal energy
in the gain region, $t_\mathrm{adv}$ increases  
further and a favorable situation for explosive runaway
ensues. Numerical simulations confirm this 
necessary (though not sufficient) condition for the onset
of a SN explosion by the neutrino-driven mechanism
\citep[see, e.g.,][]{Burasetal2006B,MarekJanka2009,Fernandez2012,Summaetal2016,Jankaetal2016},
and \citet{Mueller2016} argued that other heuristic
explosion criteria proposed in the literature, e.g.\
the antesonic condition\index{antesonic condition} of 
\citet{PejchaThompson2012},
have basically a very similar physical meaning.

Multi-dimensional flows in the postshock layer due to 
hydrodynamic instabilities such as convective overturn and the SASI
have a multitude of favorable effects for the development of
an explosion. They lead to an increase of the shock radius 
by pushing the shock farther out (see
Sect.~\ref{sec:turbulence}) and thus allow for a larger mass
in the gain layer; they stretch the dwell time of matter in
the wider gain region and
therefore increase the neutrino heating and the heating 
efficiency \emph{for a fixed value of the neutrino-heating 
functional}\index{heating functional}
(i.e., of the product of luminosity and mean 
squared energy of $\nu_e$ plus $\overline{\nu}_e$;
Eqs.~\ref{eq:totlum} and \ref{eq:meane2})
compared to the 1D case; they reduce energy 
losses from the gain layer through the re-emission of
neutrinos because neutrino-heated, high-entropy
matter can become buoyant and rise outward
to cooler regions farther away from the gain radius; and
they additionally reduce the heating time scale by raising
the total energy of the gas in the gain layer due to
kinetic energy of non-radial flows and turbulent motions.
All these effects in combination facilitate the onset of
explosions in more than one spatial 
dimension for lower values of the heating 
functional than in spherical symmetry (see also 
Sect.~\ref{sec:critlum}) and enable neutrino-driven
explosions to become more energetic for given neutrino
luminosity and spectra.

\subsection{Energetics of Neutrino-driven Explosions}
\label{sec:expenergy}

The fact that runaway shock 
expansion\index{runaway shock expansion} sets in when the 
neutrino heating and dwell time scales become equal
means that the postshock matter expands away from the 
gain radius as soon as it has absorbed an energy roughly
equal to its binding energy in the gravitational potential
of the NS. This implies that at least some of the matter 
in the gain layer begins to acquire positive total 
(i.e., internal plus kinetic plus gravitational)
energy\index{total energy} $e_\mathrm{tot}$.
As a consequence ---unless a very intense neutrino
flash were able to superheat the matter in the gain
layer within a short instant of time, creating an 
electron-positron pair-plasma fireball\index{fireball}---,
neutrinos do \emph{not} deliver the net energy of the
SN blast, although their energy deposition is crucial
to enable the onset of the explosion.

Rather than being transferred by neutrino interactions,
the excess energy that the SN ejecta will 
possess after their escape from the gravitational
trough of the compact remnant, is provided
by the release of nuclear binding energy in the
recombination\index{nuclear recombination}
of free nucleons to $\alpha$-particles
and iron-group nuclei.
This recombination sets in when the postshock matter 
expands to a radius exceeding roughly 
$R_\mathrm{diss} \sim 240$\,km (Eq.~\ref{eq:rdiss}
with $M\sim 1.5\,M_\odot$), i.e., at a distance from
the neutrinosphere where neutrino heating has already
become inefficient. Nuclear 
recombination\index{nuclear recombination} releases 
$E_\mathrm{nuc} = 7$--8.8\,MeV per nucleon, depending 
on whether helium or iron-group nuclei form. Numerical
simulations, however, show that effectively $\gtrsim5\,$MeV
per nucleon contribute to the net positive energy that
matter can carry away to infinity.

Therefore the explosion energy of neutrino-driven SNe
depends crucially on the amount of material that is 
expelled by neutrino heating and releases its nuclear
recombination energy during outward expansion. At the
onset of the explosion, the mass in the gain layer can
be estimated as
\begin{eqnarray}
M_\mathrm{g} &=& \int_{R_\mathrm{g}}^{R_\mathrm{s}}
\mathrm{d}V\,\rho(\qvec{r}) \sim
\frac{\dot M\,R_\mathrm{s}^{3/2}}{\sqrt{2\,G\,M}}\,\,
\beta_\rho\,\ln\left(\frac{R_\mathrm{s}}{R_\mathrm{g}}\right)
\label{eq:mgain1}\\
&\approx& 6.3\times 10^{-3}\,M_\odot\,
\left(\frac{\dot M}{0.2\,M_\odot/\mathrm{s}}\right)\!
\left(\frac{R_\mathrm{s}}{200\,\mathrm{km}}\right)^{\! 3/2}\!
\left(\frac{M}{1.5\,M_\odot}\right)^{\!-1/2} ,
\label{eq:mgain2}
\end{eqnarray}
where we made use of Eqs.~(\ref{eq:vfree},\ref{eq:rhopre}),
assumed a density profile following 
$\rho(r)\propto r^{-3}$ in the 
radiation-dominated\index{radiation dominated}, nearly
isentropic\index{isentropic} (because convectively mixed) 
postshock layer \citep{Janka2001A}, and applied
$\beta_\rho\ln(R_\mathrm{s}R_\mathrm{g}^{-1})\approx 7$
and typical values also for the other quantities in 
Eq.~(\ref{eq:mgain2}). With a recombination energy
$E_\mathrm{nuc}$ between $\sim$5\,MeV and at most
8.8\,MeV per nucleon and a mass $M_\mathrm{heat}$ for 
the neutrino-heated ejecta, we therefore estimate an energy
release of
\begin{eqnarray}
E_\mathrm{rec} &=& E_\mathrm{nuc}\,
\left(\frac{M_\mathrm{heat}}{m_\mathrm{u}}\right) 
\approx (9.6\,...\,17.0)\times 10^{50}\,\mathrm{erg}\,
\left(\frac{M_\mathrm{heat}}{0.1\,M_\odot}\right) 
\label{eq:erecom1} \\
&\approx& (6.0\,...\,10.7)\times 10^{49}\,\mathrm{erg}\,
\left(\frac{M_\mathrm{g}}{6.3\times 10^{-3}\,M_\odot}\right) \nonumber\\
&\approx& (6.0\,...\,10.7)\times 10^{49}\,\mathrm{erg}\,
\left(\frac{\dot M}{0.2\,M_\odot/\mathrm{s}}\right)\!
\left(\frac{R_\mathrm{s}}{200\,\mathrm{km}}\right)^{\! 3/2}\!
\left(\frac{M}{1.5\,M_\odot}\right)^{\!-1/2} \!\!\!\! ,
\label{eq:erecom2}
\end{eqnarray}
where we introduced Eq.~(\ref{eq:mgain2}) in the second 
and third line. From Eq.~(\ref{eq:erecom2}) it is evident
that the mass in the gain layer at the beginning of the
explosion is not sufficient to explain explosions significantly
more energetic than about $10^{50}$\,erg. A SN energy of
$10^{51}$\,erg requires a mass of $\sim$0.1\,$M_\odot$ of
neutrino-heated ejecta. However, when the
explosion sets in, additional matter ---typically several
times the original mass of the gain layer--- expands from 
the cooling layer through the gain radius into the gain 
layer, for which reason the mass in the gain layer rises
steeply once the SN shock takes off. 
Moreover, multi-dimensional simulations point to
a long period (possibly lasting for more than a second) 
of continued accretion after the onset of the explosion,
during which a considerable fraction of the accreted 
matter is re-ejected after having absorbed energy from 
neutrinos in the vicinity of the gain radius
\citep{MarekJanka2009,Mueller2015,Bruennetal2016,Muelleretal2016B}. 
Finally, also the neutrino-driven wind\index{neutrino-driven wind}
that is expelled by neutrino 
heating from the surface of the new-born NS contributes to 
the mass of neutrino-heated ejecta and to the explosion 
energy through its wind power.
The energy of neutrino-driven explosions can 
therefore be expressed as
\begin{equation}
E_\mathrm{exp} \approx E_\mathrm{nuc}\, 
\left(\frac{M_\mathrm{heat}}{m_\mathrm{u}}\right) 
+ E_\mathrm{burn} - E_\mathrm{bind} 
\label{eq:eexp}
\end{equation}
with
\begin{equation}
M_\mathrm{heat} = f\,M_\mathrm{g} + 
\dot M\, t_\mathrm{acc}\,\eta_\mathrm{ej} + M_\mathrm{wind}\,.
\label{eq:heatmass}
\end{equation}
In Eq.~(\ref{eq:heatmass}), $f\ge 1$ is a numerical factor
accounting for the additional mass shifting from the cooling
layer into the gain layer at the onset of the explosion, 
$t_\mathrm{acc}$ is the period of time over which accretion 
with (an approximately constant) rate $\dot M$ continues 
beyond the start of the explosion,
$\eta_\mathrm{ej}\le 1$ is the fraction of the accreted matter
that gets re-ejected after having absorbed energy from neutrinos,
and $M_\mathrm{wind}$ is the mass of the neutrino-driven wind
blown off the NS surface. For simplicity, in Eq.~(\ref{eq:eexp})
the effective recombination\index{nuclear recombination}
energy per nucleon (i.e., after
subtracting the remaining gravitational binding effects) is 
assumed to be the same for all three components of the
neutrino-heated ejecta, although there can be some variation
in time as well as between the different components. The term
$E_\mathrm{burn}$ adds the extra energy provided by explosive
nuclear burning in progenitor layers that are heated and 
(directly) ejected by the outgoing SN shock, and $E_\mathrm{bind}$
corrects for the gravitational binding energy\index{binding energy}
of these layers.
Equation~(\ref{eq:eexp}) is approximate also because it assumes
that there is a well defined transition between the phase of 
simultaneous accretion and mass re-ejection on the one hand
and direct outward acceleration of shock-heated matter on the
other hand.

The different terms in Eqs.~(\ref{eq:eexp}) and (\ref{eq:heatmass})
are strongly dependent on the progenitor star, with the mass in 
the gain layer given by Eqs.~(\ref{eq:mgain1},\ref{eq:mgain2})
as a function of the mass-accretion rate, $\dot M$, which itself
depends on the compactness\index{core compactness}
of the progenitor core. The neutrino-wind\index{neutrino-driven wind}
mass $M_\mathrm{wind}$ scales steeply with the neutrino luminosity 
of the new-born NS and its mass, because $\dot M_\mathrm{wind} \propto
L_\nu^a\,M^{-b}$ with $a = 2.4\,...\,2.5$ and $b = 2$ for Newtonian
gravity \citep{QianWoosley1996} and $b = 2.7\,...\,3.1$ for 
general-relativistic potential \citep{Thompsonetal2001}.
In the case of O-Ne-Mg-core progenitors a non-negligible fraction of
the explosion energy comes from the neutrino-driven wind
(third term on the rhs in Eq.~\ref{eq:heatmass}), whereas massive 
iron-core progenitors with large core-compactness values obtain
most of their explosion energy from 
the second term on the rhs of Eq.~(\ref{eq:heatmass}).

Energy from explosive nuclear fusion\index{nuclear fusion}
of silicon and oxygen to nickel releases at most
about 0.8\,MeV per nucleon and thus provides an energy of
\begin{equation}
E_\mathrm{fusion} \lesssim 0.8\,\mathrm{MeV}\,
\left(\frac{M_\mathrm{Ni}}{m_\mathrm{u}}\right) \approx
1.54\times 10^{50}\,\mathrm{erg}\,
\left(\frac{M_\mathrm{Ni}}{0.1\,M_\odot}\right)\,.
\label{eq:efusion}
\end{equation}
In core-collapse SNe nuclear fusion never contributes more than
roughly 10\% of the explosion energy:
low-mass progenitors eject very little nickel
($\lesssim$$10^{-2}$\,$M_\odot$, some of which stems from nucleon
recombination\index{nuclear recombination}
in expanding, neutrino-heated plasma) so that less than
$\sim$$10^{49}$\,erg are produced by explosive burning, while in
high-mass progenitors less than about $\sim$$10^{50}$\,erg result
from nuclear burning (SN~1987A\index{SN~1987A}, 
for example, was the explosion of
a 15--20\,$M_\odot$ star with an energy of 
$\sim$$1.3\times 10^{51}$\,erg and a $^{56}$Ni mass of about
0.07\,$M_\odot$).

In neutrino-powered SNe neutrinos transfer the
energy to matter that expands away from the heating layer 
as soon as it has absorbed enough energy to overcome its 
gravitationally bound state. Therefore the explosion energy
scales roughly with the mass $M_\mathrm{heat}$ of the 
neutrino-heated matter (Eqs.~\ref{eq:eexp},\ref{eq:heatmass}).
For this reason the order of magnitude of the
explosion energy is, very approximately, also
determined by the gravitational binding
energy of the progenitor layers around the mass cut
that separates ejecta and compact remnant. (Note that the
collapse of these layers proceeds nearly 
adiabatically.\index{adiabatic})
This insight explains the wide spread of observationally
inferred energies of core-collapse SNe from less than
$\sim$$10^{50}$\,erg for the lowest-mass SN progenitors
to more than $\sim$$10^{51}$\,erg for high-mass stars
like the progenitors of SN~1987A\index{SN~1987A} and 
Cassiopeia~A\index{Cassiopeia~A}
(excluding hypernovae\index{hypernova}
with their more extreme energies).
It also allows one to understand why the energy scale 
of core-collapse SNe
is similar to that of Type~Ia SNe\index{Type~Ia SN}, 
whose energies are 
of the order of the gravitational binding energy of the 
exploding white dwarfs\index{white dwarf}. 
In both cases one considers 
degenerate stellar objects or cores, and in both cases
the gravitational binding energy is dominated by matter
near the surface of these degenerate bodies
\citep[see also][]{Burrows2013}. 

The expansion of neutrino-heated matter away from the
region of strong heating provides a natural
feedback mechanism\index{feedback!mechanism}
that regulates the explosion energy of
neutrino-powered SNe to be of the order of the gravitational
binding energy of the progenitor shells around the mass cut. 
This is a distinct difference from the case of 
hypernovae\index{hypernova},
whose much larger energy scale points to a fundamentally
different explosion mechanism that does not only lack such 
a regulating feedback\index{feedback!regulating}, 
but which is also much more efficient 
than the neutrino mechanism, and in which it is not stellar
ejecta that store the energy from the beginning. Since
neutrino-antineutrino annihilation\index{neutrino-antineutrino annihilation}
was found not to be
effective enough to create a 
pair-plasma fireball\index{pair-plasma fireball}, the
energy reservoir of hypernova explosions is probably
the huge rotational energy of a NS spinning near its
break-up limit\index{break-up limit}
or of a black hole-accretion torus\index{black hole-accretion torus} 
configuration, whose rotation energy can be highly 
efficiently tapped by strong magnetic 
fields\index{magnetic field}. The energy
at the beginning is therefore electromagnetic energy, not
thermal energy of baryonic matter.
An interesting question is whether SN cases exist where
neutrinos \emph{and} magnetic fields both contribute to
powering the explosion.

In this context it is still necessary to answer the 
question how energetic neutrino-driven explosions can be.
Is the neutrino-driven mechanism able to explain SN
explosions with observed energies around $10^{51}$\,erg 
or more? 1D hydrodynamic simulations with parametric 
neutrino ``engines''\index{neutrino engine}
as well as analytic-parametric 
treatments suggest that neutrino-driven explosions could
be as energetic as $\sim$$2\times 10^{51}$\,erg
(Sect.~\ref{sec:progexprem} and Fig.~\ref{fig:janka-landscape}).
Reliable answers of these questions, however, can only 
be obtained by 3D simulations with a self-consistent 
modeling of all physics ingredients.
From the discussion in this Section it is clear that
a crucial requirement for powerful explosions is that
the SN blast has to be launched sufficiently early, i.e.\
at a time when the mass-accretion rate of the shock is
still sufficiently high to guarantee that the gain
layer contains an appreciable amount of matter and that
a sufficiently large gas mass can be channeled through
the neutrino-heating region also by post-explosion 
accretion and re-ejection
(Eqs.~\ref{eq:mgain1},\ref{eq:mgain2},\ref{eq:eexp},\ref{eq:heatmass}).

\subsection{The Role of Non-radial Flows --- ``Turbulence''}
\label{sec:turbulence}

As discussed in Sect.~\ref{sec:competition}, multi-dimensional
effects, i.e., non-radial mass motions in the postshock
layer, have a variety of explosion-supporting effects.
The growth conditions and saturation properties of such
non-radial, turbulent flows\index{turbulence} associated with
convection and the SASI\index{standing-accretion-shock instability (SASI)} 
have been a matter
of vivid debate. A concise recent review of the relevant
arguments and literature can be found in \citet{Mueller2016}.

Buoyancy-driven convective instability is fostered by the 
negative entropy gradient established by neutrino heating
in the gain layer. Its growth rate in the linear regime is
of the order of the 
Brunt-V\"ais\"al\"a frequency\index{Brunt-V\"ais\"al\"a frequency} 
$\omega_\mathrm{BV}$: 
\begin{equation}
\omega_\mathrm{BV}^2 = g\,\left(\frac{1}{\rho}\,
\frac{\partial\rho}{\partial r} - 
\frac{1}{\rho\,c_\mathrm{s}^2}\,\frac{\partial P}{\partial r}
\right) 
\label{eq:bvfrequency}
\end{equation}
($P$, $\rho$, $c_\mathrm{s}$, and $g\approx G\,M\,r^{-2}$ 
are the local pressure,
density, sound speed, and gravitational acceleration, 
respectively; convective 
instability\index{convective instability} holds when
$\omega_\mathrm{BV}^2 > 0$). Outward rise of buoyant plumes
in the inward moving accretion flow of the postshock layer,
however, requires a somewhat stricter condition to be fulfilled
\citep[unless initial perturbations are 
sufficiently large;][]{Foglizzoetal2006}:
\begin{equation}
\chi \equiv \int_{R_\mathrm{g}}^{R_\mathrm{s}}
\mathrm{d}r\,\,\frac{\omega_\mathrm{BV}}{|v_r(r)|}
\sim \frac{t_\mathrm{adv}}{t_\mathrm{conv}}
\gtrsim 2\,...\,3 \,.
\label{eq:chiparameter}
\end{equation}
Here, all quantities ($\omega_\mathrm{BV}$, $v_r$, 
gain radius $R_\mathrm{g}$, and shock radius $R_\mathrm{s}$)
refer to the angle- and time-averaged averaged mean flow
\citep{Fernandezetal2014}, $t_\mathrm{adv}$ is given
by Eq.~(\ref{eq:tadvec}), and $t_\mathrm{conv}\sim
\langle\omega_\mathrm{BV}^{-1}\rangle$
is the convective growth time scale, volume averaged over
the gain layer. 

The linear growth rate of SASI shock deformation modes
due to an amplifying advective-acoustic 
cycle\index{advective-acoustic cycle} in the
accretion flow between stalled accretion shock and
NS surface is given by
\citep{Foglizzoetal2007,Schecketal2008}:
\begin{equation}
\omega_\mathrm{SASI} = \frac{\ln|{\cal Q}|}{t_\mathrm{cyc}}\,.
\label{eq:omegasasi}
\end{equation}
Here, ${\cal Q}$ is the cycle efficiency ($|{\cal Q}| > 1$
for SASI growth) and
\begin{equation}
t_\mathrm{cyc} = \int_{R_0}^{R_\mathrm{s}}
\frac{\mathrm{d}r}{|v_r(r)|} +
\int_{R_0}^{R_\mathrm{s}} \frac{\mathrm{d}r}{c_\mathrm{s}(r)}
\sim t_\mathrm{adv}
\label{eq:cycletime}
\end{equation}
the duration of the cycle as the sum of sound-travel time (second,
sub-dominant term because $c_\mathrm{s}(r)\gg |v_r(r)|$) and 
advection time between shock radius and a cycle-coupling
radius $R_0$, which is located in the flow-deceleration region
between NS radius ($R_\mathrm{ns}$) and gain radius. (The last
inequality is therefore very approximative since $R_0\sim R_\mathrm{g}$).
According to Eq.~(\ref{eq:omegasasi}) the SASI growth rate scales
inversely with the cycle period. SASI activity is therefore 
expected to be strongest during retraction phases of the
accretion shock, whereas its development is less favored when 
the shock expands and $\chi\gtrsim 3$ becomes fulfilled for 
efficient growth of convection.

The impact of these hydrodynamic instabilities on the
shock propagation and the neutrino-driven mechanism depends on the
saturation amplitude\index{saturation amplitude}
of turbulent velocity fluctuations\index{turbulence!velocity fluctuations},
volume-averaged over the gain layer (indicated by the angle
brackets):
\begin{equation}
\langle\delta v\rangle =
\sqrt{\langle(v_r - \bar v_r)^2\rangle
+ \langle(v_\theta - \bar v_\theta)^2\rangle
+ \langle(v_\phi - \bar v_\phi)^2\rangle} 
= \sqrt{\frac{2\,E_\mathrm{turb,g}}{M_\mathrm{g}}} \,,
\label{eq:deltav}
\end{equation}
where $\bar v_r$, $\bar v_\theta$, and $\bar v_\phi$ are angular 
averages of the $r$, $\theta$, and $\phi$ components of the 
velocity,
respectively, over spherical shells, and $E_\mathrm{turb,g}$
is the turbulent kinetic energy in the gain layer.
\cite{Mueller2016} argued, and presented evidence, for similar
saturation amplitudes, $\langle\delta v\rangle \sim
K\,|\langle v_r\rangle|$ with a constant $K$ in the
same ballpark for both convective 
and SASI\index{standing-accretion-shock instability (SASI)}
regimes despite different saturation mechanisms for
the two instabilities.

In the case of convection, saturation occurs when the 
volume-integrated net neutrino-heating rate, 
$Q_\nu=Q_\nu^+-Q_\nu^-$, and the convective luminosity
in the gain region roughly balance each other, which implies
a quasi-stationary state of buoyant driving and turbulent 
dissipation \citep{Murphyetal2013}. Based on this assumption,
\cite{MuellerJanka2015} and \cite{Mueller2016} deduced a 
scaling relation between $\langle\delta v\rangle$ 
and the average net neutrino-heating
rate, $q_\nu=q_\nu^+-q_\nu^-$, in the gain region: 
\begin{equation}
\langle\delta v\rangle = 
C\,\left[\frac{q_\nu}{m_\mathrm{u}}\,
(R_\mathrm{s}-R_\mathrm{g})\right]^{1/3}
= C\,\left[\frac{Q_\nu}{M_\mathrm{g}}\,
(R_\mathrm{s}-R_\mathrm{g})\right]^{1/3} 
\label{eq:vturb}
\end{equation}
(with $C$ being a numerical coefficient that depends on 
the dimension and the velocity components included in 
the definition of $\langle\delta v\rangle$).
\cite{Mueller2016} showed that this can be converted
to the form $\langle\delta v\rangle\sim
K\,|\langle v_r\rangle|$ with $K$ of
the order of unity.

In the case of the SASI, a similar scaling law is 
obtained by assuming that the SASI amplitude saturates
when the kinetic energy of coherent SASI motions is 
dissipated in turbulent flows through parasitic
instabilities\index{instability!parasitic} 
such as Kelvin-Helmholtz\index{instability!Kelvin-Helmholtz}
shear vortices
\citep{Guiletetal2010}. By equating the SASI growth
rate (Eq.~\ref{eq:omegasasi}) with the average shear
rate in the gain layer, \citet{Mueller2016} derived
\begin{equation}
\langle\delta v\rangle\sim \ln|{\cal Q}|
\,|\langle v_r\rangle|\,,
\label{eq:vturb2}
\end{equation}
which seems to capture the basic dynamics of 
the postshock flow in multi-dimensional simulations
for a roughly constant value of the quality factor 
$|{\cal Q}| > 1$.

These considerations are only qualitative 
order-of-magnitude estimates. The relative importance
of convection and the SASI in building up and storing 
non-radial kinetic energy in the postshock layer,
their detailed roles in the revival of the stalled
SN shock, and the exact differences between
2D and 3D dynamics remain topics of vivid debate
(for reviews, see \citealt{Jankaetal2016},
\citealt{Mueller2016}, and \citealt{Foglizzoetal2015};
for examples of conflictive points of view, see also
\citealt{Burrowsetal2012,Fernandez2015,CardallBudiardja2015}).
Controversies are often based on simplified numerical models,
but conclusive answers for the SN-core dynamics on the
way to a successful explosion require fully self-consistent
3D simulations.

The leverage of different effects associated with
non-radial mass motions
in the gain layer on the behavior of the SN shock (see
Sect.~\ref{sec:competition}) must be pinpointed by
a careful quantitative analysis of the properties of 
the postshock turbulence\index{turbulence} and its consequences for
energy and momentum transport, turbulent pressure,
or dissipative heating
\citep{MurphyMeakin2011}, for example by means of
a Reynolds decomposition of the momentum 
stress\index{turbulence!Reynolds stress}
tensor $\rho\,\qvec{v}\otimes\qvec{v}$
\citep{Murphyetal2013,CouchOtt2015,Radiceetal2016}.

Instead, \citet{MuellerJanka2015} proposed a simple, 
schematic framework to account for the overall impact
of postshock turbulence on the shock evolution
in a (quasi-)stationary accretion situation, which proved
to be quite successful in direct comparison to 2D and
3D simulations in the convective as well as SASI regimes. 
Motivated by the recognition that the squared turbulent 
Mach number, 
\begin{equation}
\langle\mathrm{Ma}^2\rangle =
\frac{\langle \delta v^2 \rangle}{\langle c_\mathrm{s}^2\rangle} =
\frac{2E_\mathrm{turb,g}/M_\mathrm{g}}{\langle c_\mathrm{s}^2\rangle}
\label{eq:mach}
\end{equation}
with
\begin{equation}
\langle c_\mathrm{s}^2 \rangle =
\frac{1}{M_\mathrm{g}}\,\int_{R_\mathrm{g}}^{R_\mathrm{s}}
\mathrm{d}V\,\rho(\qvec{r})\,c_\mathrm{s}^2(\qvec{r}) 
\end{equation}
being the mass-weighted average of the squared sound speed
in the gain layer,
plays a pivotal role because it reflects the violence of 
aspherical mass motions in the gain layer and determines
the amplitude of shock oscillations, they considered
the effect of turbulent stresses\index{turbulence!stresses} 
as an additional 
isotropic pressure contribution: $P_\mathrm{turb} \approx
\rho\,\langle\delta v^2\rangle$ 
\citep[see also][]{Murphyetal2013,CouchOtt2015,Radiceetal2015}.
This turbulent pressure
can be expressed in terms of the gas pressure $P$ as
$P_\mathrm{turb}\approx\rho\,c_\mathrm{s}^2\langle\mathrm{Ma}^2\rangle
\approx \frac{4}{3}\,\langle\mathrm{Ma}^2\rangle\,P$, because 
$P = \Gamma^{-1}\rho\,c_\mathrm{s}^2$
with $\Gamma\approx \frac{4}{3}$ in the 
radiation-dominated\index{radiation dominated} 
environment of the gain layer.

Replacing the gas pressure $P$ by $P+P_\mathrm{turb} = 
(1 + \frac{4}{3}\langle\mathrm{Ma}^2\rangle)\,P$ on the lhs of 
the equality relation of Eq.~(\ref{eq:postpress}) and applying
the gain condition (Eq.~\ref{eq:gainradius}) in the form
$P_\mathrm{g}^{3/2} \propto L_\nu\langle E_\nu^2\rangle\,
R_\mathrm{g}^{-2}$ for the radiation-dominated plasma in the
gain layer \citep[$P\propto T^4$, $\rho\propto
T^3 \propto r^{-3}$;][]{Janka2001A}, \cite{MuellerJanka2015}
derived a relation for the shock-stagnation radius:
\begin{equation}
R_\mathrm{s}\propto\frac{\left(L_\nu\langle E_\nu^2\rangle\right)^{4/9}
R_\mathrm{g}^{16/9}}{\dot{M}^{2/3}\,M^{1/3}}\,\,
\xi_\mathrm{turb}^{2/3} \,.
\label{eq:rshock}
\end{equation}
Here the factor
\begin{equation}
\xi_\mathrm{turb} = 1 + \frac{4}{3}\langle\mathrm{Ma}^2\rangle \ge 1
\label{eq:xiturb}
\end{equation}
accounts for the increase of the average shock radius caused
by the additional turbulent pressure\index{turbulence!pressure} 
support of the shock against
the ram pressure of the infalling preshock matter. With
$\xi_\mathrm{turb} = 1$ the 1D case is recovered \citep{Janka2012}.
Similarly, centrifugal support\index{centrifugal support}
associated with rapid rotation\index{rotation} can 
lead to an expansion of the shock. This can 
be taken into account by introducing another correction  
factor $\xi_\mathrm{rot}^{-2/3}$ in Eq.~(\ref{eq:rshock}) with
$\xi_\mathrm{rot}$ being defined as
\begin{equation}
\xi_\mathrm{rot} = \sqrt{1-\frac{j_0^2}{2G\,M\,R_\mathrm{s}}}\le 1\,,
\label{eq:xirot}
\end{equation}
where $j_0$ is the specific angular momentum (averaged on
spherical shells) of matter ahead of the shock \citep[for the 
derivation, see][]{Jankaetal2016}.
Also perturbations in the convective silicon and oxygen-burning
shells of the pre-collapse progenitor can have a healthy influence
on the shock-stagnation radius by enhancing the turbulent 
activity in the postshock layer \citep{CouchOtt2013,MuellerJanka2015}. 
Corresponding modifications of Eq.~(\ref{eq:rshock}) were discussed by 
\citet{Muelleretal2016}.

\begin{figure}[!]
\sidecaption
\includegraphics[scale=1.0]{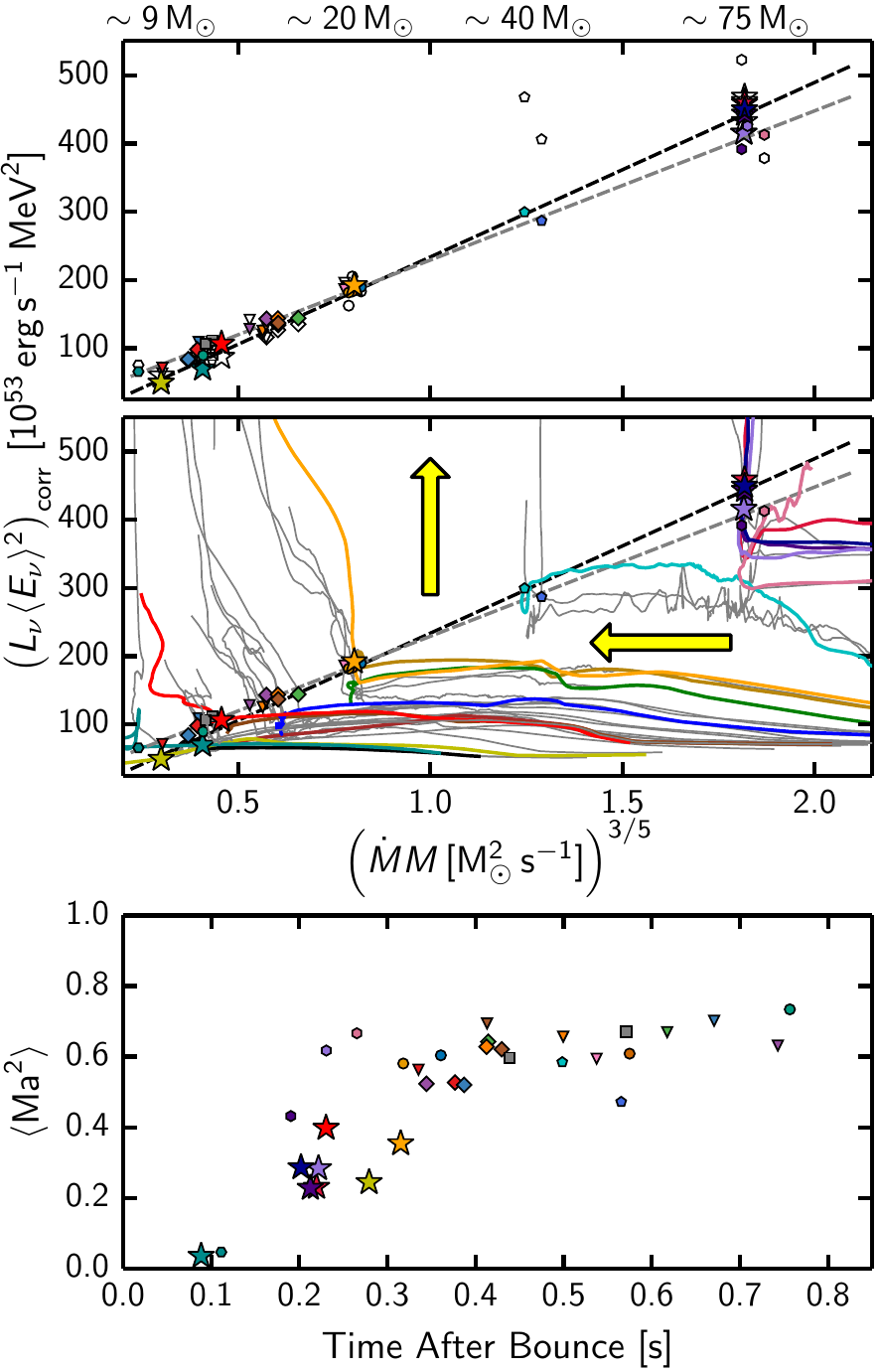}
\caption{Critical condition for the heating functional
at the threshold for runaway shock expansion.
The upper two panels display by dashed lines
the critical relation between 
$\left(L_\nu\langle E_\nu^2\rangle\right)_\mathrm{crit,corr}$
and $\left(\dot{M}\,M\right)^{3/5}$ at the onset of
shock runaway (Eq.~\ref{eq:critcorr}). The black dashed line 
was obtained by a least-squares fit to the critical points of
a set of non-rotating and rotating 3D models (stars), the gray 
dashed line depicts the least-squares fit for a set of 2D models,
also with and without rotation (all non-star, smaller symbols).
Open symbols correspond to the locations of the uncorrected values
of $\left(L_\nu\langle E_\nu^2\rangle\right)_\mathrm{crit}$
(Eq.~\ref{eq:lcrit}). 
The middle panel shows the evolution tracks (the post-bounce 
evolution 
proceeds from right to left, indicated by yellow arrows)
of exploding and non-exploding 2D (thin, gray lines) and 3D
models (thick, colored lines). Exploding models cross the 
corresponding critical line by turning upward.
The bottom panel depicts average squared turbulent Mach numbers
in the gain layer (Eq.~\ref{eq:mach}) at the time when runaway
shock expansion begins. (Figure courtesy of Alexander Summa)
}
\label{fig:janka-criticalcurve}
\end{figure}

\subsection{Universal Critical Luminosity Condition}
\label{sec:critlum}

In Sect.~\ref{sec:massivesne} the road to explosion was
described as a tight competition of preshock ram pressure of
the infalling matter and postshock heating by the neutrino
luminosity of the new-born NS. For both competing effects the
mass-accretion rate $\dot M$ of the stalled shock is of 
pivotal importance. \citet{BurrowsGoshy1993} found (by
semi-analytic analysis in 1D) that stationary accretion 
solutions are not possible for electron-flavor neutrino 
luminosities above a critical limit $L_{\nu,\mathrm{crit}}(\dot M)$
\citep[see also][]{YamasakiYamada2006,PejchaThompson2012}.
\citet{Fernandez2012}, applying 1D hydrodynamic simulations, and 
\citet{Janka2012} with simple analytic considerations showed 
that this critical luminosity\index{critical luminosity} condition
is roughly compatible with the time-scale 
criterion\index{time-scale criterion}
$t_\mathrm{adv}/t_\mathrm{heat}\ge 1$ for runaway shock
expansion. Using approximations for both time scales
based on Eqs.~(\ref{eq:tadvec}) and (\ref{eq:theat}), 
blackbody\index{blackbody}
assumptions ($L_\nu\propto R_\mathrm{ns}^2T_\nu^4$) 
for the neutrino emission \citep[following][]{BurrowsGoshy1993},
and Eq.~(\ref{eq:rshock}) with 
$\xi_\mathrm{turb} = 1$ and $R_\mathrm{g}\propto R_\mathrm{ns}$ 
for the shock-stagnation radius,
\citet{Janka2012} derived for the critical luminosity limit:
\begin{equation}
L_{\nu,\mathrm{crit}}(\dot M) \propto
\beta_\rho^{-2/5}\,\dot M^{2/5}\,M^{4/5} \,.
\label{eq:lcrit0}
\end{equation}
This relation reproduces the functional behavior obtained
by \cite{BurrowsGoshy1993} very well. The numerical factor
of the scaling relation becomes 
$(5\,...\,6)\times 10^{52}\,\mathrm{erg\,s}^{-1}$ for
individual luminosities of $\nu_e$ or $\overline{\nu}_e$, 
when $\beta_\rho\sim 10$, $\dot M = 1\,M_\odot$/s, and
a NS mass of $M=1.5\,M_\odot$ are used, and slightly varies
with the choice of other involved parameters.

\citet{MuellerJanka2015} generalized the critical luminosity
condition, based on the time-scale criterion 
$t_\mathrm{adv} = t_\mathrm{heat}$ for the onset of shock 
runaway, to a critical condition for the 
heating functional\index{heating functional}
$L_\nu\langle E_\nu^2\rangle$, including also the effects of
turbulent stresses\index{turbulence!stresses}
in the postshock flow by employing
Eq.~(\ref{eq:rshock}) for the 
shock-stagnation radius\index{shock-stagnation radius}.
Further adding the possible effects of 
rotation\index{rotation} into the
treatment, \citet{Jankaetal2016} obtained:
\begin{equation}
\left(L_\nu \langle E_\nu^2\rangle\right)_\mathrm{crit}\propto
\left(\dot{M}\,M\right)^{3/5}\left|\bar{e}_\mathrm{tot,g}\right|^{3/5}
R_\mathrm{g}^{-2/5}\,\xi_\mathrm{turb}^{-3/5}\,\xi_\mathrm{rot}^{6/5}
\equiv \left(\dot{M}\,M\right)^{3/5}\xi_\mathrm{g}\,,
\label{eq:lcrit}
\end{equation}
with $\bar{e}_\mathrm{tot,g}$ being the average,
mass-specific binding energy in the gain layer:
\begin{equation}
\bar{e}_\mathrm{tot,g} = \frac{E_\mathrm{tot,g}}{M_\mathrm{g}}\,.
\end{equation}
In Eq.~(\ref{eq:lcrit}) the time-dependent quantity 
$\xi_\mathrm{g}$, defined as
\begin{equation}
\xi_\mathrm{g}\equiv\left|\bar{e}_\mathrm{tot,g}\right|^{3/5}
R_\mathrm{g}^{-2/5}\,\xi_\mathrm{turb}^{-3/5}\,\xi_\mathrm{rot}^{6/5}\,,
\label{eq:xi}
\end{equation}
subsumes all gain-layer related properties.
Following \citet{Summaetal2016}, 
$\xi_\mathrm{g}$ can be used to correct $L_\nu \langle E_\nu^2\rangle$
for variations of the time evolution of gain radius,
binding energy, nonradial (turbulent) postshock flows, and rotation,
which cause time- and model-dependent variations of the critical
condition in addition to the basic dependence on the NS mass
$M$ and the shock-accretion rate $\dot{M}$. 
Doing so yields a universal relation
for the heating functional at the threshold for shock runaway:
\begin{equation}
\left(L_\nu \langle E_\nu^2\rangle\right)_\mathrm{crit,corr}\equiv
\frac{1}{\xi_\mathrm{g}/\xi_\mathrm{g}^*}
\left(L_\nu \langle E_\nu^2\rangle\right)_\mathrm{crit}
\propto \left(\dot{M}\,M\right)^{3/5}\,.
\label{eq:critcorr}
\end{equation}
An arbitrary constant $\xi_\mathrm{g}^*$ normalizes
the correction factor relative to a chosen reference
model, for which $\xi_\mathrm{g}^*$ is evaluated at the time when
$t_\mathrm{adv}/t_\mathrm{heat} = 1$.

Figure~\ref{fig:janka-criticalcurve} displays the critical
points in the 
$\left(L_\nu \langle E_\nu^2\rangle\right)_\mathrm{crit,corr}$-$\left(\dot{M}\,M\right)^{3/5}$-plane 
for a large set of full-scale 2D and 3D SN simulations in a
wide mass range from the low-mass end ($<$10\,$M_\odot$) to 
heavy progenitors around 75\,$M_\odot$.
All points line up nicely on straight curves, suggesting that
Eq.~(\ref{eq:critcorr}) captures the basic physics of shock runaway
very well. Moreover, the best-fit relations (least-squares fits) for 
2D and 3D models (grey and black dashed lines, respectively) are
nearly identical, suggesting that Eq.~(\ref{eq:critcorr}) defines
a universal critical condition that holds for non-rotating as well
as rotating core-collapse conditions (nearly) independently of 
dimension.

The bottom panel in Fig.~\ref{fig:janka-criticalcurve} shows 
the squared turbulent Mach number $\langle\mathrm{Ma}^2\rangle$,
averaged over the gain layer (Eq.~\ref{eq:mach}), 
at the time when runaway shock
expansion sets in. All successful 3D models show runaway shock
expansion rather early, and some of them exhibit the tendency
to do so even slightly earlier than the corresponding 2D cases.
The fact that they do so despite possessing 
considerably lower values of 
$\langle\mathrm{Ma}^2\rangle$ at the onset of criticality
seems to imply a considerably higher critical heating functional 
$\left(L_\nu \langle E_\nu^2\rangle\right)_\mathrm{crit}$
(Eq.~\ref{eq:lcrit}) in 3D, since in this case
$\xi_\mathrm{turb}$ is lower (Eq.~\ref{eq:xiturb}).
(Note that this difference would not show up in the critical
curve in Fig.~\ref{fig:janka-criticalcurve}, because the
critical heating functional there is normalized by applying the
factor $\xi_\mathrm{g}^{-1}$.) Simulations, however, do not
comply with this expectation but instead show that the 
critical conditions of 2D and 3D models are rather close,
a finding which points to subtle hydrodynamical effects that  
can aid explosions in 3D despite lower overall values of 
$\langle\mathrm{Ma}^2\rangle$ in the gain layer
\citep[see the discussion in sect.~3.3.4 of ][]{Mueller2016}.
One possibility is that in 3D kinetic energy in the gain layer
is efficiently stored in spiral SASI 
modes\index{standing-accretion-shock instability (SASI)!spiral modes} 
\cite{Fernandez2015}.
Such coherent flows do not behave like turbulent velocity
fluctuations and, consequently,
do not add to $\xi_\mathrm{turb}$ as defined by us. Instead,
they decrease the specific binding energy of the gain layer,
$|\bar{e}_\mathrm{tot,g}|$ and therefore also
$\xi_\mathrm{g}$ according to Eq.~(\ref{eq:xi}). This effect 
on the critical condition is similar to the influence of
moderately rapid rotation, in which case centrifugal effects
in the infall region ahead of the SN shock are negligibly small
(therefore $\xi_\mathrm{rot} \approx 1$; Eq.~\ref{eq:xirot}),
but the rotational kinetic energy in the postshock region 
lowers the value of $|\bar{e}_\mathrm{tot,g}|$ 
\citep{Jankaetal2016}.

\section{Astrophysical Implications of Neutrino-driven explosions}
\label{sec:astroimplications}

The understanding of the neutrino-driven mechanism and the
hydrodynamical modeling of explosions have meanwhile matured 
enough to make predictions of observables and to study 
implications of neutrino-powered SN. In this Section some
recent developments in this context will be summarized.

\subsection{Neutrino and Gravitational-wave Signals}
\label{sec:nugwsignals}

Both neutrinos and gravitational waves play an important role
as messengers that can directly probe the dynamics and, in
the case of neutrinos, also the thermodynamics of the SN core.

There is a variety of effects that can imprint characteristic
features on the luminosity evolution and the spectral properties
of the neutrino signal, such as the prominent $\nu_e$ burst 
produced by electron captures at the moment of
shock breakout\index{shock breakout!electron-neutrino burst}
from the neutrinosphere\index{neutrinosphere}; the subsequent 
accretion luminosities\index{accretion luminosity}
of $\nu_e$ and $\overline{\nu}_e$ during the 
shock-stagnation phase, which are determined by the
mass accretion rate $\dot M$ and therefore by the compactness of
the progenitor core; and the cooling and deleptonization emission
of neutrinos and antineutrinos of all flavors 
during the Kelvin-Helmholtz phase\index{Kelvin-Helmholtz phase}
of the new-born NS, which depend
on the mass of the compact remnant and its high-density equation
of state with possible phase transitions in the super-nuclear 
regime.

Similarly, time-dependent centrifugal deformation induced by 
rapid rotation as well as violent mass motions associated with
convective overturn\index{convective overturn}
and SASI\index{standing-accretion-shock instability (SASI)}
activity in the neutrino-heating 
layer and convection inside of the NS
lead to variations of the mass-quadrupole moment and therefore
gravitational-wave\index{gravitational waves} radiation.
While the reader is referred to the special chapters on neutrinos
and gravitational waves from core-collapse SNe for more 
information and details, two interesting phenomena are briefly
mentioned here, because they are relevant for the pre-explosion
diagostics of neutrino-driven SNe in the case of neutrino and
gravitational-wave measurements connected to a future galactic SN.

Large-amplitude SASI\index{standing-accretion-shock instability (SASI)}
sloshing and spiral motions
of the postshock layer induce fluctuations of the mass-accretion
rate of the NS and, correspondingly, show up as pronounced, 
quasi-periodic modulations of the radiated neutrino luminosities 
and mean energies with amplitudes of more than 10\% and 
phase-synchronized variations of up to 1\,MeV, respectively
\citep{Tamborraetal2013,Tamborraetal2014}. Though dependent
on the viewing angle, these variations with a typical  
frequency of $t_\mathrm{cyc}^{-1}$ (see Eq.~\ref{eq:cycletime})
are likely to be detectable by IceCube\index{IceCube}
and Hyper-Kamiokande\index{Hyper-Kamiokande} 
for a SN anywhere in the Galaxy, if SASI activity precedes
the onset of the explosion in this event. Similarly, also
the gravitational-wave emission\index{gravitational waves}
is modulated with characteristic
spectral peaks around frequencies $t_\mathrm{cyc}^{-1}$ 
and $2\,t_\mathrm{cyc}^{-1}$ \citep{Andresenetal2016}, which, in
the lucky case of a SN that is not too distant, could be measured 
by existing interferometric instruments \citep{Kurodaetal2016}.

In the 3D SN simulations by the Garching group a new, stunning
feature was discovered in the neutrino emission: a lepton-number
self-sustained asymmetry 
(LESA)\index{lepton-number self-sustained asymmetry (LESA)}, 
in which the lepton-number
emission ($\nu_e$ minus $\overline{\nu}_e$) develops a large
dipole component some 150--200\,ms after bounce, whose amplitude
can exceed the monopole of the lepton-number flux, and whose
direction remains rather stationary or drifts only slowly on
time scales much longer than those that are typical of
mass motions due to SASI 
activity and convective overturn \citep{Tamborraetal2014b}.
While the hemispheric difference of the lepton-number flux is
huge (even to the extent that the $\nu_e$ and $\overline{\nu}_e$
fluxes can dominate in opposite hemispheres), the individual 
fluxes of $\nu_e$ and $\overline{\nu}_e$ exhibit hemispheric
differences only of order 10\%, and the total $\nu_e$ plus
$\overline{\nu}_e$ flux as well as the total neutrino flux
(including heavy-lepton neutrinos) feature large-scale 
directional variations only on the level of percent.

This new LESA phenomenon is present in all 3D simulations of
the Garching group, but the amplitude of the lepton-number 
emission dipole\index{lepton-number emission dipole}
is smaller in the case of rotation\index{rotation} and even more
suppressed when the NS spins faster. LESA can be traced back
to a neutrino-hydrodynamical instability in the convective
layer inside of the NS \citep{Jankaetal2016}. It could have
important consequences for 
neutrino-flavor oscillations\index{neutrino-flavor oscillations}
in SN cores, SN neutrino detection, neutrino-induced
SN nucleosynthesis\index{nucleosynthesis}, and 
NS kicks\index{neutron-star kick}.

\begin{figure}[!]
\sidecaption[t]
\includegraphics[scale=.30]{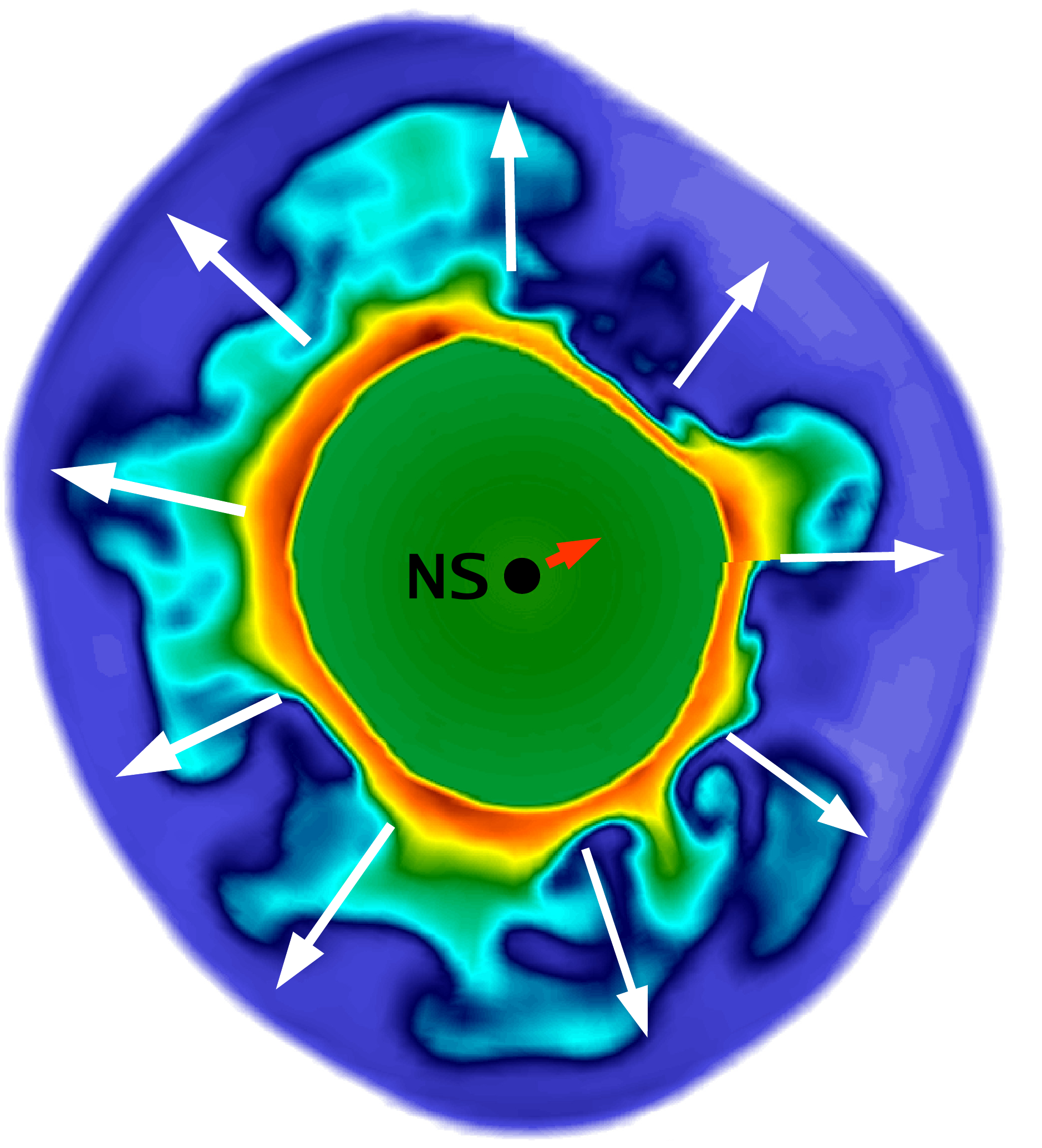}\hspace{7pt}
\includegraphics[scale=.30]{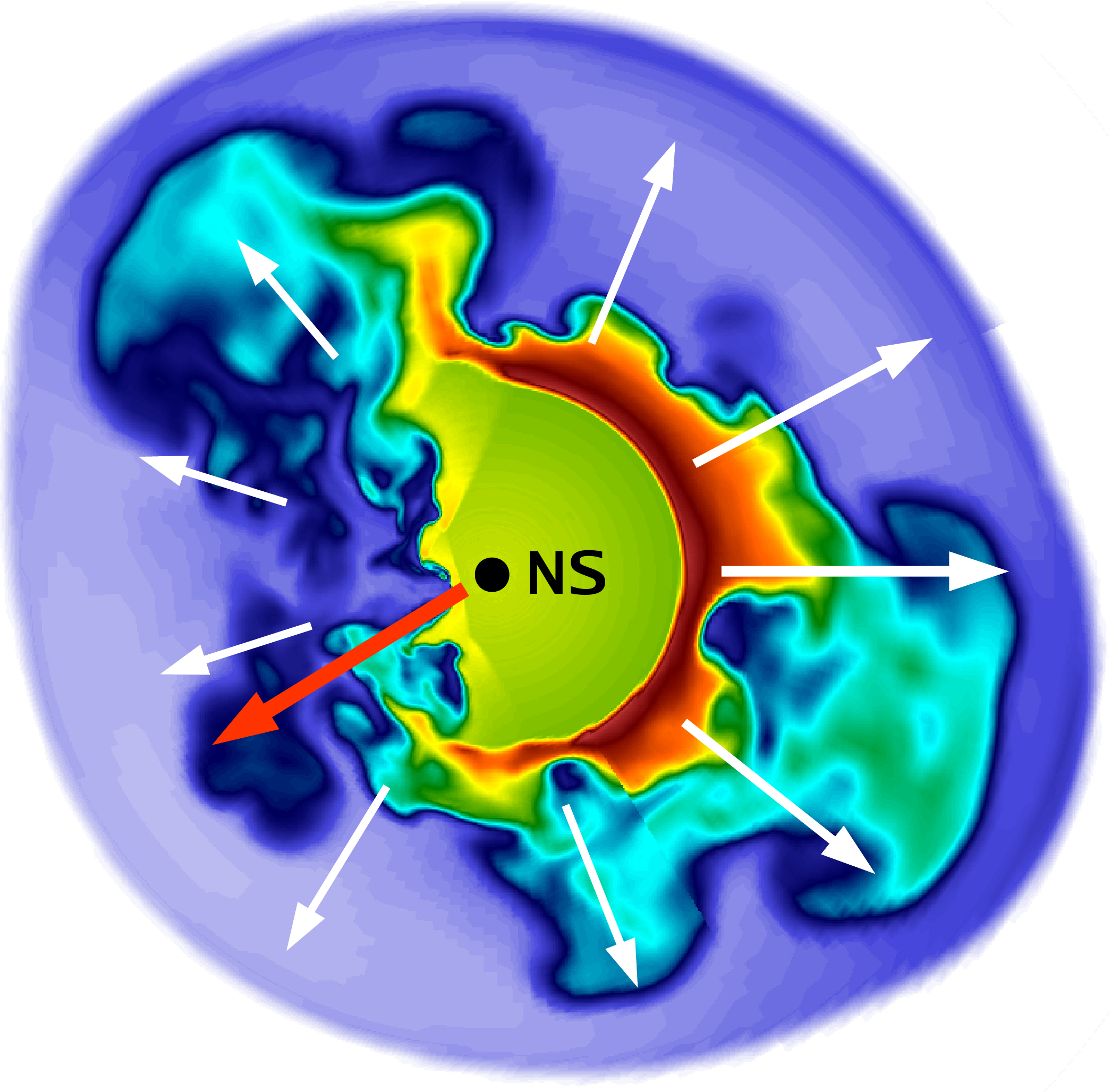}
\caption{NS kick by asymmetric SN explosions. The color
coding shows entropy distributions in cross sectional planes of
3D simulations of neutrino-driven explosions for
two different 15\,$M_\odot$ stars about 1.4\,s
after core bounce \citep{Wongwathanaratetal2013}. Dark blue are
low-entropy, dense ejecta, bright blue, green, yellow, and red
are high-entropy, neutrino-heated ejecta that partly rise in
low-density Rayleigh-Taylor plumes. The white arrows indicate
the momentum of the expanding ejecta, the red arrow the NS kick
direction and magnitude. The NS is gravitationally pulled by the
clumps of dense, slow ejecta opposite to the direction of the
stronger blast-wave expansion.
{\em Left:} Nearly spherical explosion with small NS kick
($\sim$100\,km\,s$^{-1}$).
{\em Right:} Highly aspherical explosion with large dipolar and
quadrupolar ejecta inhomogeneities and a stronger shock
expansion towards the right side. Here the NS receives a large
kick ($\sim$550\,km\,s$^{-1}$)
mediated by the gravitational attraction of the less rapidly
expelled, dense clumps on the left side.
}
\label{fig:janka-nskick}
\end{figure}

\begin{figure}[!]
\begin{center}
\includegraphics[scale=.22]{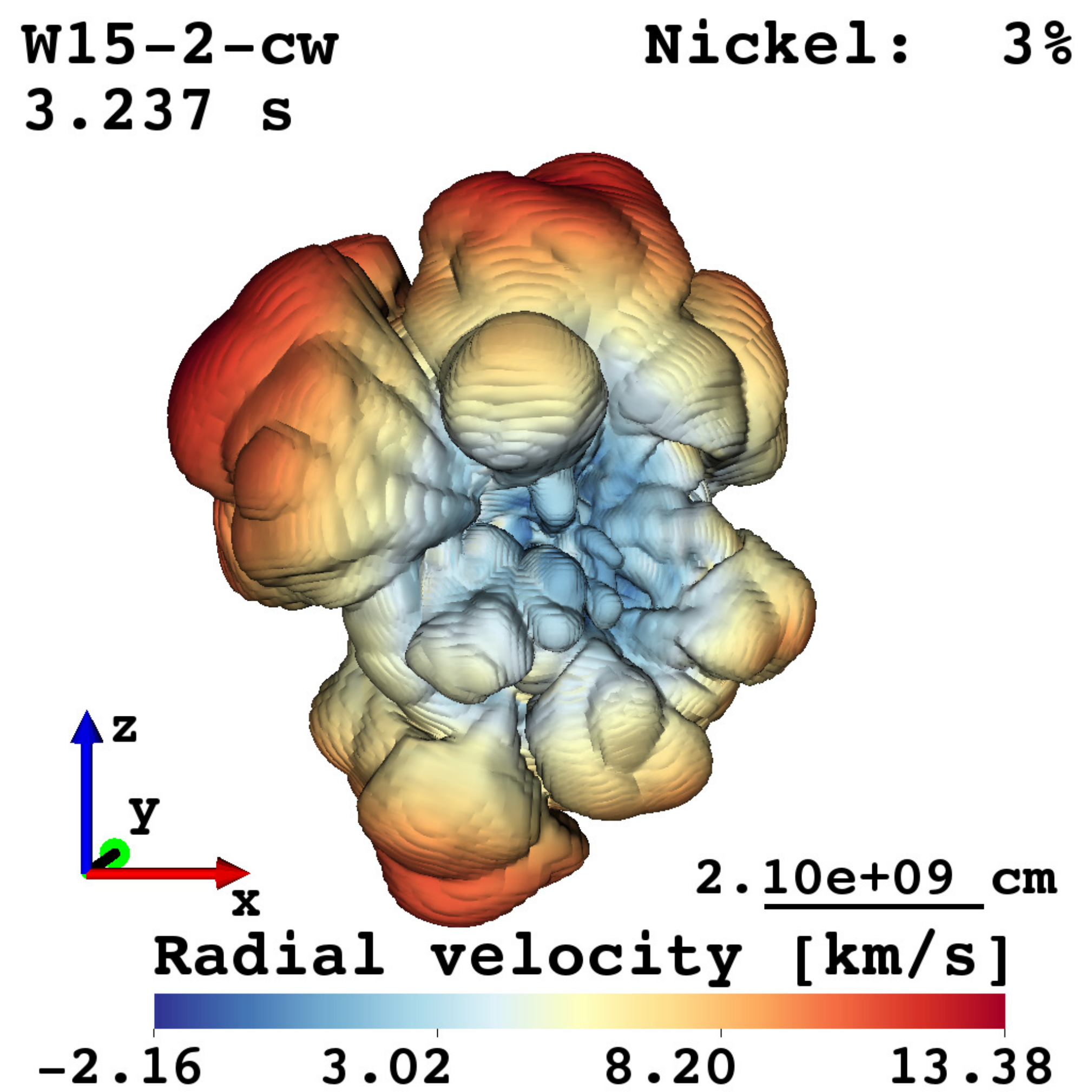}\hspace{2pt}
\includegraphics[scale=.22]{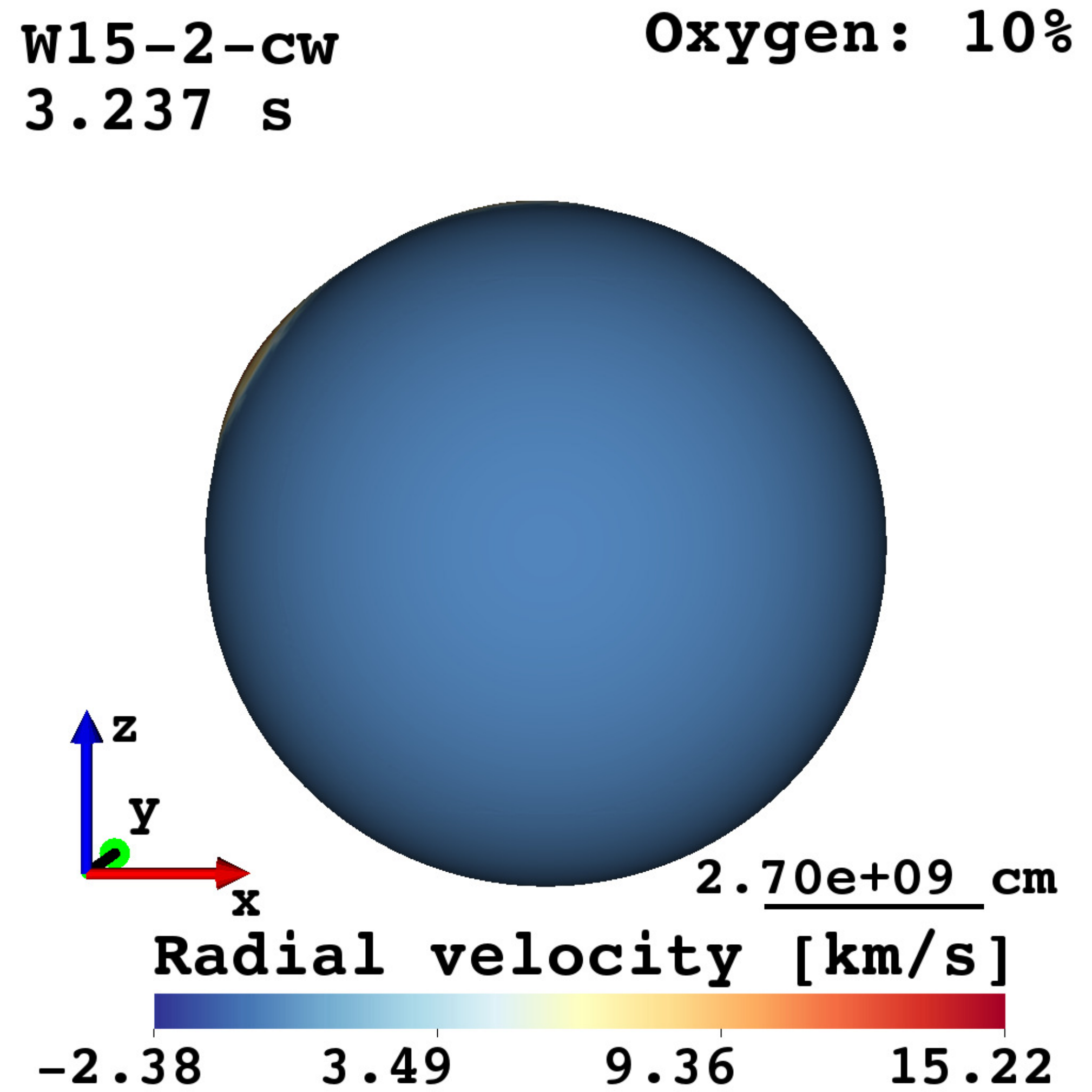}\\
\includegraphics[scale=.22]{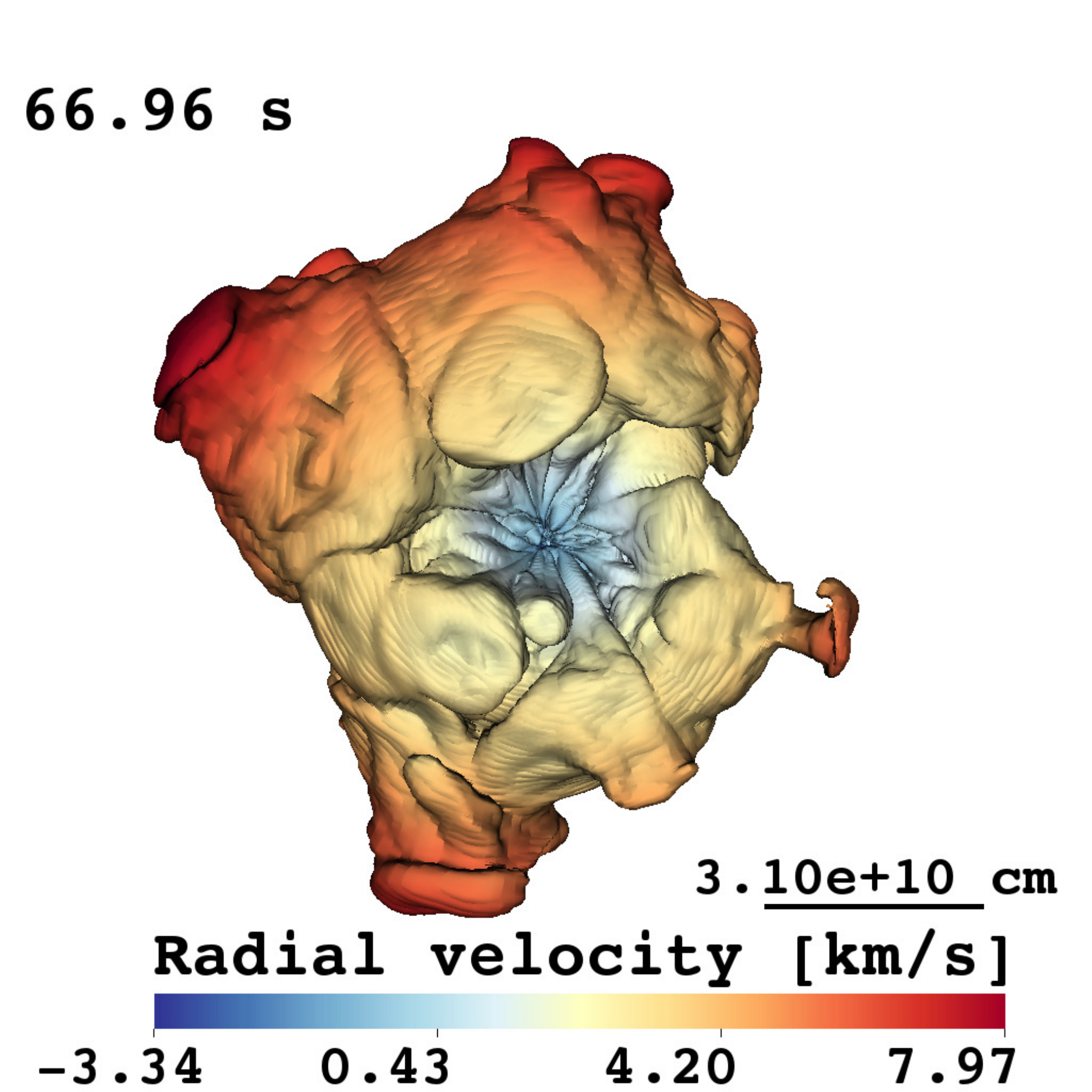}\hspace{2pt}
\includegraphics[scale=.22]{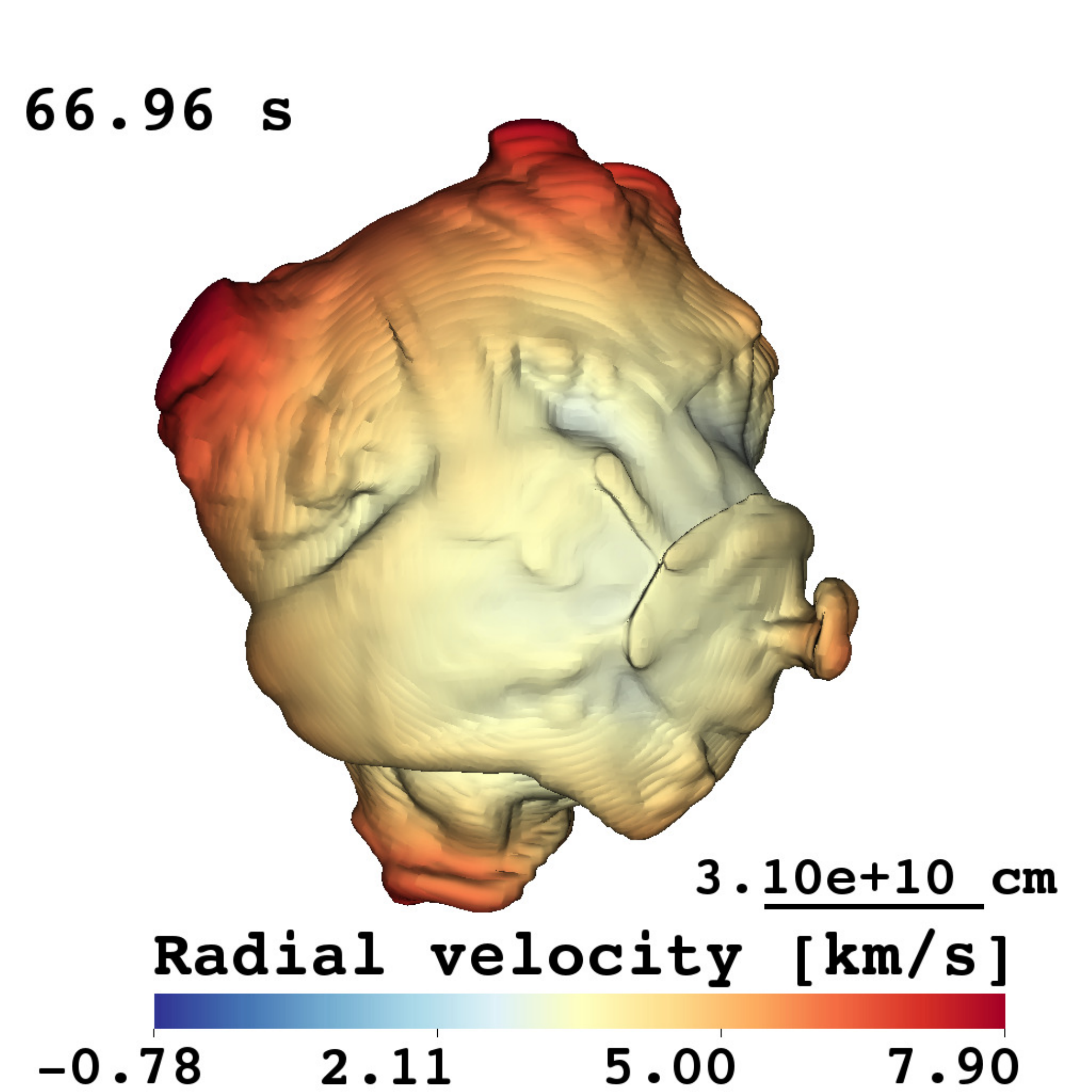}\\
\includegraphics[scale=.22]{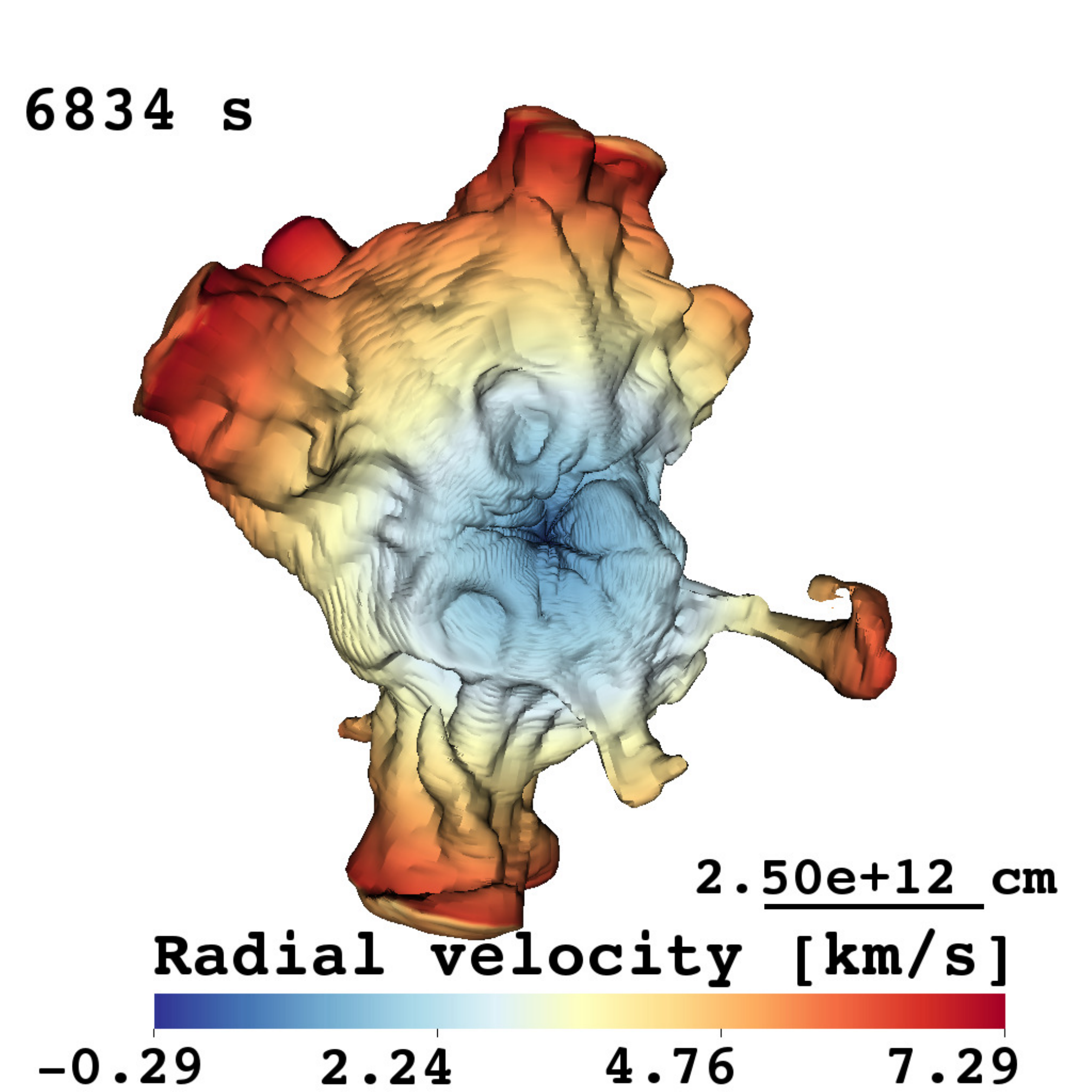}\hspace{2pt}
\includegraphics[scale=.22]{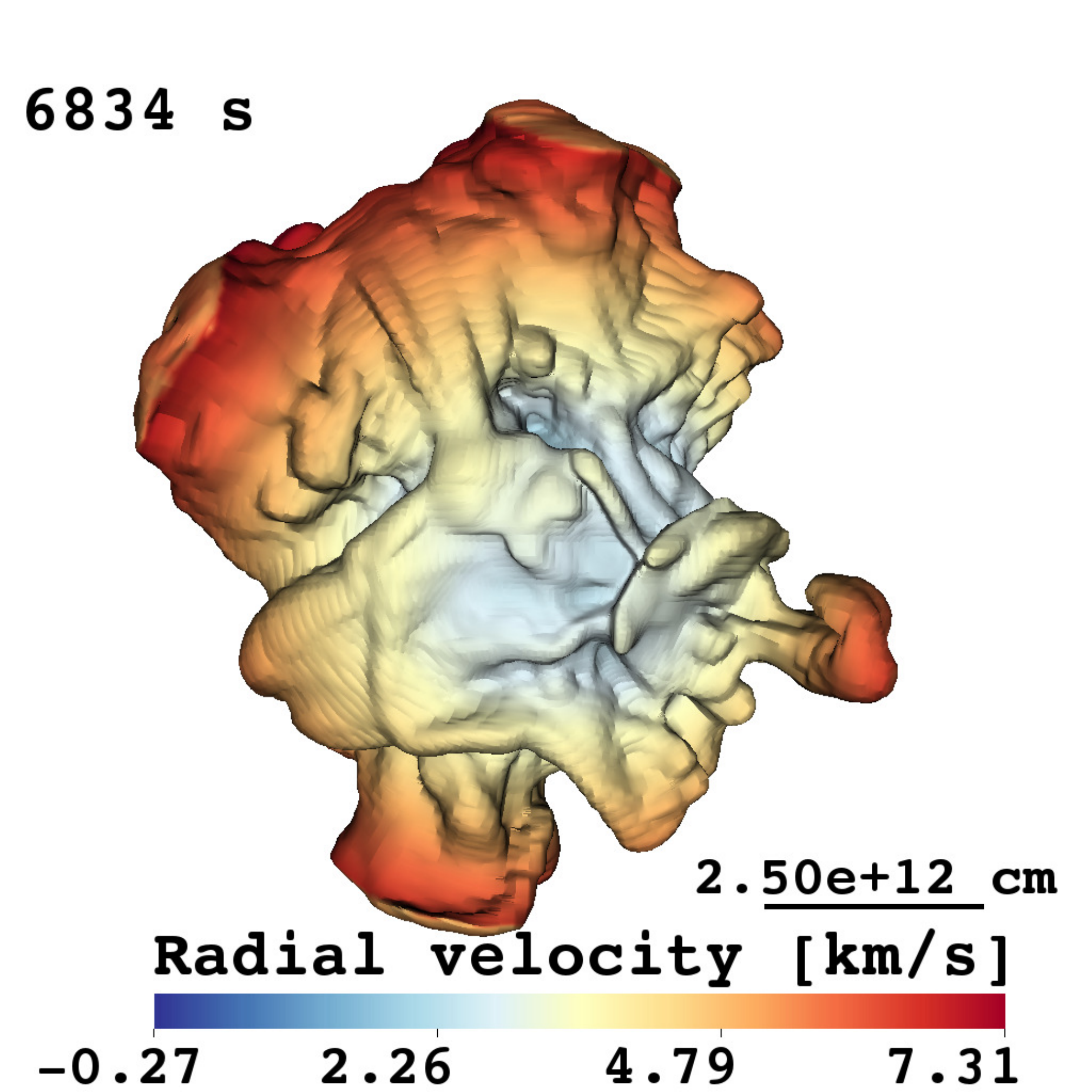}\\
\includegraphics[scale=.22]{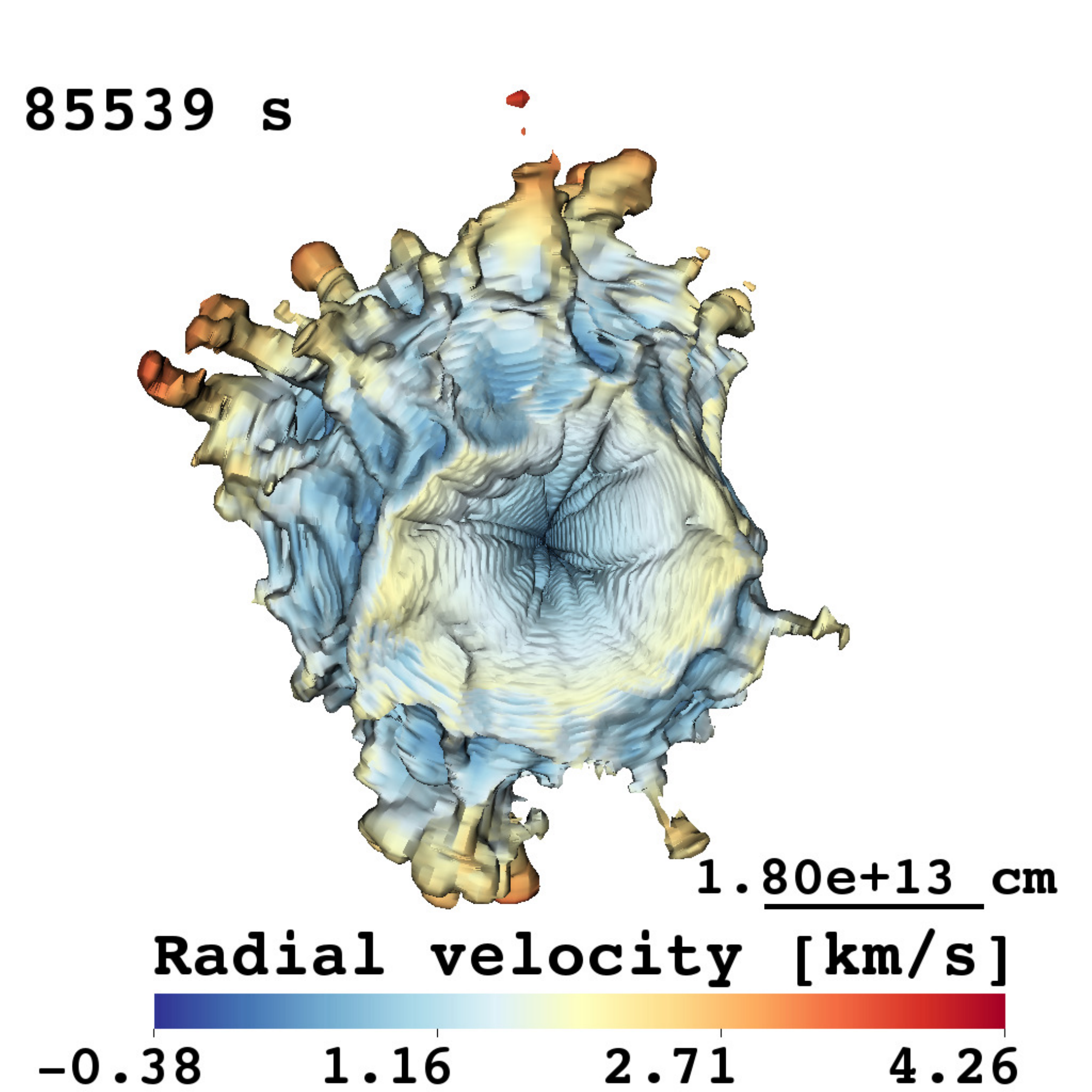}\hspace{2pt}
\includegraphics[scale=.22]{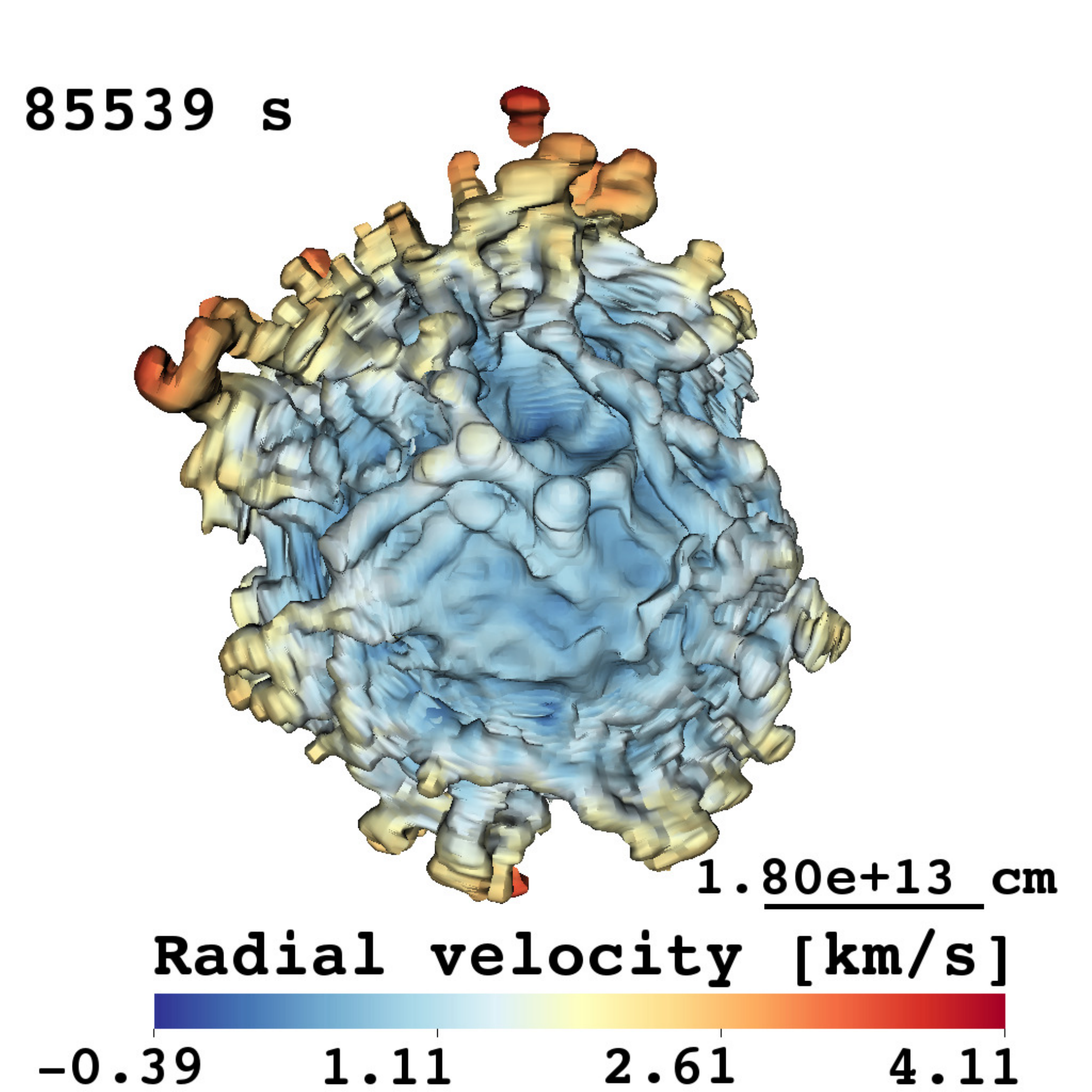}
\end{center}
\caption{The first day of a SN. The images from a 3D
simulation of a neutrino-driven explosion of a
15\,$M_\odot$ red supergiant (SN~IIP) progenitor
\citep{Wongwathanaratetal2015}
show isosurfaces of constant mass fractions of
$^{56}$Ni (3\%; {\em left}) and oxygen (10\%; {\em right}).
The color coding visualizes the radial velocity as given
by the color bars.
{\em Top:} Shortly before the SN shock crosses the C-O/He
composition interface.
{\em Second row:} Shortly before shock passage through He/H
composition interface. The reverse shock from the C-O/He
interface has already compressed the nickel plumes.
{\em Third row:} At the time when the reverse shock from
the He/H interface hits the most extended nickel fingers.
{\em Bottom:} At shock breakout from the stellar surface.
The large-scale high-entropy bubbles of the beginning
explosion ({\em top}) fragment to smaller-scale structures.
(Figures courtesy of Annop Wongwathanarat)
}
\label{fig:janka-snlongtime}
\end{figure}

\begin{figure}[b]
\sidecaption
\includegraphics[scale=0.10]{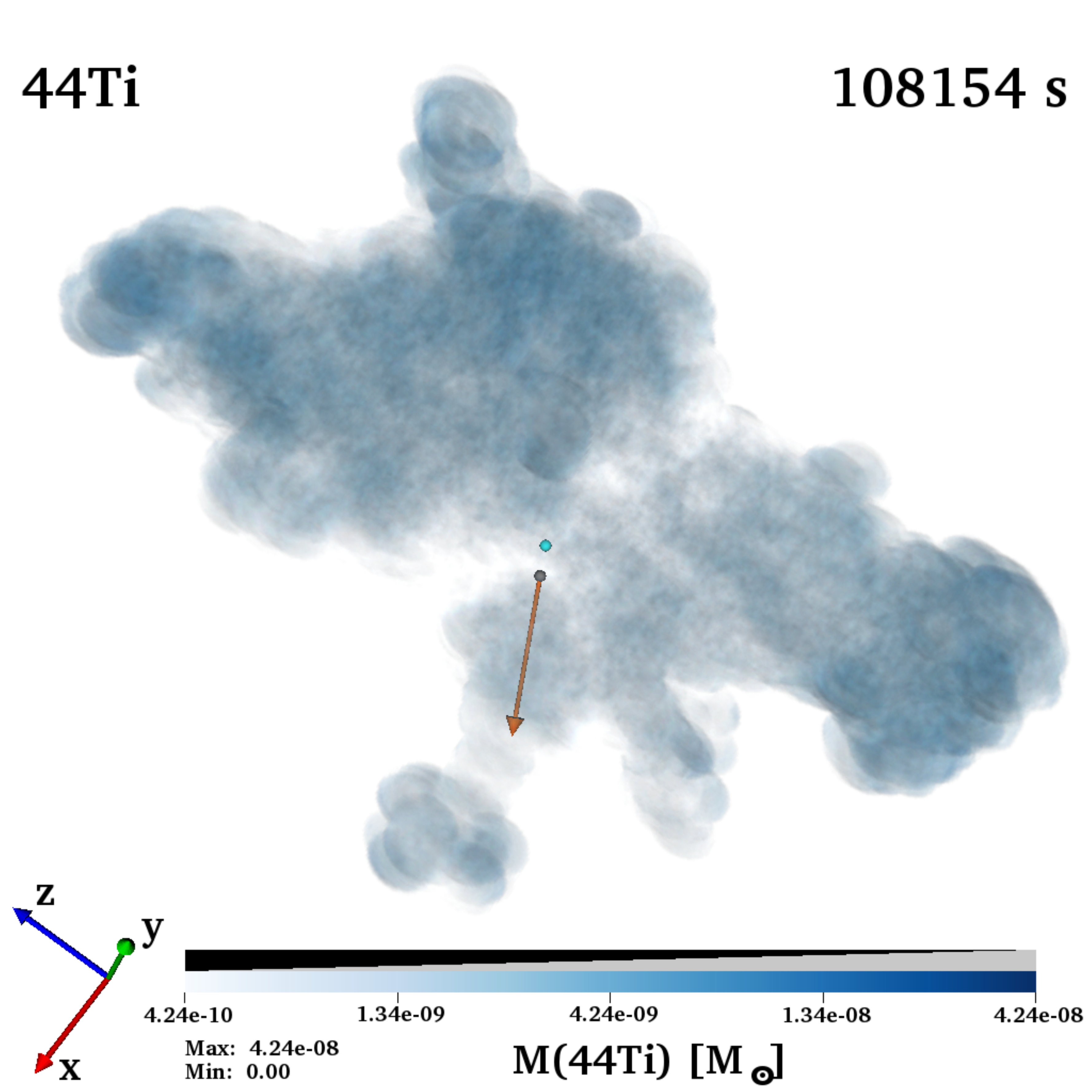}\hspace{4pt}
\includegraphics[scale=0.10]{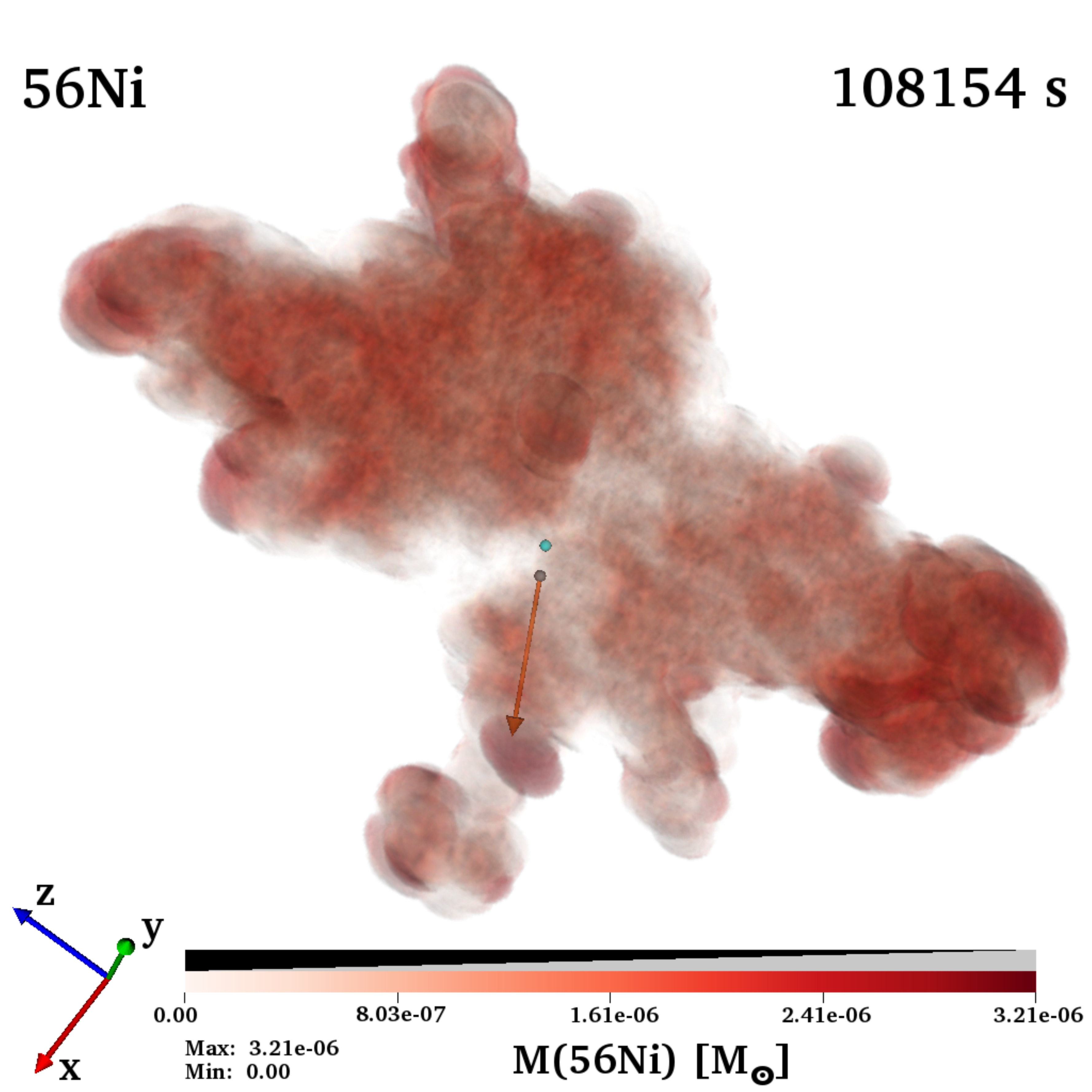}\hspace{4pt}
\includegraphics[scale=0.20]{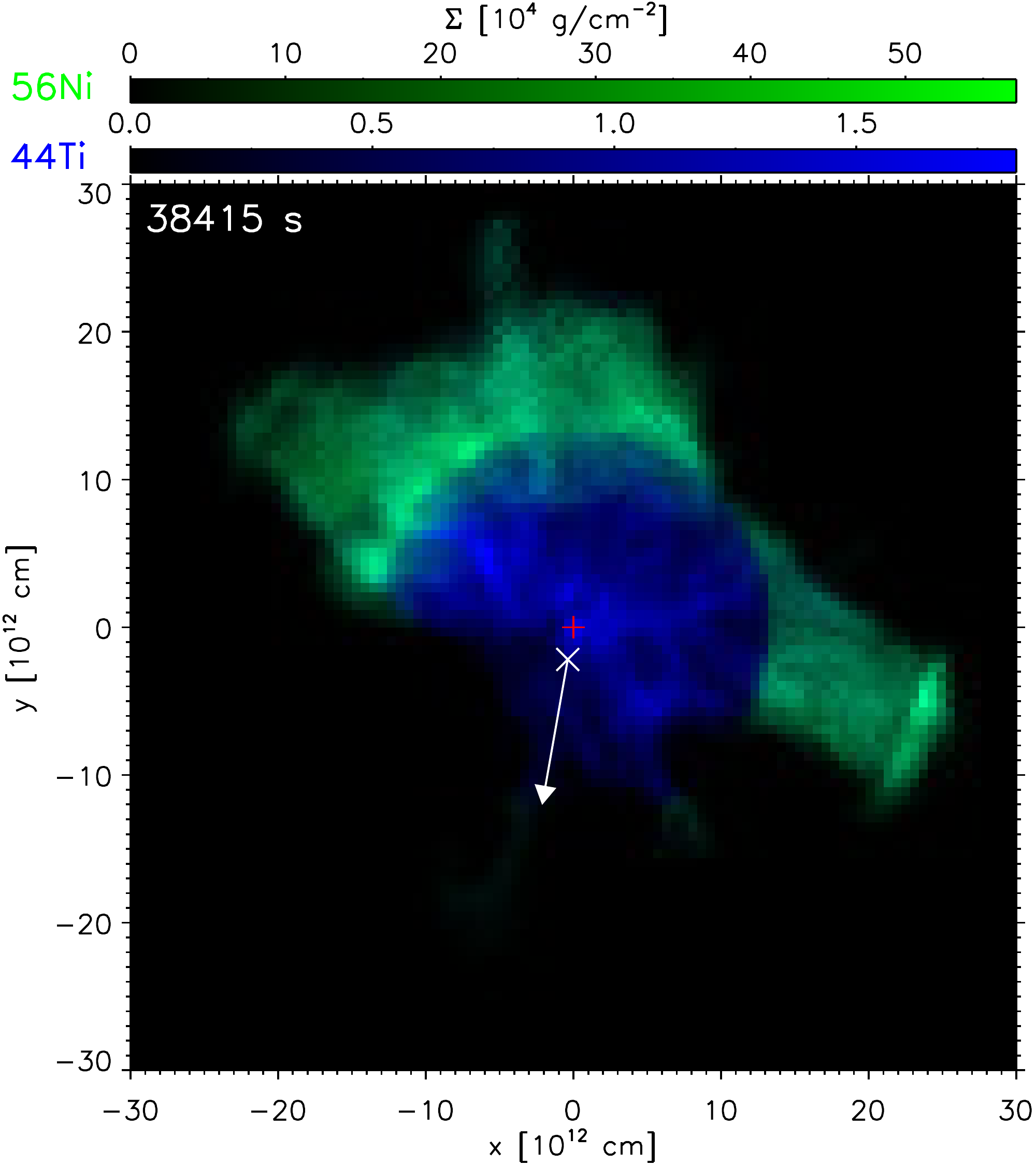}
\caption{NS kick and explosive SN nucleosynthesis in a 3D
simulation of a highly asymmetric neutrino-driven explosion
of a 15\,$M_\odot$ SN~IIb progenitor \citep[the same explosion
model as shown in 
Fig.~\ref{fig:janka-snlongtime} but with a nearly completely 
stripped hydrogen envelope;][]{Wongwathanaratetal2013,Wongwathanaratetal2016}.
The SN blast ejects 
heavy elements from silicon to the iron group, which are
explosively produced in shock-heated and neutrino-heated
matter, preferentially in the direction of the stronger
explosion, i.e.\ opposite to the NS kick direction. The latter
is indicated by the red or white arrows pointing downward.
The ray-tracing images display the distributions of radioactive
$^{44}$Ti ({\em left}) and $^{56}$Ni ({\em middle}) more
than a day after the initiation of the SN outburst, when the 
ejecta expand essentially homologously. The largest fingers
that are present at this late time originate from the 
biggest high-entropy bubbles of neutrino-heated matter that
dominated the asymmetry at the beginning of the explosion.
The {\em right
image} shows $^{56}$Ni (green) and $^{44}$Ti (blue) in
combination, but assuming that only the outer 50\% of the
nickel mass are visible. The similarity to the NuSTAR 
$^{44}$Ti map of the Cassiopeia~A remnant
\citep{Grefenstetteetal2014} is striking.
Iron (most of which is the stable decay product of $^{56}$Ni)
can be observed only in the reverse-shock heated
outer shell of the SN remnant, while $^{44}$Ti is seen by
its decay activity also in the inner, unshocked volume,
concentrated mostly in the hemisphere opposite to the NS kick.
(Figure courtesy of Annop Wongwathanarat)
}
\label{fig:janka-casa}
\end{figure}

\subsection{Neutron Star Kicks and Anisotropic Nucleosynthesis}
\label{sec:nskicks}

Because of momentum conservation in the rest frame of the
progenitor star, asymmetric mass ejection of a SN or asymmetric
neutrino emission can lead to a recoil acceleration of the NS
that is left behind by the explosion
\citep[e.g.,][]{JankaMueller1994,Laietal2001}.

The hydrodynamic instabilities that precede and foster the onset
of neutrino-driven SN explosions are associated with a considerable
redistribution of matter and energy around the nascent NS.
Corresponding asymmetries of the mass distribution or velocity
field of the innermost SN ejecta exert anisotropic hydrodynamic
(i.e., gas-pressure mediated) and gravitational forces that 
push or pull the compact remnant in the direction opposite to
the stronger
explosion (Fig.~\ref{fig:janka-nskick}). Gravitation as a
long-range interaction is very efficient in accelerating the NS
over time scales of several seconds and therefore much longer than
asymmetric accretion is likely to last after the SN blast has
taken off. Even only a small hemispheric mass-ejection difference
$\Delta m$, expelled on the one side and missing on the other 
in an otherwise homogeneous ejecta shell, can yield an
appeciable NS kick, $v_\mathrm{ns}$, by its asymmetric gravitational
attraction \citep[``gravitational tug-boat mechanism'' for
NS acceleration;][]{Wongwathanaratetal2013}\index{gravitational tug-boat mechanism}:
\begin{equation}
v_\mathrm{ns} \approx \frac{2G\Delta m}{r_\mathrm{i}\,v_\mathrm{s}}
\approx 540\,\frac{\mathrm{km}}{\mathrm{s}}\,\,
\left(\frac{\Delta m}{10^{-3}\,M_\odot}\right)
\left(\frac{r_\mathrm{i}}{100\,\mathrm{km}}\right)^{\!-1}
\left(\frac{v_\mathrm{s}}{5000\,\mathrm{km/s}}\right)^{\!-1} ,
\label{eq:kickest}
\end{equation}
where $r_\mathrm{i}$ and $v_\mathrm{s}$ are the initial radius and
the expansion velocity of the shell, respectively.
A hemispheric mass-ejection asymmetry as small 
as $10^{-3}\,M_\odot$, expanding with a
constant velocity of 5000\,km\,s$^{-1}$, can thus pull the NS to a
velocity of 540\,$\mathrm{km\,s}^{-1}$. 

3D simulations by \citet{Wongwathanaratetal2010,Wongwathanaratetal2013}
showed that the explosion asymmetries that grow naturally from 
small initial seed perturbations due to convective and SASI 
activity in the postshock layer (for an example, see
Fig.~\ref{fig:janka-snlongtime}) are sufficiently large 
to explain NS kicks\index{neutron-star kick} of
several 100\,km\,s$^{-1}$ and, most likely, even beyond 
1000\,km\,s$^{-1}$ 
\citep[as already obtained in 2D simulations by][]{Schecketal2006}. 
Neutrino-driven
SN explosions are therefore able to explain the observed 
space velocities of typically $\sim$200--500\,km\,s$^{-1}$ of 
the majority of young pulsars
\citep[see, e.g.,][]{Arzoumanianetal2002,Hobbsetal2005}.

\citet{Janka2016} argued that because of momentum conservation
a simple relation exists between the
NS kick\index{neutron-star kick} velocity on the one hand and
the momentum-asymmetry of the ejecta, $\alpha_\mathrm{ej}$,
and the explosion energy, $E_\mathrm{exp}$, of the SN on
the other hand:
\begin{equation}
v_\mathrm{ns} = 211\,\frac{\mathrm{km}}{\mathrm{s}}\,\,\zeta\,
\left(\frac{\alpha_\mathrm{ej}}{0.1}\right)
\left(\frac{E_\mathrm{exp}}{10^{51}\,\mathrm{erg}}\right)
\left(\frac{M}{1.5\,M_\odot}\right)^{-1} ,
\label{eq:vns}
\end{equation}
where $\zeta$ is a numerical factor of order unity.
Equation~(\ref{eq:vns}) means that the
NS kick\index{neutron-star kick} grows roughly linearly with the
explosion energy (or, alternatively, with the relevant
ejecta mass, $M_\mathrm{ej}$)
and with the explosion asymmetry
$\alpha_\mathrm{ej}$. Both dependences are easy to
understand: a more asymmetric and more powerful explosion
is able to impart a larger kick to the NS.
The momentum asymmetry parameter $\alpha_\mathrm{ej}$ is
determined by the 
stochastic growth of hydrodynamic instabilities in the
postshock layer, which trigger the onset of an asymmetric
explosion. Kick velocities in excess of 1000\,km\,s$^{-1}$
require $\alpha_\mathrm{ej} \gtrsim 0.5$ for all other
factors in Eq.~(\ref{eq:vns}) being unity, which is a rare
case but within reach of some published explosion models
\citep[e.g.,][]{Schecketal2006,Wongwathanaratetal2013}.

Since in the described hydrodynamical kick scenario the NS 
velocity vector points away from the stronger explosion, 
3D models predict enhanced explosive 
nucleosynthesis\index{nucleosynthesis} 
in the hemisphere opposite to the direction of the 
NS kick\index{neutron-star kick}
\citep{Wongwathanaratetal2013}. This concerns
mostly chemical elements between $^{28}$Si and the 
iron-group, which are significantly or exclusively
produced during the SN outburst in the innermost
shock-heated and neutrino-heated ejecta. The hemispheric
asymmetry of the element distribution increases with the
magnitude of the NS kick. Indeed, high-resolution mapping
with the NuSTAR X-ray telescope reveals that the spatial
distribution of $^{44}$Ti in the 
Cassiopeia~A\index{Cassiopeia~A} SN remnant
is fully compatible with this prediction
\citep{Grefenstetteetal2014}. Inspecting the observational
information of $^{44}$Ti and iron in combination, the 
geometry and many morphological features of this
young gas remnant exhibit amazing resemblance to results
that can be obtained in 3D simulations of neutrino-driven
explosions \citep[see Fig.~\ref{fig:janka-casa} 
and][]{Wongwathanaratetal2016}.

\begin{figure}[!]
\sidecaption[t]
\includegraphics[scale=.25]{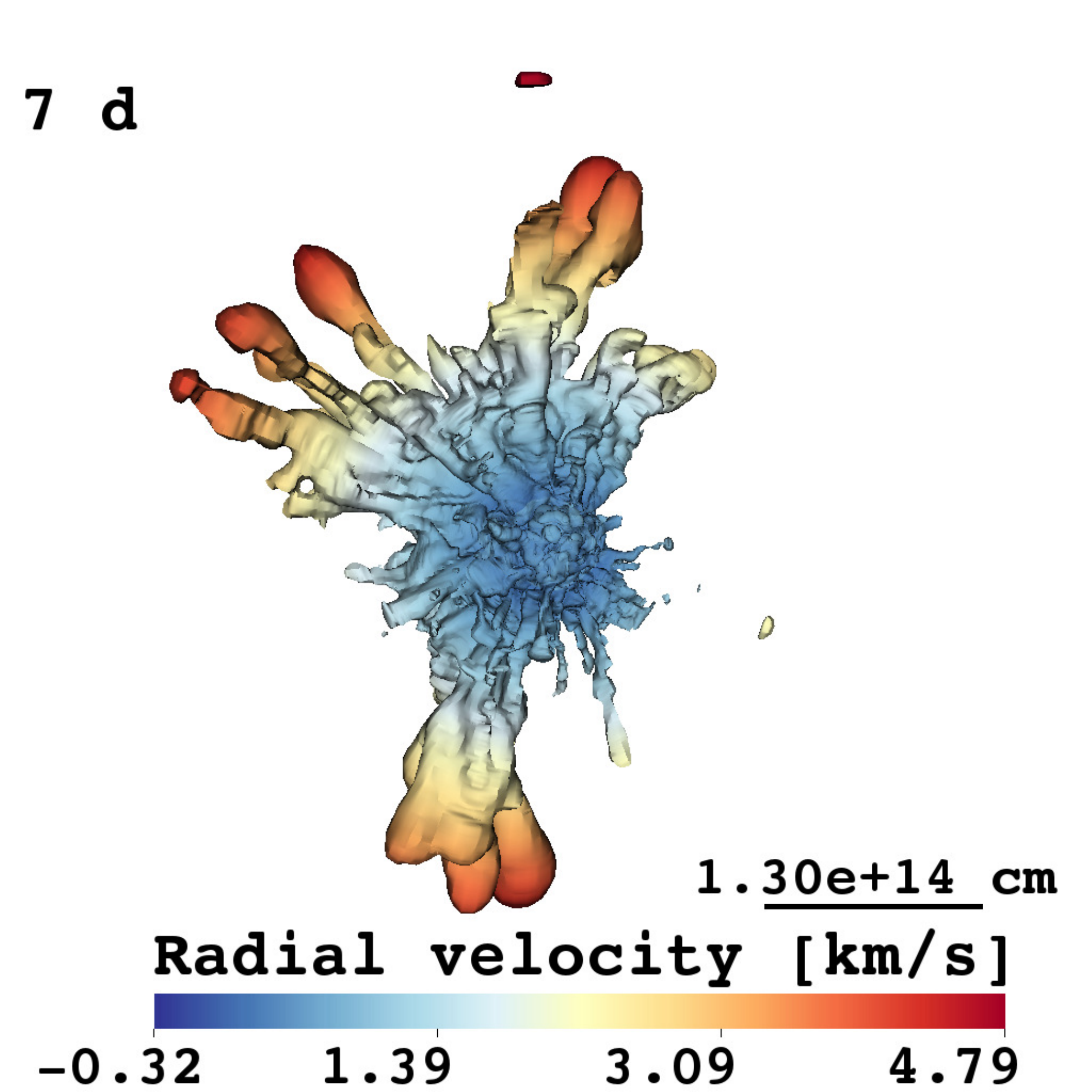}\hspace{2pt}
\includegraphics[scale=.25]{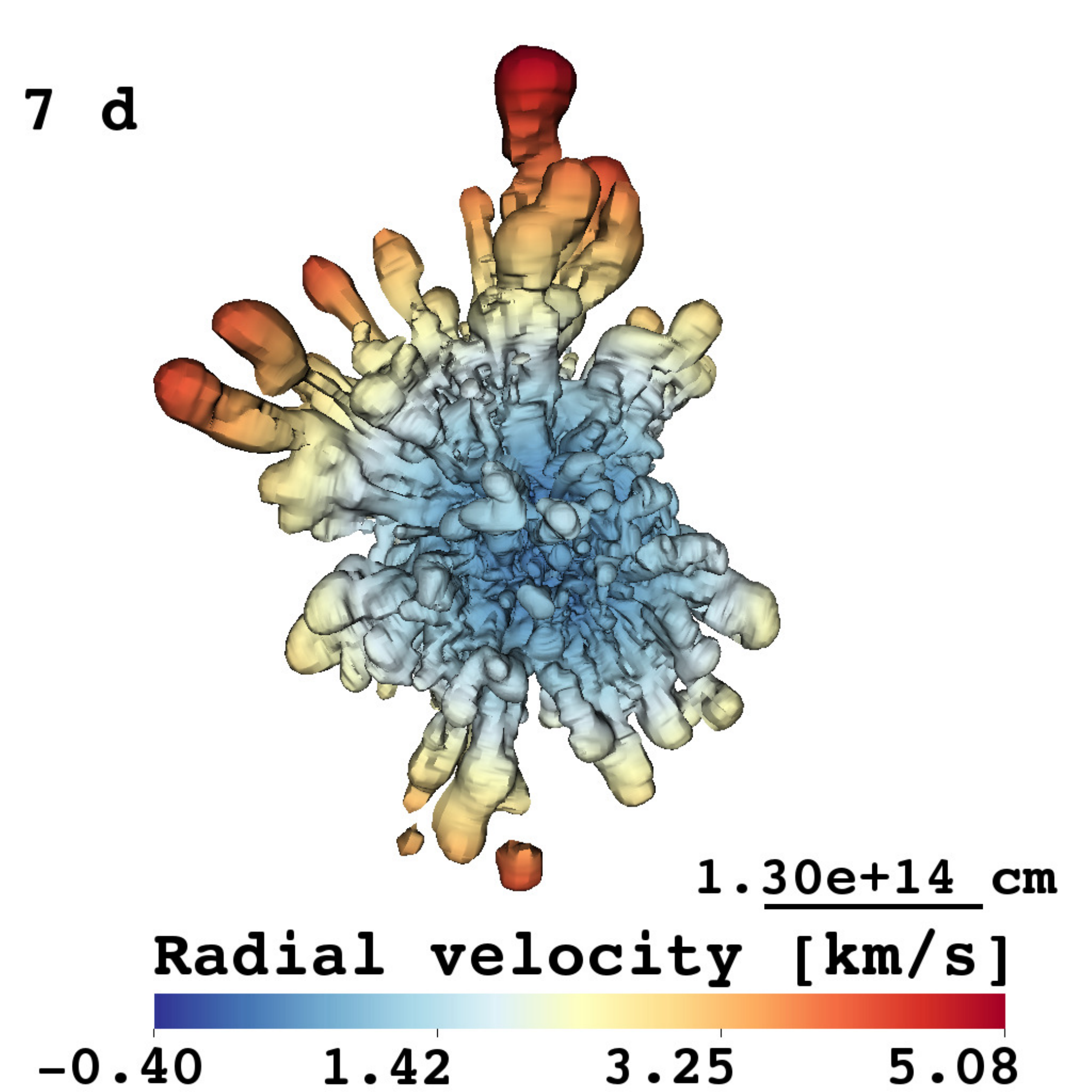}\\
\includegraphics[scale=.25]{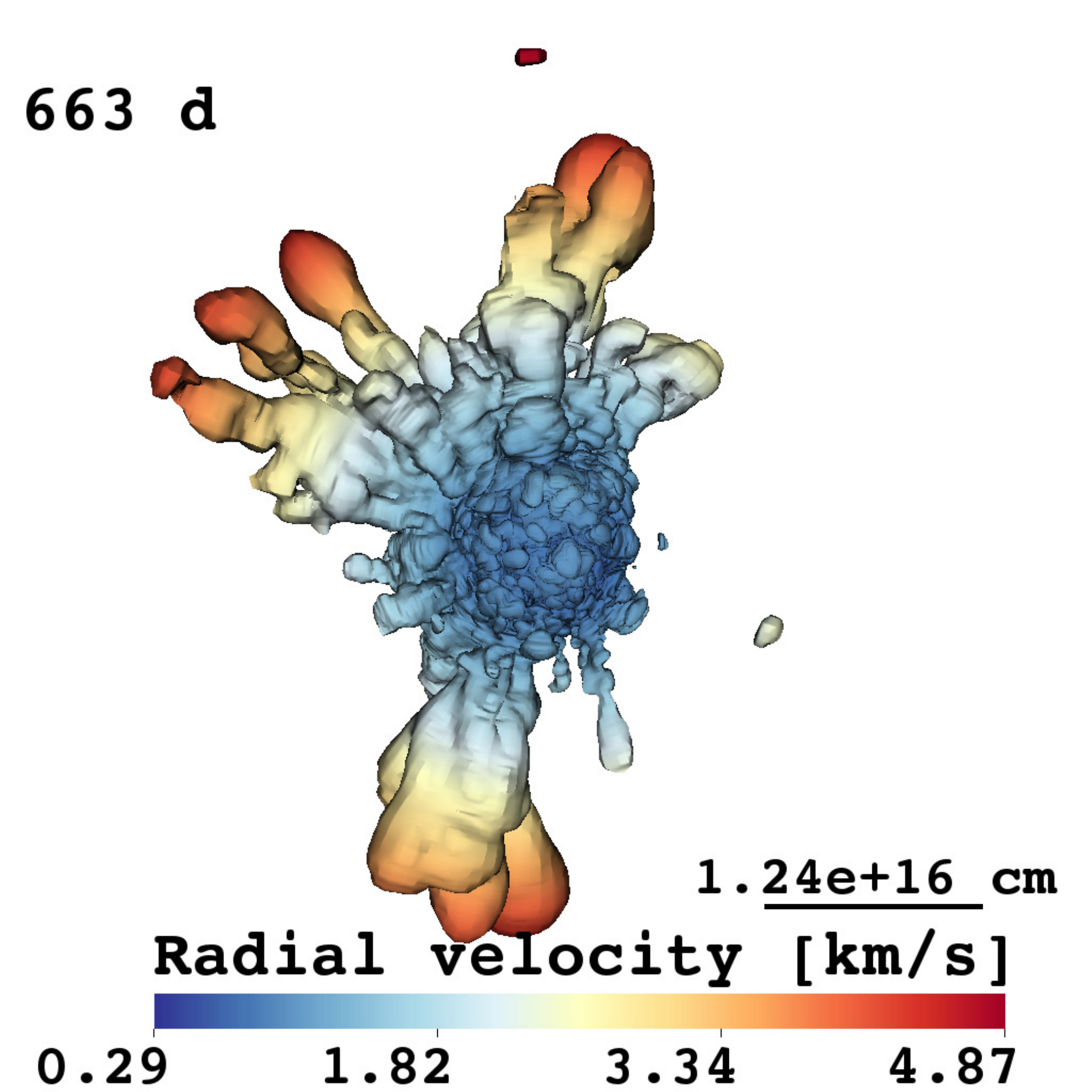}\hspace{2pt}
\includegraphics[scale=.25]{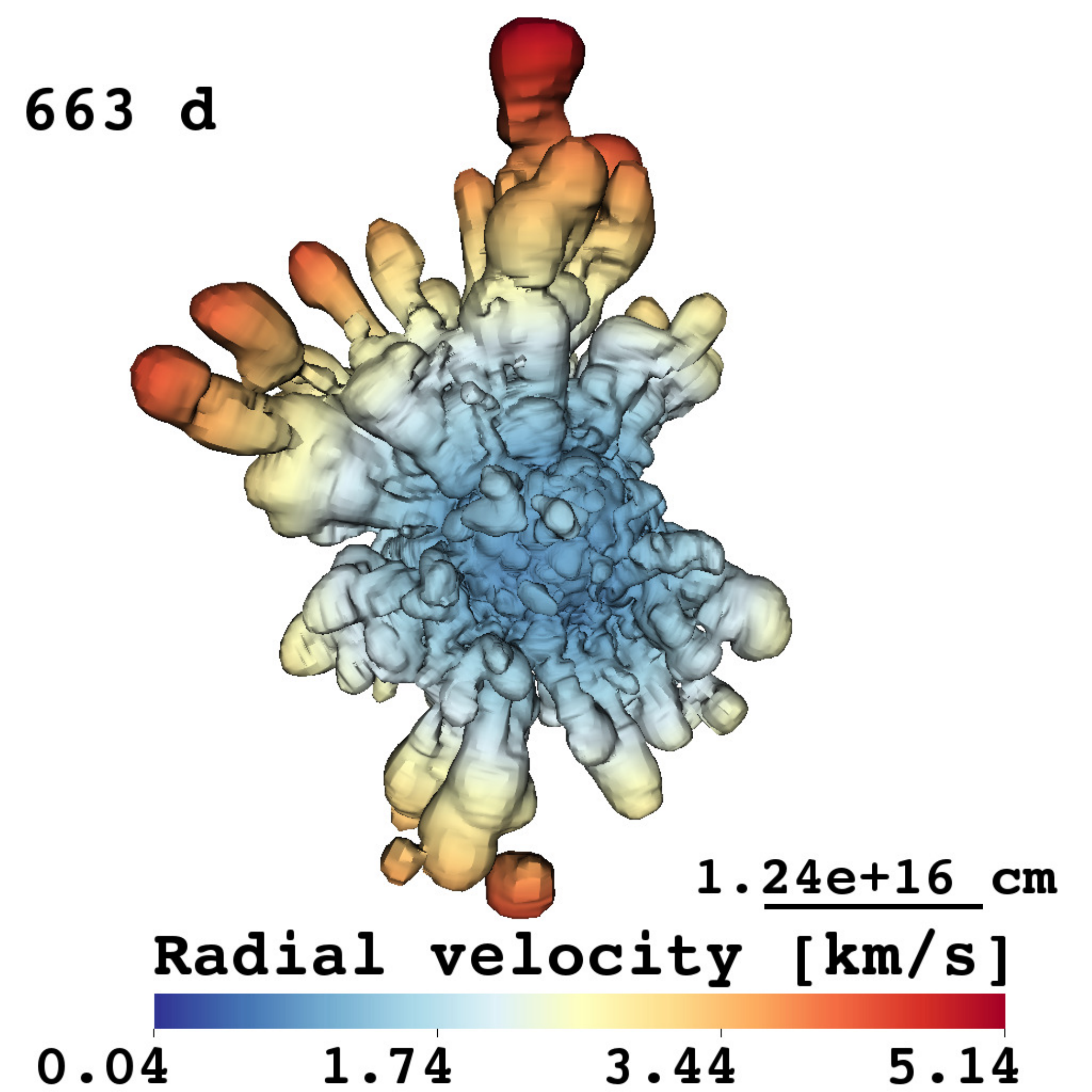}
\caption{The first two years of a Type IIP SN. The images are 
from a 3D simulation that continues the neutrino-driven
explosion model shown in Fig.~\ref{fig:janka-snlongtime}
from shock breakout to almost two years later.
Isosurfaces of constant mass fractions of 3\% for $^{56}$Ni
({\em left}) and 10\% for $^{16}$O ({\em right}) are shown
(analog to Fig.~\ref{fig:janka-snlongtime}). The color coding
denotes radial velocities.
{\em Top:} Structure after the red supergiant has reached
the phase of homologous expansion. Notice the
significant changes of the geometrical structure and velocity 
distribution compared to the moment of
shock breakout in the bottom images of 
Fig.~\ref{fig:janka-snlongtime}.
{\em Bottom:} After two years of radioactive heating by
$^{56}$Ni decay to stable $^{56}$Fe. There is minor boosting
of the fastest ejecta in the most extended fingers, but quite 
significant long-time acceleration of the bulk of the ejecta 
at velocities below $\sim$2000\,km\,s$^{-1}$.
(Figure courtesy of Michael Gabler)
}
\label{fig:janka-snlongtime2}
\end{figure}

\subsection{Explosion Asymmetries and Large-scale Mixing}
\label{sec:expasymm}

Hydrodynamic instabilities\index{instability!hydrodynamic}
do not only play a crucial role 
during the first second of neutrino-driven SN explosions
but also when the shock wave moves out from the stellar
core through the mantle and envelope of the disrupted star.
Large-scale radial mixing is induced by the growth of 
Rayleigh-Taylor (RT)\index{instability!Rayleigh-Taylor}
mushrooms at the C-O/He and He/H 
composition-shell interfaces\index{composition-shell interface}
after the passage of the shock.
These mixing instabilities destroy the onion-shell structure
of the progenitor star by carrying 
radioactive nuclei\index{radioactive nuclei} and
iron-group elements from the innermost layers into the 
helium and hydrogen shells and, reversely, displace
hydrogen from the fast stellar envelope inward into the 
slowly expanding metal core. The presence of such transport
processes was for the first time concluded from observations
of SN~1987A\index{SN~1987A}, 
where it was needed to explain the unexpectedly
early detection of X-rays and $\gamma$-rays from the decay
of radioactive isotopes, the smoothness and large width of 
the light-curve\index{light curve}
peak, and a variety of velocity-dependent
spectral-line features that pointed to 
clumpiness\index{ejecta!clumpiness} and
inhomogeneities in the SN ejecta
\citep[for a review, see][]{Arnettetal1989}.

The growth of the RT instabilities near the C-O/He and He/H
composition-shell interfaces is triggered by crossing
density and pressure gradients that are associated with local
density maxima in the postshock flow \citep{Chevalier1976}. 
These build up when the shock decelerates in the 
flattening density profiles within composition shells after
phases of acceleration in regions with steeper density 
gradients near the shell interfaces\index{composition-shell interface}. 
While the outer edges of the density maxima become RT
unstable\index{instability!Rayleigh-Taylor}, their inner edges
steepen into strong reverse shocks\index{shock!reverse}, 
which decelerate and
compress the neutrino-heated and metal-rich inner ejecta
when encountering them \citep{Kifonidisetal2003}.

\citet{Kifonidisetal2006}
showed that the RT growth rate is considerably boosted 
by the initial asymmetries that support the onset of
neutrino-driven explosions, and \citet{Hammeretal2010}
and \citet{Wongwathanaratetal2015} demonstrated,
performing 3D simulations of neutrino-driven explosions
continuously from core bounce until shock 
breakout\index{shock!breakout} at the
stellar surface, that this interaction of initial and
secondary instabilities can facilitate the penetration of
$^{56}$Ni and $^{44}$Ti with high velocities 
(up to more than 4000\,km\,s$^{-1}$ for the fastest clumps)
deep into the hydrogen envelope\index{hydrogen envelope}
as well as inward mixing of significant amounts of 
hydrogen to velocities as low as $\sim$100\,km\,s$^{-1}$.
For blue supergiant\index{blue supergiant}
progenitors with compact, small
helium cores ($\sim$4--4.5\,$M_\odot$) both effects
in combination are efficient enough to produce a good 
match of the wide, dome-like shape of the 
light-curve\index{light curve}
maximum observed for SN~1987A\index{SN~1987A} 
\citep{Utrobinetal2015}.

\citet{Wongwathanaratetal2015}, however, found that the 
interaction of initial and 
secondary instabilities\index{instability!secondary} 
depends extremely sensitively, and in a subtle way,
on the detailed density 
structure of the progenitor star, which determines the
acceleration and deceleration phases of the outgoing
shock and therefore the RT growth rates in the unstable
layers as well as the time when the 
reverse shocks\index{shock!reverse} 
collide with the material of the expanding metal core.
While especially the reverse shock from the He/H interface
decelerates the metal-rich ejecta, the secondary RT
instabilities enable mixing at the shell interfaces
and lead to a fragmentation of the large-scale asymmetries
of the early explosion into smaller fingers and filaments
(see Fig.~\ref{fig:janka-snlongtime}).

Nevertheless, the largest and most powerful plumes created
at the beginning of the neutrino-driven explosion trigger
also the strongest growth of the RT instabilities at the
composition-shell interfaces, and thus they shape the final
morphology of the SN explosion and of the developing
SN remnant\index{supernova remnant} 
(see Figs.~\ref{fig:janka-casa} and \ref{fig:janka-snlongtime2}).
The initial explosion asymmetries that are intrinsically
connected to the neutrino-driven explosion mechanism
must therefore be included in self-consistent hydrodynamic
3D simulations of the SN blast wave from its initiation to 
the late stages of homologous expansion, if SN models
shall be compared to the observational properties of SN 
explosions and SN remnants. These initial asymmetries
can either grow stochastically from (small-scale) random
fluctuations, or they might be triggered by large-scale
asymmetries in the convective silicon and oxygen 
burning\index{convective burning} 
shells of the progenitor at the onset of core collapse
\citep{BurrowsHayes1996,ArnettMeakin2011,CouchOtt2013,MuellerJanka2015,Muelleretal2016,Mueller2016}.

\begin{figure}[!]
\sidecaption
\includegraphics[scale=0.225]{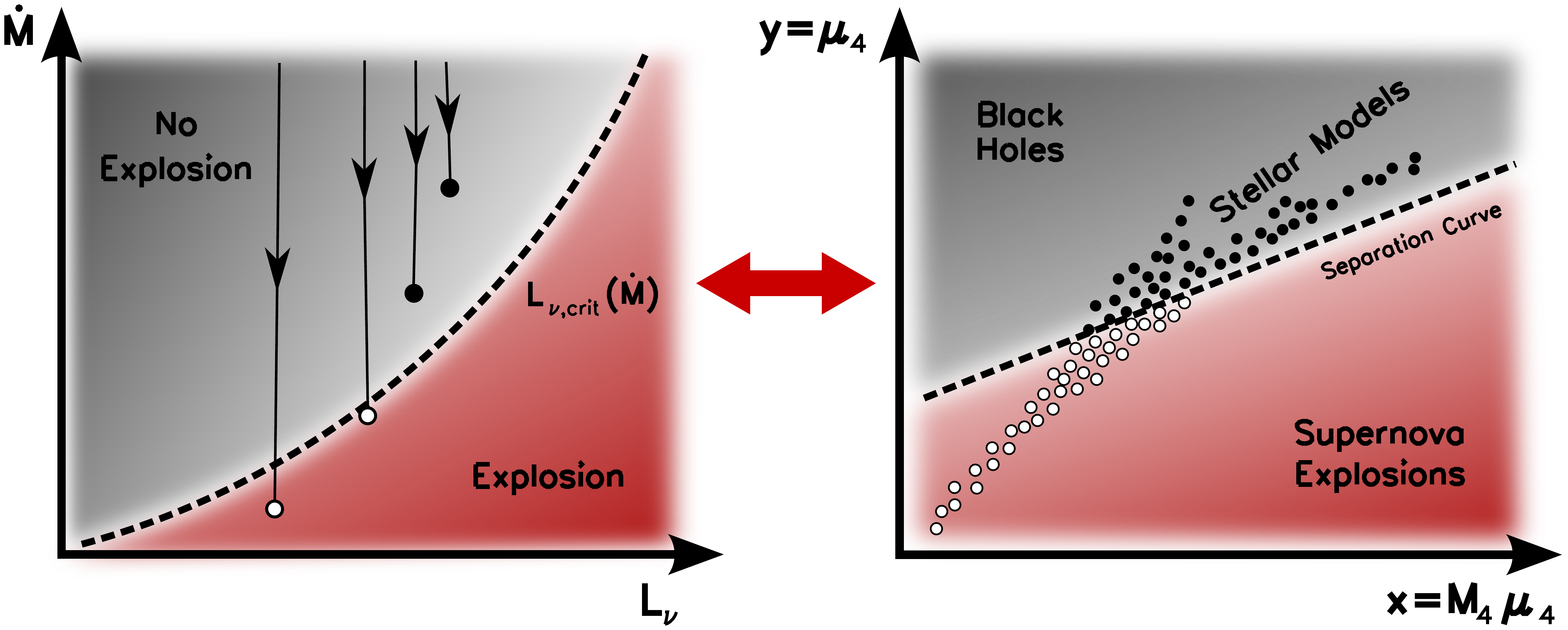}
\caption{Two-parameter criterion for the ``explodability'' of
stars. There is a direct correspondence between the 
$L_\nu$-$\dot M$ plane with the critical neutrino-luminosity
curve $L_{\nu,\mathrm{crit}}(\dot M)$ ({\em left}) and the
2D-plane spanned by the progenitor-specific parameters of
$M_4\mu_4$ and $\mu_4$, with a separation line between 
exploding and non-exploding cases ({\em right}).
In the left plot post-bounce evolution
paths of successfully exploding models (white circles) and 
non-exploding models (black circles) are schematically indicated,
corresponding to white and black circles for pre-collapse 
stellar models in the right plot.
Evolution paths of successful cases cross the critical line 
at some point and the accretion ends after the explosion has
taken off. In contrast, the tracks of failing cases never
reach the critical condition for launching a runaway 
expansion of the shock. The symbols in the left plot
mark ``optimal points'' relative to the critical curve,
corresponding to the stellar conditions described by the
parameters $M_4$ and $\mu_4$ at the radius where the dimensionless
entropy per baryon is $s=4$. This location is usually 
decisive for the successful explosion or failure of a 
progenitor, because the accretion rate drops strongly outside.
\citep[Figure from ][$\copyright$ The American Astronomical 
Society]{Ertletal2016}
}
\label{fig:janka-2parameters}
\end{figure}

\subsection{Progenitor-explosion-remnant Connection}
\label{sec:progexprem}

The compactness parameter\index{compactness parameter}
$\xi_M$ (Eq.~\ref{eq:compactness}) 
of the progenitor core ($M$ is typically chosen between 
1.5\,$M_\odot$ and 2.5\,$M_\odot$)
is able to provide crude information about the destiny of 
a collapsing massive star: explosion as a SN for low 
compactness or collapse to a black hole (BH) for compactness
values above a certain threshold \citep{OConnorOtt2011}.
However, neither the best choice of this threshold value 
nor of the mass $M$ is a priori clear. Also other criteria
such as the envelope binding energy 
\citep[e.g.\ at 1.5\,$M_\odot$;][]{Burrowsetal2016}
or the mass derivative of the binding energy outside
2\,$M_\odot$ \citep{Sukhboldetal2016} may serve as 
similarly good single-parameter criteria for the stellar
``explodability''.

Another problem of the compactness parameter as such a 
criterion is the fact that its physical significance for
the readiness of a star to explode is not obvious. A large
compactness value implies a high mass accretion rate
$\dot M$ of the new-born NS. But, as discussed in
Sect.~\ref{sec:massivesne}, a high mass accretion rate 
has two competing effects, namely, a large 
ram pressure\index{ram pressure} ahead
of the shock, which impedes shock expansion, and a
high accretion luminosity of $\nu_e$ and $\overline{\nu}_e$,
which enhances the heating behind the shock and increases
the chance of an explosion. The 
compactness\index{compactness parameter} $\xi_M$ as
a single-parameter criterion does not reflect this 
ambiguity.

For these reasons \citet{Ertletal2016} introduced a new
criterion involving two parameters that are able to 
capture the two rivalling effects as the underlying 
physics of the neutrino-driven explosion mechanism.
The first parameter is the normalized mass inside a
dimensionless entropy per nucleon of $s=4$, 
\begin{equation}
M_4\equiv m(s=4)/M_\odot\,,
\label{eq:ms4}
\end{equation}
which usually is close to the inner boundary of the
silicon-enriched oxygen shell\index{oxygen shell!silicon-enriched}. 
When this location with its
density and entropy jump arrives at the stalled shock during
core collapse, an explosion is most likely to set in.
The second parameter is the mass derivative at this 
radial position,
\begin{equation}
\mu_4\equiv\left.\frac{\mathrm{d}m/M_\odot}{\mathrm{d}r/1000\,\mathrm{km}}\right|_{s=4}\,.
\label{eq:dmdr}
\end{equation}
Both of the parameters $M_4$ and $\mu_4$ are determined 
from the pre-collapse structure of the progenitor star.

When the progenitor mass shell $m(r)$ ($r$ measures the radius
in the progenitor) has collapsed to the stalled shock, one
can choose $M=m(r)$ as a suitable proxy of the NS mass,
because the layer between NS surface and shock contains only
little mass compared to the NS itself.
A rough measure of the mass-accretion rate $\dot M$ of the
shock is then given by the mass-derivative
$m'(r)\equiv\mathrm{d}m(r)/\mathrm{d}r=4\pi r^2\rho(r)$ at the
progenitor radius $r$. This is the case
because $m'(r)$ can be directly linked
to the free-fall accretion rate of matter collapsing into the 
shock from initial radius $r$ \citep{Ertletal2016}:
\begin{equation}
\dot M=\frac{\mathrm{d}m}{\mathrm{d}t_\mathrm{ff}}
=\frac{\mathrm{d}m(r)}{\mathrm{d}r}
\left(\frac{\mathrm{d}t_\mathrm{ff}(r)}{\mathrm{d}r}\right)^{-1}
\approx\frac{2}{3}\,\frac{r}{t_\mathrm{ff}}\,m'(r)\,,
\label{eq:macc}
\end{equation}
where $t_\mathrm{ff}=\sqrt{\pi^2\,r^3/[8G\,m(r)]}$ 
is the free-fall time scale\index{free-fall time scale}
\citep{WoosleyHeger2015b}.
Inspection of large sets of stellar models reveals an
extremely tight linear correlation between 
$m'(r)|_{s=4}\propto \mu_4$ and $\dot M$ at the 
time the $s=4$ interface falls to the 
accretion shock\index{accretion shock}.
This proves that $\dot M\propto m'(r)$ defines the main
dependence between the lhs and the rhs of Eq.~(\ref{eq:macc}),
while the factor $r\,t_\mathrm{ff}^{-1}\propto\sqrt{m(r)\,r^{-1}}
\propto\sqrt{\overline{\rho}\,r^2}$ (with $\overline{\rho}$
being the average density inside of $r$) has only
a secondary influence.
Similarly, at the time when $M_4$ arrives at the shock,
one finds an extremely tight linear relation
between $m(r)\,m'(r)|_{s=4}\propto M_4\mu_4$ and the $\nu_e$
plus $\overline{\nu}_e$ luminosity, $L_\nu$, of the NS,
for which the accretion luminosity\index{accretion luminosity} 
$L_\nu^\mathrm{acc}\propto G\,M\,\dot M\,R_\mathrm{ns}^{-1}$
accounts for the dominant contribution varying between
different progenitors.
This means that the product $M\,\dot M\propto M_4\mu_4$
captures the main progenitor dependence 
of the neutrino emission. The dependence on the NS radius
is weak since $R_\mathrm{ns}$ varies weakly between
different progenitors at the time when the explosion sets
in \citep{Ertletal2016}. 
 
According to the concept of a 
critical luminosity\index{critical luminosity} limit
$L_{\nu,\mathrm{crit}}(\dot M)$ for stationary accretion 
solutions (Eq.~\ref{eq:lcrit0} and Sect.~\ref{sec:critlum}),
exploding models must pass this threshold luminosity for
shock runaway, while non-exploding models should stay below
this limit (Fig.~\ref{fig:janka-2parameters}, left panel;
see also Fig.~\ref{fig:janka-criticalcurve}, middle panel).
Because at the onset of the explosion tight correlations 
exist between $L_\nu$ and $M_4\mu_4$ on the one hand and
$\dot M$ and $\mu_4$ on the other hand, it may be expected that 
the critical luminosity curve $L_{\nu,\mathrm{crit}}(\dot M)$
has a correspondence in the $M_4\mu_4$-$\mu_4$ plane
(see Fig.~\ref{fig:janka-2parameters}).
Exactly such a separation line between exploding and 
non-exploding models is found in the huge set of hundreds 
of hydrodynamical SN simulations performed by 
\citet{Ertletal2016}. 
A fraction of 97\% of all models were found to be
classified correctly by the 
two-parameter criterion\index{two-parameter criterion} with
respect to their explosion behavior. A similarly high 
success rate was reported by \citet{Muelleretal2016B}, 
using a completely different, analytical 
approach for modeling the SN explosions including a
parametric description of multi-dimensional effects.

\begin{figure}[b]
\sidecaption
\includegraphics[scale=0.70]{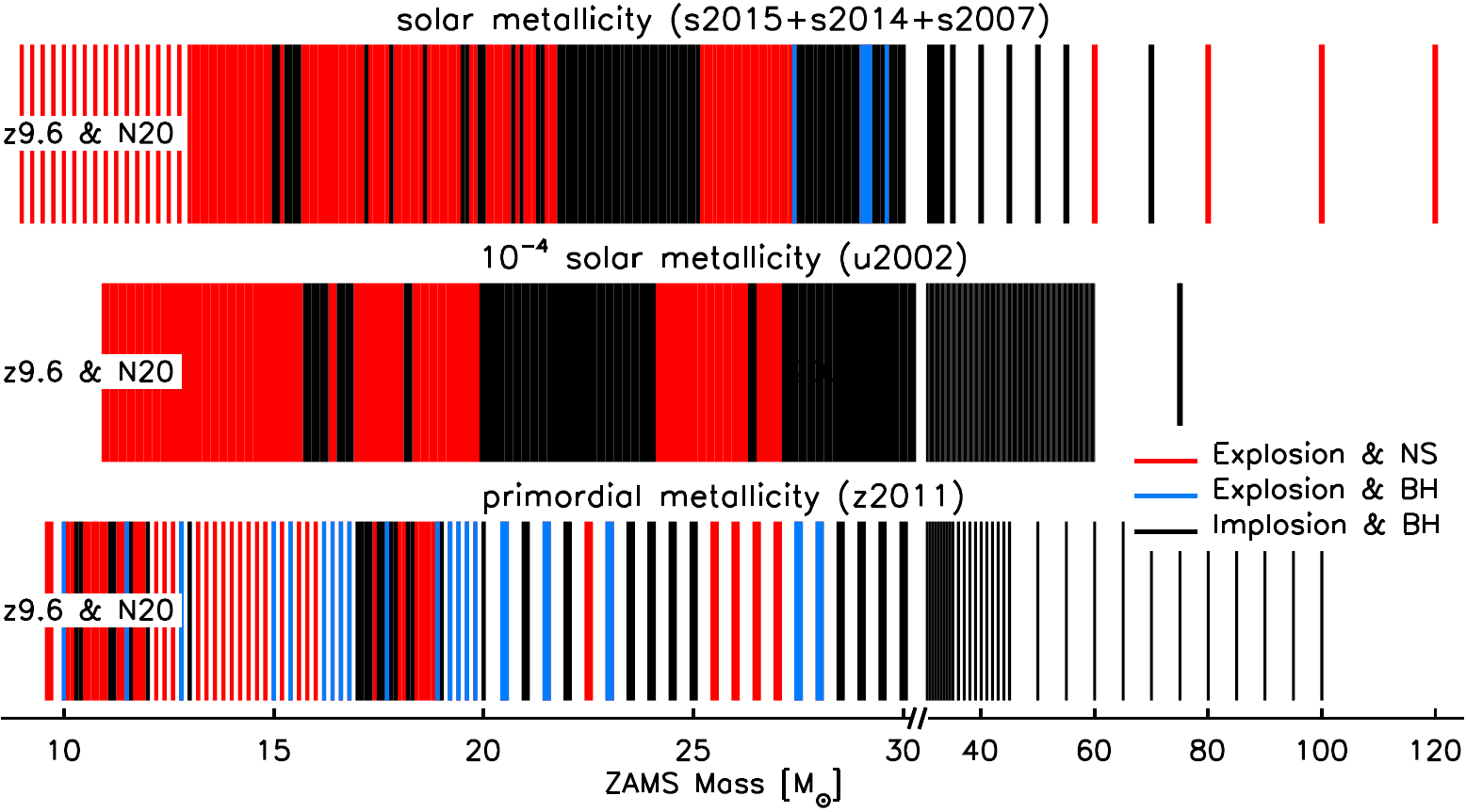}
\caption{NS and BH formation cases as function of progenitor
ZAMS mass, based on 1D simulations with a calibrated neutrino
``engine'' \citep[for more details of the modeling approach,
see][]{Ertletal2016,Sukhboldetal2016}.
The {\em upper row} displays results for the compilation of
solar-metallicity progenitors used by \citet{Sukhboldetal2016},
the {\em middle row} ultra metal-poor ($10^{-4}$
solar metallicity) models (set u2002) between 11.0\,$M_\odot$
and 75.0\,$M_\odot$ from \citet{Woosleyetal2002}, and
the {\em bottom row} zero-metallicity models (set z2011)
between 9.6\,$M_\odot$ and 100.0\,$M_\odot$
from \citet{HegerWoosley2010} for the stars above and including
10.3\,$M_\odot$ and from A.~Heger (2015, private communication)
for the stars with lower masses. Red vertical bars indicate
successful explosions with NS formation, black bars BH formation
without SN explosion, and blue bars fallback SNe where BHs form
due to massive fallback, which leads to more than 3\,$M_\odot$
of baryonic matter in the compact remnant. The rugged landscape
of alternating intervals of NS and BH formation events is a 
consequence of non-monotonicities in the pre-collapse structure
of the progenitors as discussed in Sect.~\ref{sec:stellarcores}.
(Figure courtesy of Thomas Ertl)
}
\label{fig:janka-ns+bh}
\end{figure}

\begin{figure}[!]
\sidecaption
\includegraphics[scale=0.70]{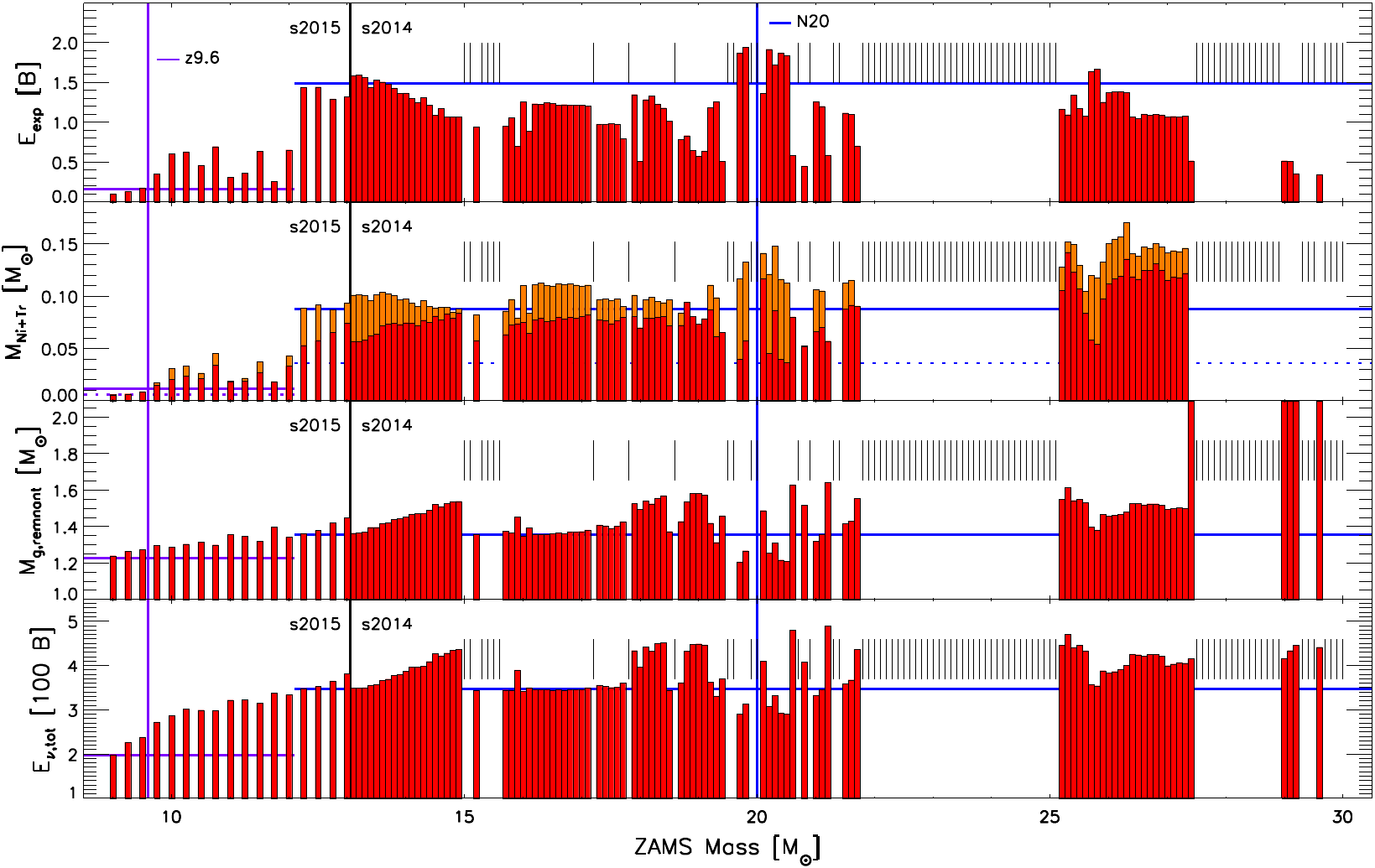}
\caption{Explosion and remnant properties as functions of progenitor
ZAMS mass for the compilation of single-star, solar-metallicity
progenitor models considered by \citet{Sukhboldetal2016}.
The results are based on 1D simulations with a calibrated neutrino
``engine'' \citep[for more details, see][]{Ertletal2016,Sukhboldetal2016}.
{\em Top panel:} explosion energy in bethe ($1\,\mathrm{B} = 10^{51}$\,erg);
{\em second row:} explosively produced and expelled 
$^{56}$Ni in shock-heated
and neutrino-heated ejecta (red) with uncertainty range indicated
by orange segments of the histogram bars;
{\em third row:} gravitational mass of the NS;
{\em bottom:} total energy release of the new-born NS in neutrinos.
In the $^{56}$Ni mass and NS mass fallback is taken into account.
The vertical black lines in the upper part of
each panel indicate cases of direct BH formation
without explosion. Five low-energy explosions above 27\,$M_\odot$
are fallback SNe that produce BHs by massive fallback (see also
Fig.~\ref{fig:janka-ns+bh}).
(Figure courtesy of Thomas Ertl)
}
\label{fig:janka-landscape}
\end{figure}

The SN simulations for the large grid of progenitors
of different 
masses between 9\,$M_\odot$ and 120\,$M_\odot$ (partly in
steps of 0.1\,$M_\odot$) and different metallicities were
not computed in 3D, which is presently not feasible. Instead,
the calculations were performed in spherical symmetry (1D),
using a parametrized neutrino ``engine''\index{neutrino engine}
that attempts to 
reproduce the essential physics of the neutrino-driven 
mechanism. The values of free parameters in the engine
model were calibated such that basic observational properties
(explosion energies and nickel yields) of the well studied 
cases of SN~1987A\index{SN~1987A}
on the high-mass side (testing progenitors
in the 15--20\,$M_\odot$ range) and of 
Crab\index{Crab} near the low-mass
end ($\sim$9--10\,$M_\odot$) were reproduced with suitable
progenitor models; in the case of SN~1987A also consistency
with constraints set by the neutrino detection was requested
\citep[see][]{Uglianoetal2012,Ertletal2016,Sukhboldetal2016}.

Applying this neutrino-engine treatment\index{neutrino engine}
to the progenitor
sets for different metallicities, a variegated landscape 
is obtained, with islands of non-exploding stars alternating
with intervals of successful explosions
(Figs.~\ref{fig:janka-ns+bh} and \ref{fig:janka-landscape}).
This astonishing result is, on the one hand, connected to the
non-monotonic variations of the pre-collapse structure with
the ZAMS mass\index{zero-age-main-sequence (ZAMS) mass}, 
which is reflected by the core 
compactness\index{core compactness}
(see Sect.~\ref{sec:stellarcores}).
On the other hand it is also a consequence of the tight 
competition between shock-confining ram pressure and
shock-pushing neutrino heating, which is characteristic of
the neutrino-driven mechanism and which makes successful
explosions sensitive to differences in the time-dependent
mass-accretion rate as discussed in
Sect.~\ref{sec:competition}. It is reassuring that this
result is not specific to the 1D explosion
modeling of \citet{Uglianoetal2012} and \citet{Ertletal2016}, 
but a rugged landscape was also found by 
\citet{PejchaThompson2015} for one of their
model sets and a different approach for determining 
successful cases. Moreover, a similar behavior was also observed
by \citet{Muelleretal2016B}, who employed a parametric-analytical
description for the effects of simultaneous mass accretion
and mass ejection as a genericly multi-dimensional 
phenomenon during the post-explosion phase.
Because they studied a ten times
finer grid of progenitors (i.e., models with a spacing of 
only 0.01\,$M_\odot$), they did not find mass intervals
of pure BH formation\index{black-hole!formation}
below 20\,$M_\odot$ but reported
considerable fluctuations in the probability of exploding
models within mass intervals of 0.5\,$M_\odot$ width.
This lends support to the suggestion by 
\citet{Clausenetal2015} of a probabilistic description 
of NS versus BH formation.

All of these results allow for the interesting conclusion
that BH birth seems to be possible not only in the collapse
of progenitor stars above a certain mass limit,
but it can occur with significant probability
even below a ZAMS mass of 20\,$M_\odot$. BH formation 
therefore seems to be more likely than previously thought,
maybe as frequent as 25--45\% of all core-collapse events
\citep{Sukhboldetal2016}.

Of course, it is by no means clear that such an
extreme sensitivity of the explosion to details of the
progenitor structure and time-dependent mass-accretion rate will 
survive in fully self-consistent
3D simulations, nor is it clear whether the corresponding
pre-collapse properties are a robust feature of the progenitor
evolution, whose modeling is still performed in 1D with
parametric recipes to treat convection and its associated
effects.

The explosion calculations for these large sets of stellar models
based on the neutrino-driven mechanism allow for many interesting
conclusions on the progenitor-explosion-remnant systematics.
Comparison of such results with observations can serve as
viability check for progenitors as well as explosion physics
\citep[for details, see][]{Uglianoetal2012,PejchaThompson2015,Ertletal2016,Sukhboldetal2016,Muelleretal2016B}.

Since the explosion energy scales with the mass of
neutrino-heated ejecta (see Sect.~\ref{sec:expenergy}) 
and determines the mass of shock-heated ejecta, neutrino-driven
explosions naturally yield a fairly tight correlation 
between energy and $^{56}$Ni mass as suggested by observations
\citep{Sukhboldetal2016,Muelleretal2016B}. Moreover,
also the ejecta mass and the explosion energy show
indications of a correlation \citep[see][]{Muelleretal2016B}.
The predicted distribution of 
NS-birth masses\index{neutron star!birth mass} is, overall,
compatible with observed NS masses (which would have to be
converted to NS birth masses for a proper comparison).
Similarly, also the BH masses\index{black-hole!mass} 
from the models overlap with
the mass distribution of observed BHs in binaries, provided
one assumes the hydrogen envelope of the collapsing star
gets unbound because of the NS mass decrement associated
with neutrino losses before BH formation \citep{Sukhboldetal2016}.
The lack of observed cases between the maximum measured 
NS mass (around 2\,$M_\odot$)
and the minimum discovered BH mass ($\sim$5\,$M_\odot$) 
naturally emerges in the model sets
of \citet{Uglianoetal2012} and \citet{Ertletal2016},
because fallback SNe\index{fallback supernova}
(whose NSs would accrete solar
masses of fallback material and thus could fill the gap)
are extremely rare for the investigated 
solar-metallicity progenitors (their relative importance
is considerably higher only in the metal-free progenitor
set; see 
Figs.~\ref{fig:janka-ns+bh} and \ref{fig:janka-landscape}).

\citet{Sukhboldetal2016} found at most 6--8\% of the
SN~IIP explosions to be connected to progenitors more massive
than 20\,$M_\odot$ when they used SN~1987A-calibrated neutrino 
engines that allow for population-integrated SN nucleosynthesis 
yields to be in overall agreement with chemogalactic constraints. 
This may offer an explanation for the rarity of observed 
high-mass SNe \citep[see][]{Smartt2015}.

These results for large sets of neutrino-driven explosions
in comparison to observations are assuring, but they are 
still based on rather simple, approximative
modeling recipes (either by 1D hydrodynamical or semi-analytic
treatments). Full 3D explosion simulations will be needed
for more solid conclusions, but tentative studies with 2D
calculations \citep{Nakamuraetal2015} should be taken with
great caution, because 2D modeling tends to produce artificial 
explosions (fostered by the presence of a symmetry axis)
and thus is prone to 
overestimating the readiness of stars to explode. Also
the single-star progenitor\index{single-star progenitor}
sets will ultimately have to
be supplemented by stellar models from 
binary evolution\index{binary evolution}.
Moreover, it is not clear whether rotation\index{rotation}
and magnetic fields\index{magnetic fields}
can be ignored for stars above $\sim$20\,$M_\odot$,
which rotate relatively faster at the onset of core collapse
\citep{Hegeretal2005} so that rotational field amplification 
during stellar collapse might lead to a non-negligible
role of magnetorotational 
effects\index{instability!magnetorational} in the SN explosions.

\section{Conclusions}

The theoretical understanding of the physics that plays
a role in the neutrino-driven explosion mechanism 
is meanwhile far advanced, and the picture how the 
diverse ingredients concur has consolidated
(Sect.~\ref{sec:mechanismphysics}).
Hydrodynamical simulations with increasingly
better treatment of the neutrino transport and other
relevant physics have become feasible
in full 3D, but they still require improvements
(Sect.~\ref{sec:multidmodels}).
With growing insight into the essential aspects that
play a role in the mechanism, also from the progenitor
side, predictions of observational consequences are
now possible by means of hydrodynamical explosion models
in different dimensions, but also by some (semi-)analytic
approaches (Sect.~\ref{sec:astroimplications}).

In spite of this remarkable progress, the 3D simulations
are not yet finally conclusive with respect to the 
possibility of explaining explosions for wider sets of 
progenitors, with respect to what 
magnitudes of explosion energies can be achieved by
the neutrino-driven mechanism, and with respect to
the reasons for
differences in the evolution of the stalled shock
observed by different groups. For sure, the present
generation of 3D simulations calls for further 
advancements. These include the need of better grid
resolution and a reduced influence of discretization
noise in the treatments by some groups; the 
replacement of remaining, different approximations
in the transport methods of all groups by more rigorous,
fully three-dimensional descriptions; upgrades of the
neutrino opacities to state-of-the-art implementations
in most of the currently applied 3D codes; and, last but
not least, a deeper understanding of the importance and
possible role of pre-collapse perturbations in the
convective burning shells for a wide range of 
progenitors \citep[for a review, see][]{Jankaetal2016}.

The long-lasting struggle for robust neutrino-driven
explosions\index{neutrino-driven explosion}, which should prove to 
be less sensitive to ``details'' of the input physics
or methods applied, still goes on. Because of the 
difficulties to obtain an ultimately satisfactory 
solution, some sceptics want to see growing
justification for their concerns that advocates 
of the mechanism are on the wrong track and a 
paradigm shift is needed.

However, there are arguments in favor of neutrino-driven
explosions that should be
taken seriously by opponents. There are considerable
successes of neutrino-driven SN models such as
the fully self-consistent explosions of low-mass
progenitors in the $\sim$9--10\,$M_\odot$ range
(Sect.~\ref{sec:lowmassexp}) as well as achievements in
reproducing, explaining, and
even predicting observed phenomena in SNe and SN remnants
(Sect.~\ref{sec:astroimplications}). There is also
a blatant lack of convincing alternatives
that could account for the far majority of SN explosions.
Magnetorotational explosions\index{magnetorotational explosion}, 
a hot candidate for some
sceptics of the neutrino mechanism, are excluded because
magnetic fields can reach dynamically relevant strength
only in the presence of spin rates in stellar cores
that are excluded by our present understanding of the
angular momentum evolution of stars based on stellar
models, observed spins of white dwarfs\index{white dwarf}
and NSs,
and astroseismological measurements. Stellar models as
well as the mentioned observations point to slow core 
rotation prior
to collapse for the far majority of SN progenitors.

In addition to these reasons that refer to solid facts,
there are some heuristic arguments that can be brought
forward in support of the neutrino mechanism.
Why is it so difficult to blow up massive stars? Is this
a serious concern and a fundamental problem of the 
neutrino mechanism, which disfavors its viability?
The answer is a clear ``no'' for the following reasons:
\begin{itemize}
\item
First, the existence of stellar-mass BHs, which was
splendidly confirmed by the first direct
measurements of gravitational waves\index{gravitational wave}, 
conveys a simple 
message: The SN mechanism \emph{is not ``robust''}, it must
fail in a significant number of cases, maybe more often
than previously expected.
\item
Second, the SN mechanism \emph{must be inefficient}. The 
reason is the huge discrepancy between the observationally
inferred explosion energy of core-collapse SNe and the 
gigantic amount of available energy that is released in 
the birth events
of NSs and BHs, which exceeds SN energies by more than a
factor of 100.
\item
Third, the SN mechanism \emph{must be self-regulated}, 
i.e., it must be connected to a sensible feedback effect
that quenches the energy input to the explosion as soon
as the energy of the expanding ejecta is roughly of the
order of the gravitational binding energy in the core of
the progenitor star (see Sect.~\ref{sec:expenergy}).
\end{itemize}

The neutrino mechanism is compatible with these 
requirements, different from the magnetorotational
mechanism\index{instability!magnetorotational}, 
whose energy scale is set by the available 
rotational energy of the newly formed NS instead of the
gravitational binding energy of the progenitor shells
that supply the neutrino mechanism with matter to be
heated (Sect.~\ref{sec:expenergy}). The absence of a
feedback mechanism\index{feedback mechanism} 
in the case of magnetorotational 
explosions\index{magnetorotational explosion}
can be concluded from the excessive
energy released in hypernova\index{hypernova} explosions 
(up to $\sim$$50\times 10^{51}$\,erg), and the dichotomy
between normal SNe and very rare, hyper-energetic
events points emphatically to two different explosion 
mechanisms at work in the two kinds of stellar blasts.
(Of course, there is no reason a priori why there should
not be a range of overlap, where both mechanisms in
synergy drive the explosions of stars.)

All these arguments, of course, are not rigorous and they
certainly cannot stand as a proof. But, at least, they 
make clear that the viability of the neutrino mechanism
cannot be rejected on grounds of plausibility arguments
that overinterpret and overstate still existing
weaknesses of the current numerical models.

Besides further work on the modeling side with the goal
to continuously upgrade the simulations for new physics
and better numerics, a further consolidation
of the paradigm of neutrino-driven explosions requires
more empirically testable predictions and the 
corresponding efforts on the observational side, even
before a next Galactic SN will provide measurable neutrino
and gravitational-wave\index{gravitational wave} signals. 
In the absence of a
single, pivotal observation or measurement that can
provide an undisputable proof,
one has to strive for as many tests as
possible for consistency between predictions derived
from neutrino-powered explosions and the events in nature.
Only such a collection of many individual pieces of this
highly complex phenomenon (including nucleosynthesis yields,
blast-wave energies, NS and BH mass distributions and kicks,
explosion asymmetries and mixing, progenitor-SN
connections, neutrinos and gravitational waves, etc.)
will permit a final breakthrough in assembling the
multifaceted picture of how massive stars explode.

\begin{acknowledgement}
The author is indebted to Thomas Ertl, Michael Gabler,
Alexander Summa, and Annop Wongwathanarat for providing
graphics used in this article, and to Elena Erastova and
Markus Rampp (MPCDF) for their great help in visualizing
3D simulation results. Careful reading of the chapter
and comments by Alexander Summa are acknowledged.
Research by the author was
supported by the European Research Council through an
Advanced Grant (ERC-AdG No.\ 341157-COCO2CASA), by
the Deutsche Forschungsgemeinschaft through
the Cluster of Excellence ``Universe'' (EXC-153), and
by supercomputing time from the European PRACE
Initiative and the Gauss Centre for Supercomputing.
\end{acknowledgement}
\section*{Cross-References}

\emph{Explosion Physics of Core-Collapse Supernovae}\\
\emph{Influence of Non-spherical Initial Stellar Structure on the
Core-collapse Supernova Mechanism}\\
\emph{Supernovae from Super Asymptotic Giant Branch Stars}\\
\emph{Supernovae from Massive Stars (12--100 Msun)}\\
\emph{Supernova Remnant Cassiopeia A}\\
\emph{Supernova remnant from SN1987A}\\
\emph{The Supernova -- Supernova Remnant Connection}\\
\emph{The Core-collapse Supernova -- Black Hole Connection}\\
\emph{Neutrino Emission from Supernovae}\\
\emph{Neutrino Signatures from Young Neutron Stars}\\
\emph{Neutrinos from Core-Collapse Supernovae and Their Detection}\\
\emph{Diffuse Neutrino Flux from Supernovae}\\
\emph{Gravitational Waves from Core-collapse Supernovae}

\bibliography{JankaReferences-file}

\end{document}